\documentclass[12pt]{article}
\pdfoutput=1
\usepackage{cite}
\usepackage{heck-jhep}
\usepackage{graphicx}
\usepackage{multicol}
\usepackage{amsmath}
\usepackage{amsfonts}
\usepackage{amssymb}
\usepackage[all]{xy}
\usepackage{hyperref}

\def\spc{\hspace{.5pt}}
\usepackage{amssymb, amsmath, amsopn, amsthm}
\usepackage{psfrag}
\usepackage{subfigure}

\usepackage{color}
\usepackage{epsfig}
\usepackage{graphicx}
\usepackage{epstopdf}
\usepackage{mathtools}
\usepackage{bbold}
\def\be{\begin{equation}}
\def\ee{\end{equation}}

\begin{document}

\setlength{\textwidth}{16.8cm}
\setlength{\textheight}{22.5cm}
\addtolength{\oddsidemargin}{-6mm}
\addtolength{\topmargin}{-2mm}
\addtolength{\baselineskip}{-.3mm}

\date{December 2014}
\title{\LARGE\bf  Conformal Bootstrap, Universality
  \\[-7mm] { and} \\[2mm]
Gravitational Scattering}

\institution{PU}{\centerline{Department of Physics, Princeton University, Princeton, NJ 08544, USA}}

\authors{Steven Jackson, Lauren McGough and Herman Verlinde\footnote{e-mail: {\tt srjackso@princeton.edu, mcgough@princeton.edu, verlinde@princeton.edu}}}

\def\spc{\hspace{.5pt}}


\abstract{We use the conformal bootstrap equations to study the non-perturbative gravitational scattering between  infalling and outgoing particles in the vicinity of a black hole horizon in AdS. We focus on irrational 2D CFTs with large $c$ and only Virasoro symmetry. The scattering process is described by the
matrix element of two light operators (particles) between two heavy states (BTZ black holes).
We find that the operator algebra  in this regime is (i) universal and identical to that of Liouville CFT, and (ii) takes the form of an exchange algebra, specified by an R-matrix that exactly matches with the scattering amplitude of 2+1 gravity. The R-matrix is given by a quantum 6j-symbol and the scattering phase by the volume of a hyperbolic tetrahedron. We comment on the relevance of our results to scrambling and the holographic reconstruction of the bulk physics near black hole horizons.}

\addtolength{\abovedisplayskip}{1mm}
\addtolength{\belowdisplayskip}{1mm}


\maketitle
\def\mathbi#1{\textbf{\em #1}} 
\def\som{{ \textit{\textbf s}}} 
\def\tom{{ \textit{\textbf t}}} 
\def\nom{{ \textit{\textbf n}}} 
\def\mom{{ \textit{\textbf m}}} 
\def\kom{{ \textit{\textbf k}}}  
\def\nomt{{ \textit{\textbf n}}}  
\def\momt{{ \textit{\textbf m}}}  
\def\komt{{ \textit{\textbf k}}} 
\def\la{\langle}
\def\bea{\begin{eqnarray}}
\def\eea{\end{eqnarray}}
\def\is{\! & \! = \! & \!}
\def\ra{\rangle}
\def\half{{\textstyle{\frac 12}}}
\def\cL{{\cal L}}
\def\bigll{\bigl}
\def\bigrr{\bigr}
\def\halfi{{\textstyle{\frac i 2}}}

\def\ba{\begin{eqnarray}}
\def\ea{\end{eqnarray}}





\def\ibar{{\, {\overline{i}}\, }}
\def\kbar{{\spc {\bar{k}\spc }}}

\newcommand{\rep}[1]{\mathbf{#1}}
\def\betaH{\beta}
\def\XX{\text{X}}
\def\IR{\text{IR}}
\def\UV{\text{UV}}
\def\GUT{\text{GUT}}
\def\singlet{\text{GUT}}
\def\be{\bea}
\def\ee{\eea}
\def\delbar{\overline{\partial}}
\newcommand{\smpc}{\hspace{.5pt}}
\def\ra{\bigr\rangle}
\def\la{\bigl\langle}
\def\ccdot{\!\spc\cdot\!\spc}
\def\AA{{\raisebox{-1pt}{\scriptsize \nspc $A$}}}

\def\nspc{\!\spc\smpc}
\def\uU{\mbox{\textit{\textbf{U}}}}
\def\cC{\mbox{\textit{\textbf{C}\!\,}}}
\def\bn{\mbox{\textit{\textbf{n}\!\,}}}
\def\bbn{\mbox{\scriptsize{\textit{\textbf{n}}}}}
\def\pP{\mbox{\textit{\textbf{P}\!\,}}}
\def\rR{{\textit{\textbf{R}\!\,}}}

\def\im{{\rm i}}
\def\tr{{\rm tr}}
\def\oneoverN{{1\over N}} 

\enlargethispage{\baselineskip}

\setcounter{tocdepth}{2}
\tableofcontents

\newpage
\addtolength{\baselineskip}{.25mm}
\addtolength{\parskip}{.3mm}
\renewcommand\Large{\fontsize{15.5}{16}\selectfont}

\def\La{\Bigl\langle}
\def\Ra{\Bigl\rangle}
\def\Li{\Bigl |}
\def\Ri{\Bigl |}
\def\la{\bigl\langle}
\def\ra{\bigl\rangle}
\def\li{\bigl|}
\def\ri{\bigl|}
\def\spc{\hspace{1pt}}
\def\is{\! & \! = \! & \!}
\def\cO{{\cal O}}
\def\bea{\begin{eqnarray}}
\def\eea{\end{eqnarray}}
\def\zb{\overline{z}}
\def\tZ{{\spc\widetilde{\!Z\!\spc}}}
\def\rma{{\raisebox{.15pt}{\fontsize{9pt}{0.5pt}$\rmaa$}}}
\def\rmb{{\mbox{\fontsize{8.5pt}{0.5pt}$\rmbb$}}}
\def\rmc{{\mbox{\fontsize{9pt}{0.5pt}$\rmcc$}}}
\def\rme{{\mbox{\fontsize{9pt}{0.5pt}$\rmee$}}}
\def\rmd{{\mbox{\fontsize{8.5pt}{0.5pt}$\rmcd$}}}
\def\rmf{{\mbox{\fontsize{9pt}{0.5pt}$\rmff$}}}
\def\rmo{{\mbox{\scriptsize $\rm 1$}}}
\def\rmtw{{\mbox{\scriptsize $\rm 2$}}}
\def\rmth{{\mbox{\scriptsize $\rm 3$}}}
\def\rmf{{\mbox{\scriptsize $\rm 4$}}}
\newcommand{\pd}{\partial}
\newcommand{\pdd}[2]{\frac{\pd{#1}}{\pd{#2}}}
\newcommand{\mb}{\mathbf}
\newcommand{\bs}{\boldsymbol}
\newcommand{\nn}{\nonumber}
\newcommand{\mc}[1]{\ensuremath{\mathcal{#1}}}
\newcommand{\mbb}[1]{\ensuremath{\mathbb{#1}}}
\newcommand{\imp}{\Longrightarrow}
\newcommand{\Tr}{\mbox{Tr}}
\newcommand{\fs}[1]{\slashed{#1}}
\def\nn{\nonumber}

\def\wbar{\bar{w\spc}\nspc}
\def\zbar{\bar{z}}
\def\nspc{\hspace{-1pt}}
\def\smpc{\hspace{.5pt}}
\def\sotimes{\raisebox{1pt}{\small $\smpc\otimes\smpc$}}
\def\Oplus{\mbox{\footnotesize $\bigoplus$}}
\def\plus{\raisebox{1pt}{\tiny +}}
\def\CFT{{\mbox{\fontsize{6pt}{.05pt}$\rm \spc CFT$}}}
\def\LL{{\mbox{\fontsize{6pt}{.05pt}${\rm Liouville}$}}}
\def\ybar{{\bar y}}
\def\dbar{{\overline\partial}}
\def\rmaa{
\alpha}
\def\rmbb{
\beta}
\def\rmcc{
\gamma}
\def\rmdd{
\delta}
\def\rmee{
\mu}
\def\rmff{
\nu}
\def\rmab{
\alpha\beta}
\def\rmac{
\alpha\gamma}
\def\rmbc{
\beta\gamma}
\def\pplus{{\mbox{\tiny$+$}}}
\def\mmin{{\mbox{\tiny$-$}}}
\def\eplusl{{e^\pplus{\!\!\!\nspc\smpc}_\lL}}
\def\eminr{{e^\mmin{\!\!\!\nspc\smpc}_\rR}}
\def\eplusr{{e^\pplus{\!\!\!\nspc\smpc}_\rR}}
\def\eminl{{e^\mmin{\!\!\!\nspc\smpc}_\lL}}

\def\graycylinder{\begin{tikzpicture}[mycyl/.style={cylinder, shape border rotate=45, draw, minimum width=.0mm, aspect=.4, 
anchor=south, text width=.04cm, text height=.0cm}] 
\node [mycyl , fill=lightgray, minimum height=.0mm] at (0,0) {};
\draw[fill=white] (0,.21) ellipse (1mm and .5mm);]
\end{tikzpicture}}
\def\whitecylinder{\begin{tikzpicture}[mycyl/.style={cylinder, shape border rotate=80, draw, minimum width=.0mm, aspect=.45, 
anchor=south, text width=.04cm, text height=.0cm}] 
\node [mycyl , gray,  fill=white, minimum height=.0mm] at (0,0) {};
\draw[fill=lightgray] (0,.22) ellipse (1mm and .5mm);]
\end{tikzpicture}}

\def\eplusl{{e^\pplus{\!\!\!\nspc\smpc}_\lL}}
\def\eminr{{e^\mmin{\!\!\!\nspc\smpc}_\rR}}
\def\eplusr{{e^\pplus{\!\!\!\nspc\smpc}_\rR}}
\def\eminl{{e^\mmin{\!\!\!\nspc\smpc}_\lL}}
\def\intt{{\mbox{\small $\int$}}}
\def\alph{{\mbox{\footnotesize{a}}}}
\addtolength{\abovedisplayskip}{.4mm}
\addtolength{\belowdisplayskip}{.4mm}
\addtolength{\oddsidemargin}{2mm}

\def\lL{{\mbox{\fontsize{7pt}{.05pt}\smpc\sc{l}}}}
\def\rR{{\mbox{\fontsize{7pt}{.05pt}\smpc\sc{r}}}}

\addtolength{\abovedisplayskip}{0mm}
\addtolength{\belowdisplayskip}{0mm}
\addtolength{\baselineskip}{0mm}
\addtolength{\parskip}{.2mm}

\newcommand{\fourj}[4]{\bigl[\raisebox{1pt}{\scriptsize$\begin{array}{cc}\!\! #1 \! \!&\! \!\! #2 \! \\[-.7mm] \! \! #3 \!\! &\! \!\! #4 \! \end{array}$}\!\bigr]}

\def\ddota{${\rm \ddot a}$}
\def\ddotu{${\rm \ddot u}$}

\def\ddoto{${\rm \ddot o}$}
\newcommand{\alphaj}{j}
\newcommand{\sixj}[6]{\left\{\mbox{\small$\begin{array}{ccc}\! #1 \! &\!\! #2 \! &\!\! #3 \!\nspc \\[0mm] \! #4 \! &\!\! #5 \! &\!\! #6 \!\nspc \end{array}$}\!\right\}}

\def\VVV{V}
\def\Lit{{\rm Li_2}}
\def\Ez{\nspc\smpc\alpha\nspc\smpc}
\def\Eo{\nspc\smpc\epsilon_0\nspc\smpc}
\def\Et{\nspc\smpc\omega\nspc\smpc}
\def\Ef{\nspc\smpc\beta\nspc\smpc}

\noindent
\section{Introduction and Summary}\label{intro}


In this paper we study the role and implications of the conformal bootstrap equations for the AdS${}_3$/CFT${}_2$ correspondence. 
We are particularly interested in how the {\it non-perturbative} gravity regime emerges from the conformal field theory, and vice versa.
2D  CFT has special properties, thanks to the infinite conformal symmetry and left-right factorization of its correlation functions \cite{BPZ,EV,GregNati,Cardy,Rehren}. Similarly, 2+1-D gravity is special due to the absence of a local metric excitations \cite{Deser,WG}. Both sides possess a
precise topological and analytic structure,  in the form of braiding and fusion matrices, that satisfy non-trivial consistency conditions. Our plan is to display this structure on both sides, and use it to establish a precise non-perturbative holographic dictionary. 

 The existence of this geometric dictionary is not a new  insight, and key elements were anticipated since long before the discovery of AdS/CFT \cite{Brown,HV}. 
We will not add any major technical advance to the subject: most of the formulas that we will use have been obtained over the past two decades in work by others \cite{Nati,Zamolodchikov,TeschnerRevisited,TeschnerT,Ponsot,TZ,Kashaev-Fock,Kashaev-V,Tudor,Rosly,TeschnerV}.  Our main new observations  are that 
\begin{itemize}
\item{there exists a general kinematical regime in which the operator algebra of irrational CFTs with $c\gg 1$ takes a universal form identical to that of Liouville CFT}
\item{this operator algebra can be cast in the form of an exchange algebra, specified by an R-matrix that exactly matches with the scattering amplitude of 2+1 gravity. }
\end{itemize} 

Because we have tried to make the exposition somewhat self-contained, this paper is rather long. We will therefore start with a statement of the problem and a summary of our results. Other recent papers that exploit the properties of conformal blocks to extract new insights for  holography include \cite{EikonalAdS,JP,Katz,TwoToms,Fitzpatrick}.


\noindent
\subsection{Statement of the problem}

Consider a 2D CFT with large central charge $c$ and a weakly coupled holographic dual. It has a dense spectrum of primary states $|\spc M\spc\ra$  with asymptotic level density governed by the Cardy formula. In the dual gravity description, the heavy states $\li M \ra$, with energy $\ell M > c/12$ are expected to describe BTZ black holes with metric \cite{BTZ,Carlip}
\bea
\label{btz}
ds^2 = - f(r) dt^2 + \frac 1 {f(r)} dr^2 + r^2 d\phi^2 , \qquad \quad f(r) = \frac{r^2 - R^2}{\ell^2}, \qquad R^2 = 8M\ell^2,
\eea
with $\ell$ the AdS radius and $R$ the Schwarzschild radius.

Consider the Heisenberg state obtained by acting with a local CFT operator on a primary state\footnote{For brevity, we suppress the spatial coordinate in our notation.\\[-3mm]} 
$
\li \spc \Psi \spc \ra =
 \hat{A}(t_0) \li\smpc  M \smpc \ra$.
The corresponding Schr{\ddoto}dinger state
\bea
\li \spc \Psi(t) \spc \ra \is e^{-i (t-t_0) H}
\hat{A}(t_0) \li\smpc  M \smpc \ra 
\eea
for $t<t_0$ describes a bulk geometry with a matter perturbation that is predestined to reach the boundary at time $t=t_0$. It seems reasonable to interpret this perturbation as a particle A that escapes from near the horizon of the black hole of mass $M$. One of the goals in this paper is to put this identification through some non-trivial tests and make it more explicit.

The specific test that we would like to perform is to exhibit the gravitational backreaction of particle A through its interaction with some other particle B, and vice versa. To this end, it is useful to make a spectral decomposition of the local operator $\hat{A}(t_0)$ 
\bea
\hat{A}(t_0) \li\spc  M \spc \ra \is 
\sum_{\alpha}\, \spc  
\hat{A}_{\alpha}(t_0) \spc \li\smpc M\smpc  \ra 
\eea
into components
that increase the energy of the primary state $\li M \ra$ by a specified amount
\bea
\label{aalpha}
\hat{A}_{\alpha}(t_0)\spc \li \smpc M \smpc \ra  \is \; e^{-i \alpha t_0}\, f^{A}_\alpha\;  \li \smpc M\nspc + \nspc \alpha\spc\ra
\;\; + \;\; \mbox{descendants} .  
\eea
Here $f^A_\alpha$ denotes the relevant OPE coefficient.
The descendants are terms given by products of left and right-moving Virasoro generators acting on the primary state $\li M\! + \! \alpha\ra$.\footnote{\small \addtolength{\baselineskip}{.3mm}
Note that the right-hand side of (\ref{aalpha}) depends non-trivially on $t_0$ through the contribution of the  descendant states, which takes the schematic form : $\sum_{n>0} e^{-i(n+\alpha)t_0} | M\! + \! \alpha,n\rangle.$ In our discussion, we will pay little attention to descendent states because our focus is on the bulk physics near the horizon. The Virasoro operators generate the asymptotic symmetry group of AdS space time \cite{Brown}, and thus correspond to excitations that are localized near the AdS boundary, or more acurately, excitations that deform the transition function between the bulk and the asymptotic geometry.}
We interpret this component $\hat{A}_{\alpha}(t_0)$ as describing the partial wave in which the bulk particle A has specified energy $\alpha$. We wish to confirm this interpretation, by exhibiting how the CFT encodes the full semi-classical backreaction of particle A on the black hole geometry.

\begin{figure}[t]
\begin{center}
\includegraphics[scale=.35]{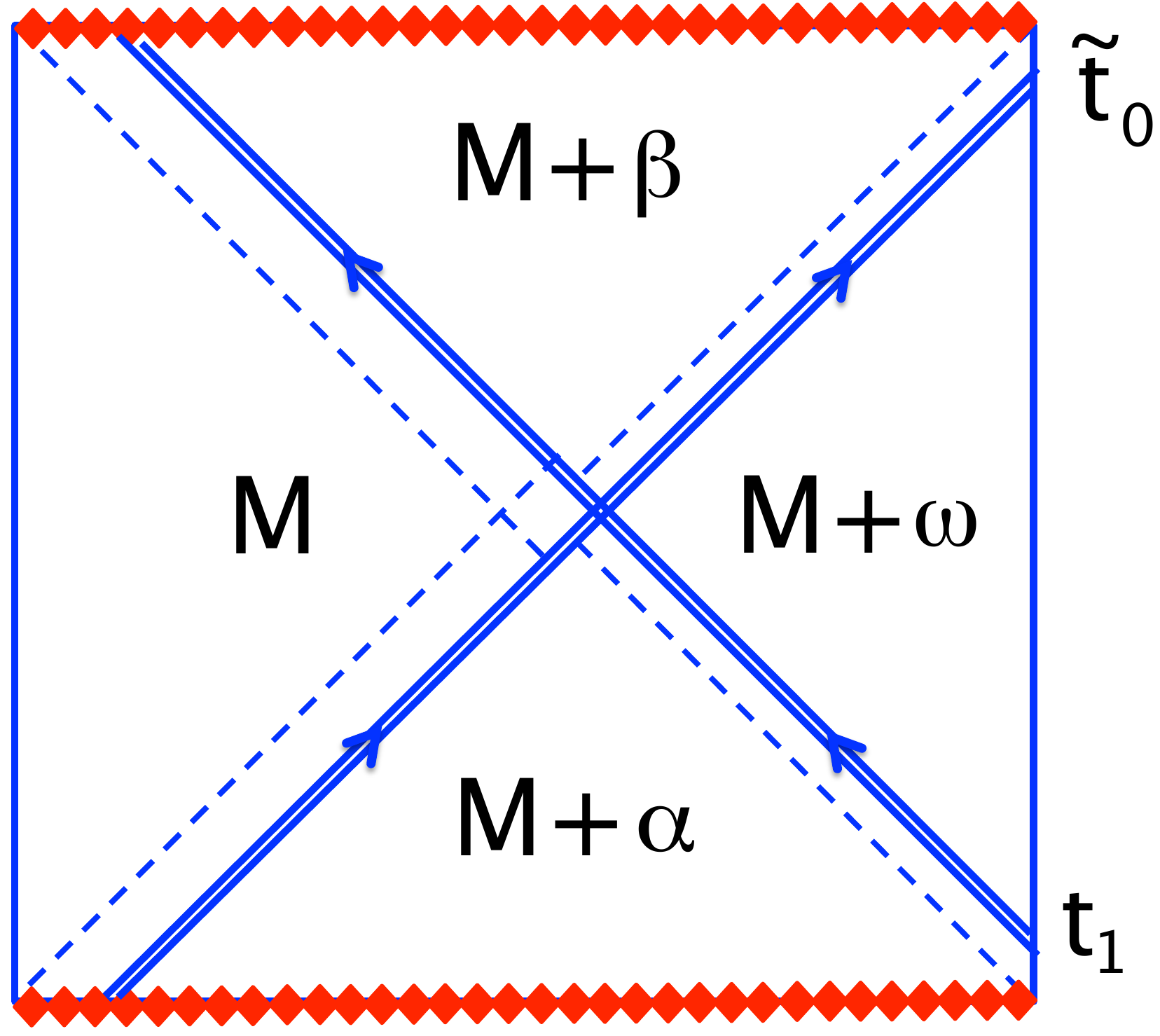}
\caption{Penrose diagram describing an eternal BTZ black hole with an infalling and outgoing matter perturbation \cite{ShSt}. The two particles collide near the horizon and affect each others' trajectory via a gravitational shock wave interaction. Alternatively (as shown in the figure) one may represent the effect of the shock wave as a shift in the location of the event horizon.}
\end{center}
\vspace{-0.2cm}
\end{figure} 

Let us now introduce \cite{ShSt} a second particle B, created by acting with another local operator $\hat{B}(t)$ at some {\it earlier} time $t_1<t_0$ 
\bea
\label{butterfly}
\hat{B}_{\omega-\alpha}(t_1)\spc  \hat{A}_{\alpha}(t_0)\spc \li \smpc M \smpc \ra \is \, e^{-i\omega t_1-i \alpha (t_0-t_1)}
 f^B_{\omega-\alpha}\,  f^A_\alpha\, \li \smpc M\nspc + \nspc \omega \smpc \ra 
\;\; + \;\; \mbox{descendants} .
\eea
Here we fixed the total energy of the bulk particles $A$ and $B$ to be $\omega$. 

The Penrose diagram of the semi-classical geometry associated to the state (\ref{butterfly}) is shown in figure~1. It shows an eternal BTZ black hole of mass $M$ with two matter pulses, one outgoing and one infalling. The stress energy of each particle induces a corresponding shift of the BTZ mass parameter across its trajectory.
The two particles collide close to the horizon, creating a future black hole region of mass $M+\beta$. The value of $\beta$ is determined~by~the collision energy of the two particles, which in turn is set by $M$, the energies $\alpha$ and $\omega$, and the time difference $t_0-t_1$ between when B departs and A arrives at the AdS boundary.

The total geometry of figure 1 is governed by 6 mass parameters
\bea
\label{massp}
M_1\spc =\, M\smpc , \ \ \  \qquad \quad \ \;  M_2\spc =\spc m_A, \ \ & & \quad M_\alpha\spc = \spc M+\alpha,\nonumber \\[-3mm]\\[-3mm]
M_{3}\spc  =\spc  M\nspc + \omega, \quad \qquad M_4\spc =\spc  m_B,\ \ & & \quad M_\beta\spc =\spc M+ \beta\spc .\nonumber
\eea
We will assume that particle $A$ and $B$ are both massless and set $m_A=m_B=0$. Our formulas in sections 2 and 3, however, remain valid for general $m_A$ and $m_B$.

Particles A and B affect each others trajectory via a gravitational shock wave interaction. Initially, the outgoing particle A hovers very close to the event horizon, until just before it escapes to infinity. 
When the infalling particle B passes through, it slightly shifts the location of the event horizon. As a result, particle A ends up moving even closer to the event horizon, or it may even end up inside the black hole. Either way, its future trajectory is strongly affected by the shift. 
 Thanks to the exponential redshift associated with black hole horizons,
 this effect grows exponentially with the time difference $t_0-t_1$ between when B falls in and when A was supposed to come out. 
In a suitable kinematical regime, this gravitational effect dominates all other interactions between the two particles. 

Gravitational shock wave interactions near black holes were studied by Dray and `t Hooft \cite{DTH} and in subsequent work \cite{Kiem}.In the context of AdS/CFT, they were recently used by Stanford and Shenker to provide new evidence that CFTs with AdS duals are fast scramblers \cite{ShSt}. 
  It was shown in \cite{Kiem} that the geometric effect of the shock wave can be captured by means of a so-called exchange algebra between infalling modes $\phi^A$ and outgoing modes $\phi^B$ of the schematic form 
  \bea
  \phi^{B}_{\omega-\alpha}\, \phi^A_\alpha \is \sum_\beta {\cal R}_{\alpha\beta} \; \phi^A_{\omega-\beta}\, \phi^B_\beta,
  \eea 
where ${\cal R}_{\alpha\beta}$ are the matrix elements of a unitary scattering operator $\hat{\cal R}$ that shifts the location of one operator by an amount proportional to the Kruskal energy of the other.

This exchange algebra is a characteristic property of infalling and outgoing modes near black horizons in any dimension \cite{Kiem}. However, gravitational interactions in 2+1 dimensions have special characteristics that make this description especially natural.  In 2+1  gravity, backreaction due to localized matter sources is described by means of simple global geometric identifications \cite{Deser,BTZ,Carlip}.  As a result,  point particles and black holes enjoy  long range interactions similar to the topological braiding relations in systems with non-abelian statistics.\footnote{\small \addtolength{\baselineskip}{.3mm} In the most well-known systems with non-abelian statistics, like the fractional quantum Hall effect, the non-abelian statistics group is compact and has finite dimensional unitary  representations.   In the gravitational setting on the other hand, the non-abelian braid statisitics group  is $SO(2,2)$, the  group of isometries of AdS${}_3$. $SO(2,2)$ is non-compact and it has a continuous series of infinite dimensional unitary representations. The R-matrices in 2+1-D gravity are therefore operators that act on an infinite dimensional Hilbert space with a  continuous spectrum. }
    The basic operation in theories with braid statistics is the R-operation, that interchanges the location of particles, or equivalently, the ordering of two operators. We therefore expect that the braid statistics of  2+1-D theories with gravitationally interacting quasi-particles can be captured by a similar type of R-operator.

How would this gravitational exchange interaction show up in the dual CFT?
It has long been known that the operator algebra of (chiral) vertex operators in 2D CFT can be cast in the form of an exchange algebra \cite{Rehren}. This property underlies the celebrated relationship between rational CFTs and 2+1-D topological field theories of particles with braid statistics \cite{WCS}. In the following we will argue, with considerable supporting evidence, that the same  algebraic structure extends to irrational CFTs with large central charge.  Specifically, we will  show that
the components of local operators $\hat{A}(t_0)$ and $\hat{B}(t_1)$ in 2D CFT must satisfy a universal exchange algebra of form
\bea
\label{exchangeR}
\hat{B}_{\omega-\alpha}(t_1)\spc  \hat{A}_{\alpha}(t_0)\spc \li \smpc M \smpc \ra  \is 
\sum_{\beta} \; {\cal R}_{\alpha\beta}\;   
\hat{A}_{\omega-\beta} (t_0) \spc \hat B_{\beta}(t_1)\spc\ri\spc M\spc \ra.
\eea
This exchange relation is an {\it exact} property of the CFT, that follows by analytic continuation from the braid relations between chiral vertex operators in the euclidean domain. It is a causal relation that  holds whenever the local operators $\hat{A}$ and $\hat{B}$ are time-like separated. 
The matrix ${\cal R}_{\alpha\beta}$ depends on  all mass parameters $M_i$
and is known as the R-matrix or braid matrix, which prescribes the monodromy properties of conformal blocks of the CFT. It encodes the value of OPE coefficients, and satisfies a number of non-trivial consistency relations analogous to Yang-Baxter equations. Finding the explicit form of the R-matrix ${\cal R}_{\alpha\beta}$ is equivalent to finding a solution to the conformal bootstrap (and thus in general a  difficult task). Hence in a concrete sense,  the exchange algebra provides a complete characterization of the CFT.

 Our goal in this paper is to derive the exact form of the R-operation and its associated exchange algebra, both from 2+1 gravity and in 2D CFT. We will find that the answers precisely match, for the simple reason that the two calculations  are essentially identical.



\noindent
\subsection{Semi-classical scattering}

How would one derive the explicit form of the scattering matrix ${\cal R}_{\alpha\beta}$ from the point of view of the gravity theory? 
As a useful warm-up exercise, we first look at the classical scattering process.
Let us pick values of the four masses $M_i$, $i=1,..,4$  in eqn (\ref{massp}), i.e. we fix the initial mass $M$ of the black hole, the masses $m_A$ and $m_B$ of the two particles, and the total energy $M+\omega$. 

From fig.~1 it is clear that once we know the initial  energy $\alpha$ of the outgoing mode A and  the initial time difference $t_\alpha = t_0-t_1$, the subsequent time evolution is uniquely determined:
 the final energy $\beta$ of particle B and the new arrival time $\tilde{t}_0$ are given functions of $\alpha$ and $t_\alpha$.
We could also reverse time and pick the final energy of the infalling mode, and the time difference $t_\beta = \tilde{t}_0-\tilde{t}_1$ between the final arrival time $\tilde{t}_0$ of particle A and the time $\tilde{t}_1$ when particle B would have originated from if particle A had not been there. The initial energy $\alpha$ and the actual departure time $t_1$ are determined as given functions of $\beta$ and $t_\beta$.

The scattering process of fig 1 for spherical matter shells was recently studied in \cite{ShSt}. We can read off the relations between the four parameters ($\alpha, t_\alpha)$ and $(\beta, t_\beta)$ from their paper. As explained in \cite{ShSt},  following \cite{DTH}, one can obtain the colliding shock wave geometry by means of a simple gluing procedure.
The radial location $r$ of the collision point is determined by a matching relation that equates the values of the $f(r)$ factors (see eqn (\ref{btz}))  of all four BTZ black hole geometries,
\bea
\label{dth}
\bigl(r^2-8(M\nspc+\nspc\alpha)\ell^2\bigr)\bigl(r^2-8(M\nspc+\nspc\beta)\ell^2\bigr) \is \bigl(r^2-8(M\nspc+\nspc\omega\bigr)\ell^2) \bigl(r^2-8M\ell^2\bigr).
\eea
We are interested in the universal behavior 
when the collision takes place very close to the horizon. In this regime, the distance between collision radius $r$ and the two Schwarschild radii $R_\alpha$ and $R_\beta$ scales exponentially with the time differences $t_\alpha$ and $t_\beta$
\bea
\label{redsh}
\qquad r^2 - 8 (M\nspc+\nspc\alpha)\smpc \ell^2 \is 4\ell^2 e^{-\kappa\smpc (t_\alpha-t_R)}, \qquad \qquad \ t_\alpha \equiv \spc t_0-t_1\nonumber \\[-2.5mm]\\[-2.5mm]\nonumber
\qquad r^2 - 8(M\nspc+\nspc\beta)\smpc \ell^2\is 4\ell^2 e^{-\kappa\smpc (t_\beta-t_R)},\qquad \qquad \ t_\beta \equiv \spc \tilde{t}_1 -\tilde{t}_0 ,
\eea
where $\kappa$ is the surface gravity and $t_R$ is a time delay, given by
\bea
\label{kappadef}
\kappa =  R/ {\ell^2}, \quad     & &\quad
\kappa \spc t_R =  \log (R^2/\ell^2).
\eea
Here we recognize the characteristic exponential redshift effect near black hole horizons.
Combining eqns (\ref{dth}) and (\ref{redsh}), one derives the following relations \cite{ShSt}
\bea
\label{relno}
\beta \is \omega - \alpha\spc +\spc  2 \spc\alpha \smpc (\omega-\alpha)
\spc e^{\kappa (t_\alpha-t_R)},\\[2mm]
\label{relnt}
\alpha \is \omega - \beta + \spc 2\spc \beta \smpc (\omega -\beta) \spc e^{\kappa (t_\beta-t_R)}.
\eea
Eqn (\ref{relno}) determines $\beta$ as a function of $\alpha$ and the time difference $t_\alpha$. We see that, due to the exponential growth in $t_\alpha$, $\beta$ quickly becomes bigger than $\omega$. Once this happens, eqn (\ref{relnt}) no longer yields a real solution for $t_\beta$. This is not surprising: as seen from fig 1, when $\beta>\omega$ the energy of mode $A$ becomes negative, indicating that its trajectory  has been shifted to behind the horizon.

Now let us replace particle A and particle B by quantum mechanical wave packets. As explained in detail in \cite{Kiem}, we
should anticipate that the second quantized mode operators $\phi_A$ and $\phi_B$ that create both asymptotic wave packets do not commute but satisfy an exchange relation.
For spherical wave-packets -- which can be simultaneously localized in time and energy -- and in the leading order semi-classical limit,
we expect that this exchange relation takes the form
\bea
\label{xch}
\phi^B_{{\omega-\alpha}}(t_1) \, \phi^A_{\alpha}(t_0) \is \spc e^{\frac{i}{\hbar}S_{\alpha\beta}} \, \phi^A_{\omega-\beta}(\smpc \tilde{t}_0\smpc) \, \phi^B_{\beta}(\tilde{t}_1).
\eea
Since $\phi^A(t_1)$ acts in the future of $\phi^B(t_0)$, this relation is in perfect accord with causality. It expresses the causal effect that  the trajectory of A, after its encounter with B, is shifted by the specified amount, relative to its original trajectory. 
The time shifts can be computed by a similar calculation as the one that gave us the relations (\ref{relno})-(\ref{relnt}). One finds that
\bea
\label{timeshift}
\tilde{t_0}-t_0 \! = \! -\frac 1 \kappa\,\log \Bigl(\frac{\omega-\beta}\alpha\Bigr),
\qquad \qquad
\tilde{t_1}-t_1 \is  -\frac 1 \kappa\,\log \Bigl(\frac{\omega-\alpha}{\beta}\Bigr).
\eea
Note that the time delay $\tilde{t}_0 -t_0$ indeed becomes infinite when $\beta$ approaches $\omega$.

The dependence of the scattering phase ${\cal R}_{\alpha\beta} = e^{\frac{i}{\hbar}S_{\alpha\beta}}$ on the initial energy $\alpha$ and final energy $\beta$ follows from the usual rules of geometric optics, via the Hamilton-Jacobi equations
\bea
\label{hj}
t_\alpha =-\frac 1 \hbar \frac{\partial S_{\alpha\beta}}{\partial \alpha},
\qquad \qquad
t_\beta \!\is \!   \frac 1 \hbar \frac{\partial S_{\alpha\beta}}{\partial \beta}. 
\eea
These equations ensure (or can be derived from the fact) that both sides of the exchange relation (\ref{xch}) have the same dependence on $\alpha$ and $\beta$.
To integrate the H-J equations, we are instructed to compute the time differences $t_\alpha$ and $t_\beta$ as functions of $\alpha$ and $\beta$.
\bea t_\alpha - t_R\! \is\!  \frac 1 \kappa \,\log \Bigl(\spc\frac{\alpha +  \beta-\omega}{2\smpc \alpha \smpc (\omega - \alpha)}\spc \Bigr), \qquad \qquad
t_\beta -t_R = \frac 1 \kappa \,\log \Bigl(\spc \frac{\alpha +  \beta-\omega }{2 \smpc \beta \smpc (\omega - \beta)\spc }\spc\Bigr).
\eea 
Standard H-J theory tells us that the result for $S_{\alpha\beta}$ is equal to the total action evaluated on the unique classical space-time trajectory that interpolates between the state where A has initial energy $\alpha$ and B has final energy $\beta$. In the regime that there exists a real trajectory, $\omega>\alpha,\beta>0$, $(\alpha+\beta)>\omega$, the above equations integrate to
\bea\label{sabshock}
 \frac{1}{\hbar} S_{\alpha\beta} \is 
\frac{1}{\kappa}\Bigl\{ \spc  \alpha \log \alpha \spc + \spc \beta\log \beta
- ( \omega\nspc -\nspc \alpha) \log(\omega\nspc -\nspc \alpha) - \spc (\Et\nspc -\nspc\Ef) \log(\Et\nspc -\nspc \Ef) \nonumber\\[2mm] & &  \qquad  -\;  (\Ez + \beta - \Et\smpc ) \log\bigl(\Ez+ \Ef  - \Et \bigr)\Bigr\} - (\Ez + \beta - \Et\smpc )  t_R .
\eea
Equivalently, $S_{\alpha\beta}$ defines the generating function of the canonical transformation between the initial and final canonical variables $(\alpha,t_\alpha)$ and $(\beta,t_\beta)$. 

In section \ref{scatteringmat} we will present a more refined and general description of this semi-classical scattering process, in terms of $SO(2,2)$ holonomy variables. We will obtain a more general exact expression for the  scattering phase $S_{\alpha\beta}$ in terms of dilogarithms that reduces to the above answer in the regime of interest. As we will see, the more general answer has a natural geometrical interpretation as  the volume of a hyperbolic tetrahedron with dihedral angles specified by the $SO(2,2)$ holonomies of the BTZ geometry and conical defects created by the particles involved in the scattering. Our main goal in the following will be to reproduce this semi-classical scattering matrix from the CFT side.


\noindent
\subsection{Scattering phase from the modular bootstrap}


 The R-matrix that specifies the exhange algebra  of chiral vertex operators in 2D CFT satisfies several consistency relations that are equivalent to the conformal bootstrap equations. The relations express the physical principles of locality and associativity of the operator algebra, and analyticity of the correlation functions. The conformal bootstrap usually refers to a set of equations for conformal blocks and OPE coefficients. The R-matrices are part of the general set of modular matrices that specify the monodromy properties of chiral correlation functions. The program of finding the monodromy matrices via their consistency relations is called the modular bootstrap.

\def\sca{\mbox{\fontsize{6.5pt}{.5pt}$\rm A$}}
\def\scb{\mbox{\fontsize{6.5pt}{.5pt}$\rm B$}}
\def\scm{\mbox{\fontsize{6pt}{.5pt}\nspc$M$}}
\def\scc{\mbox{\fontsize{6.5pt}{.5pt}$\rm C$}}

For integrable theories like rational CFTs, the R-matrices that solve all the above consistency relations are explicitly known. Our interest, however, is in irrational CFTs with large central charge and with holographic duals. These theories are not integrable and the R-matrices are not known and seemingly impossible to compute. So how can one hope to get mileage from considering the modular bootstrap for this case?

The main distinction between irrational and rational CFT is that the sum over intermediate states on the right-hand side of (\ref{exchangeR}) runs over a dense spectrum with an infinite set of labels, rather than over a small finite set. In our specific setting, the sum is dominated by intermediate states $\li M \! + \! \beta\ra$ in the densely populated neighborhood of the heavy state $\li M\ra$. It is reasonable to approximate the spectral density in this region by a continuum, and replace the sum over states $\beta$ by an integral, weighted with an appropriate measure that reflects the Cardy growth of states. This simplification, which is similar to the hydrodynamic approximation of a gas or liquid, comes with an important benefit: just like the hydrodynamic limit, it ignores irrelevant microscopic details that distinguish individual CFTs and produces a universal and more tractable set of equations that capture the essential macroscopic properties. The appearence of such a universal regime in the CFT is of course expected from the point of view of dual bulk theory, whenever one enters a kinematical regime in which gravitational effects dominate over all other interactions.

\begin{figure}[t]
\begin{center}
\raisebox{.85cm}{\includegraphics[scale=.9]{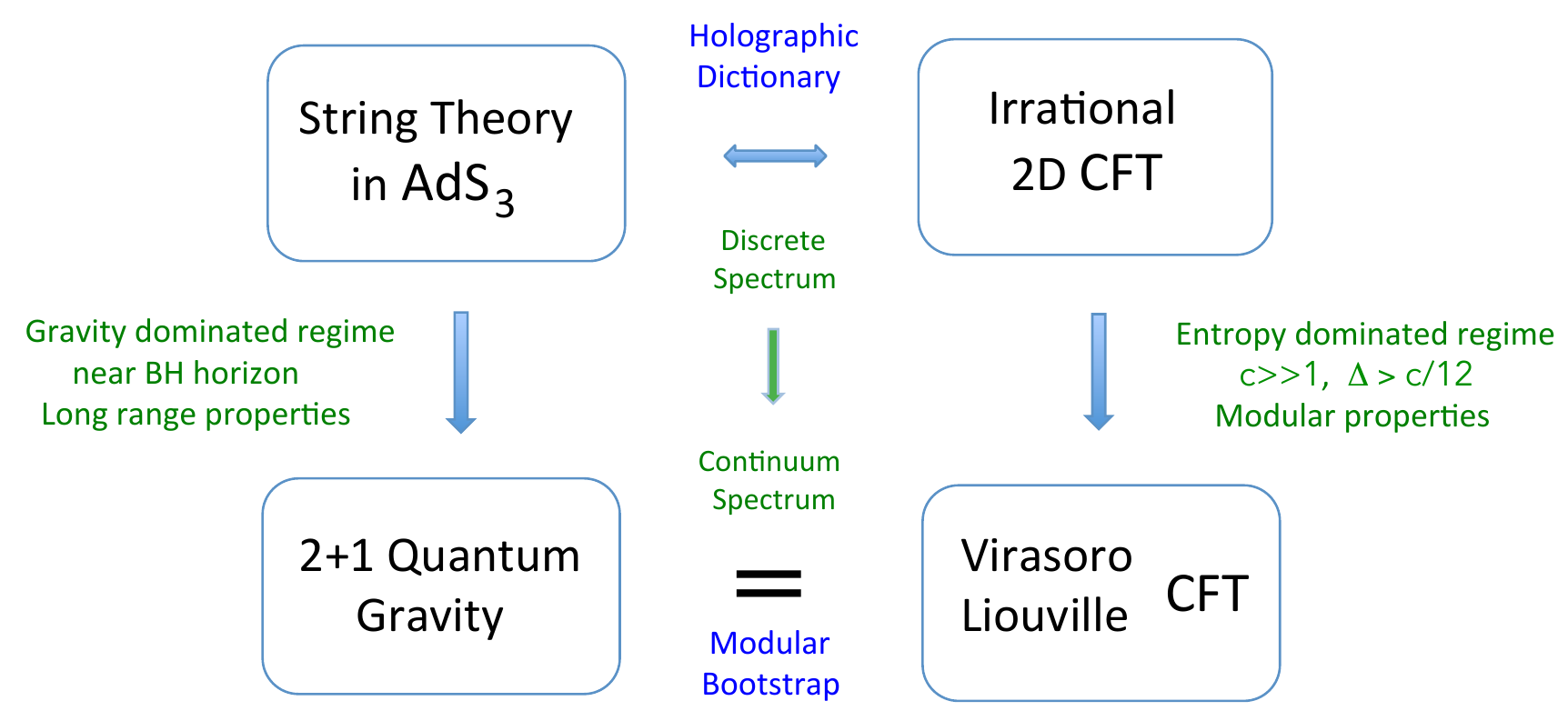}}
\vspace{-5mm}
\caption{Overview of our results. We study irrational 2D CFTs with a string theory dual on AdS${}_3$. If the CFT has large $c$, a sparse light spectrum \cite{Tom} and only Virasoro symmetry, the AdS theory is expected to have two regimes in which gravity dominates: a high energy regime close to a black hole horizon, and a low energy regime at long distances. In the CFT, the dynamics in these regimes is dominated by the high entropy part of the spectrum, and the modular properties of conformal blocks are captured by Liouville theory.  There exists a precise mathematical identification between 2+1 quantum gravity, defined as the quantization of its classical phase space, and Liouville modular geometry.}
\end{center}
\vspace{-5mm}
\end{figure}

To formulate a self-consistent set of conformal bootstrap equations in this approximation, in essence all one needs to do is to replace sums over a dicrete spectrum by integrals over a continuum. The resulting bootstrap program is equally restrictive as the discrete variant. Moreover,  through contributions by many authors, it has essentially been completed!  By now the modular geometry of Virasoro conformal blocks is fully worked out \cite{Zamolodchikov,TeschnerRevisited,Ponsot,TeschnerT,TeschnerR}. Unique and explicit expressions for the universal R-matrix, spectral density, and OPE coefficients are available and ready to be applied to the study of AdS${}_3$/CFT${}_2$. 
 
Results about Virasoro modular geometry are usually reported in the context of studies of  Liouville CFT, or equivalently,  in the context of the quantum theory of  Teichm{\ddotu}ller space\footnote{\small \addtolength{\baselineskip}{.3mm} 
Perhaps due to this categorization (and settled opinions about Liouville CFT and pure 2+1 gravity)  the universal meaning and range of applicability of these results has thus far not been widely appreciated. } -- which as we will see, beautifully connects with the quantum geometry of 2+1-D gravity.
We have collected  the basic elements of this story, and some of the explicit formulas, in sections 2, 3 and 4 and in the Appendix. The main statements are summarized as follows. 

Virasoro conformal blocks span a linear vector space, which can be endowed with a Hilbert space structure and an inner product \cite{Ponsot,TeschnerT}. The spectral density of the Hilbert space is uniquely fixed by the requirement that modular and braid operations are implemented as unitary transformations.  Crucially, as we will review in section 4, one finds that the level density for  $\Delta>c/12$ matches with the Cardy formula \cite{McGough}. 

 The fusion and braiding properties of the Virasoro conformal blocks are captured by the category of representations of the quantum group $U_q(\mathfrak{sl}(2,\mathbb{R})) \times U_q(\mathfrak{sl}(2,\mathbb{R}))$,
the $q$-deformed universal enveloping algebra associated with the non-compact group $SL(2,{\mathbb R}) \times SL(2,\mathbb{R})$ \cite{Ponsot}. 
Here each $SL(2,{\mathbb R})$ factor relates to a chiral sector of the CFT. In particular, the R-matrix which (as we will explain in section 5) governs the exchange algebra between local operators in the Lorentzian CFT \cite{Rehren},
 is given by the square (product of left times right) of the corresponding $q$-deformed $6j$-symbol
\bea
\label{qsixj}
{\cal R}_{\alpha\beta} \is e^{2\pi i (\Delta_1+ \Delta_3- \Delta_\alpha-\Delta_\beta)} \  \sixj{j_1}{j_2}{j_\alpha}{j_3}{j_4}{j_\beta}_q^2 ,
\eea

\noindent
with:\\[-14.5mm]
\bea
q \, =\, e^{i\hbar} \, = \, e^{i\pi b^2},  & & \qquad   \quad \ell M_i\, = 2\Delta_i = \, 2j_i(Q-j_i).
\eea
Here $j_i \in [0, \frac 1 2 Q] \cup \frac 1 2 Q + i {\mathbb{R}}_+$ labels the representation of $U_q(\mathfrak{sl}(2,\mathbb{R}))$, and $Q$, $b$, $\hbar$ and $\ell$ are all related and determined by the central charge $c$ via
\bea
 c \, = \, 1 + 6 Q^2, \qquad \ \  Q \! \is \! b+ b^{-1}, \qquad  \ \  \hbar = \frac{4\pi}{\ell} ,\qquad \ \ c = \frac{3\ell} 2.
\eea
Explicit formulas for the quantum $6j$-symbol are known \cite{Ponsot} but complicated (see the Appendix) and therefore perhaps not very illuminating to non-experts. However, their geometrical meaning becomes more apparent in the semi-classical  limit $\hbar \to 0$, {\it i.e.} the limit of  large central charge $c$. In the dual theory, this is the large radius limit of AdS. 

To exhibit the semi-classical behavior of the R-matrix, we write
\bea
\label{rexps}
 {\cal R}_{\alpha\beta} \is \exp\Bigl(\spc\frac {\raisebox{-2pt}{\small $i$}} {\raisebox{2pt}{\small $\hbar$}}\, S_{\alpha\beta}\spc \Bigr). 
\eea
$S_{\alpha\beta}$ has two natural geometrical interpretations. First, from the known expression of the quantum $6j$-symbols (\ref{qsixj}), one can show by direct calculation that, for small $\hbar$, $S_{\alpha\beta}$ becomes (up to trivial phase) equal to the volume of a generic hyperbolic tetrahedron \cite{Kashaev-V,Tudor,TeschnerV}
\bea
\label{tetras}
S_{\alpha\beta}\is   {\rm Vol}\Bigl(\spc T\Bigl[\mbox{\fontsize{10pt}{.5pt}$\begin{array}{ccc} \! 1\! \! &\! {2} \! &\!\nspc \raisebox{1.25pt}{\fontsize{10pt}{.5pt}$\alpha$} \! \\[-1.5mm] \! 3\! \! &\! 4 \! &\!  \nspc \raisebox{0pt}{\fontsize{10pt}{.5pt}$\!\beta$}
\!\! \end{array}$}\Bigr]\spc \Bigr)
\eea
specified by six dihedral angles $l_i$, determined by the mass parameters (\ref{massp}) via
\bea
\label{lm}
\frac{l_i} {2\pi}= \,\sqrt{8M_i}\quad \mbox{\small $i=1,3,\alpha,\beta,$} \quad & &  \quad \frac{i l_{k}}{2\pi} \, = \, \frac{\theta_{k}}{2\pi}= \sqrt{1-8\ell m_k}\quad k=2,4
\eea
This geometric interpretation of the quantum 6j-symbols was used by Murakami and Yano to derive their exact mathematical formula for the volume of the hyperbolic tetrahedra \cite{Murakami-Yano}. In the present context, this relation encourages a physical interpretation of the R-matrix in terms of gravitational scattering amplitude. In the following sections, we will confirm  this intuition and make it precise.

\begin{figure}[t]
\begin{center}
\raisebox{.8cm}{\includegraphics[scale=.7]{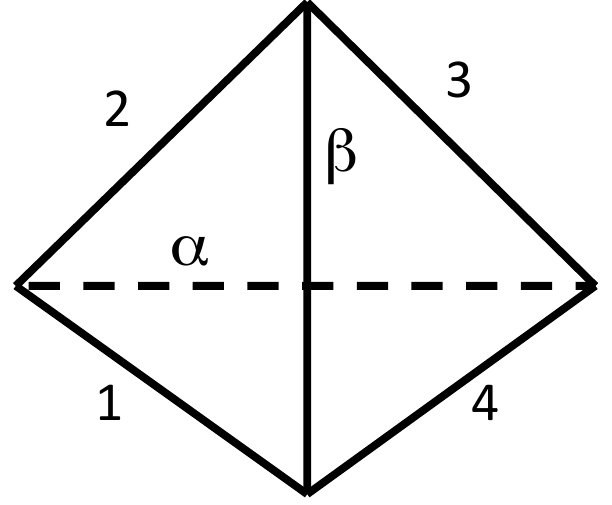}}
\vspace{-5mm}
\caption{The hyperbolic tetrahedron $T$. The dihedral angles at the six edges are determined by the six mass paramaters in (\ref{massp}) via ${l_i}/{2\pi} = \sqrt{8M_i}$.}
\end{center}
\vspace{-5mm}
\end{figure}

The relations (\ref{exchangeR}) and (\ref{rexps})-(\ref{tetras}) form a cornerstone of the
non-perturbative holographic dictionary between 2D CFT and 2+1 quantum gravity. On the CFT side, they express an exact property of Liouville theory which, according to our proposal, captures the universal behavior of generic irrational CFTs in  the regime of interest.  This proposal finds support from the gravity side, where a matching calculation of $S_{\alpha\beta}$ represents the geometric phase associated with a high energy scattering process dominated by the gravitational shock wave interaction. 


\noindent
\subsection{Scrambling behind the horizon}

Besides being a useful testing ground for studying and clarifying the appearence of gravitational interactions from CFT,  another fascinating aspect of the shockwave dynamics near black holes is that they provide a realization of the butterfly effect \cite{ShSt}. On the gravity side, it is clear how this comes about: the perturbance created by the mode B, however small in energy, can still have a dramatic effect on the future time evolution of the mode A.  If the moment of infall of B precedes the would-be-arrival of mode A by more than a critical value $t_{\rm crit}$, the mode A will never emerge at the asymptotic boundary of AdS. This is a dramatic effect. Moreover, the critical time difference grows only logarithmically in the black hole mass $t_{\rm crit} \simeq \frac 1 \kappa \log(R^2/\ell^2 \alpha)$, which is remarkably short. 

How can one understand this effect from the CFT side? At some early time, the signal A is encoded in subtle phase correlations of a strongly interacting many body state. These correlations are prearranged to produce a localized coherent effect at the prescribed time $t_0$, which is the arrival of particle A at the AdS boundary. The butterly effect highlighted in \cite{ShSt} exploits the apparent delicacy of this type of prearrangement in a strongly interacting system: it only takes a very small perturbation of the many body state to destroy the phase correlations and completely thermalize the signal. The main surprise, however, is how fast and efficient the scrambling process appears to be. 

 Thermalization is a complicated dynamical phenomenon and in general hard to quantify. The dual gravity description gives us new ways of analyzing the strongly coupled dynamics of CFTs. But rather than using gravity, it would be more satisfying to explain the effect directly within the CFT and in the process gain insight into the holographic reconstruction of the bulk physics and into the nature of AdS black hole horizons. The methods described in this paper are well suited for this particular problem. Once a CFT question can be translated in terms of Virasoro modular geometry and Liouville theory, the transition to the gravity side becomes relatively easy since both are weakly coupled in the regime of interest, $c\gg1,\Delta>\frac{c}{12}$. Hence it should be possible to gain more insight into the scrambling process with pure CFT techniques.
\begin{figure}[t]
\begin{center}
{\includegraphics[scale=.47]{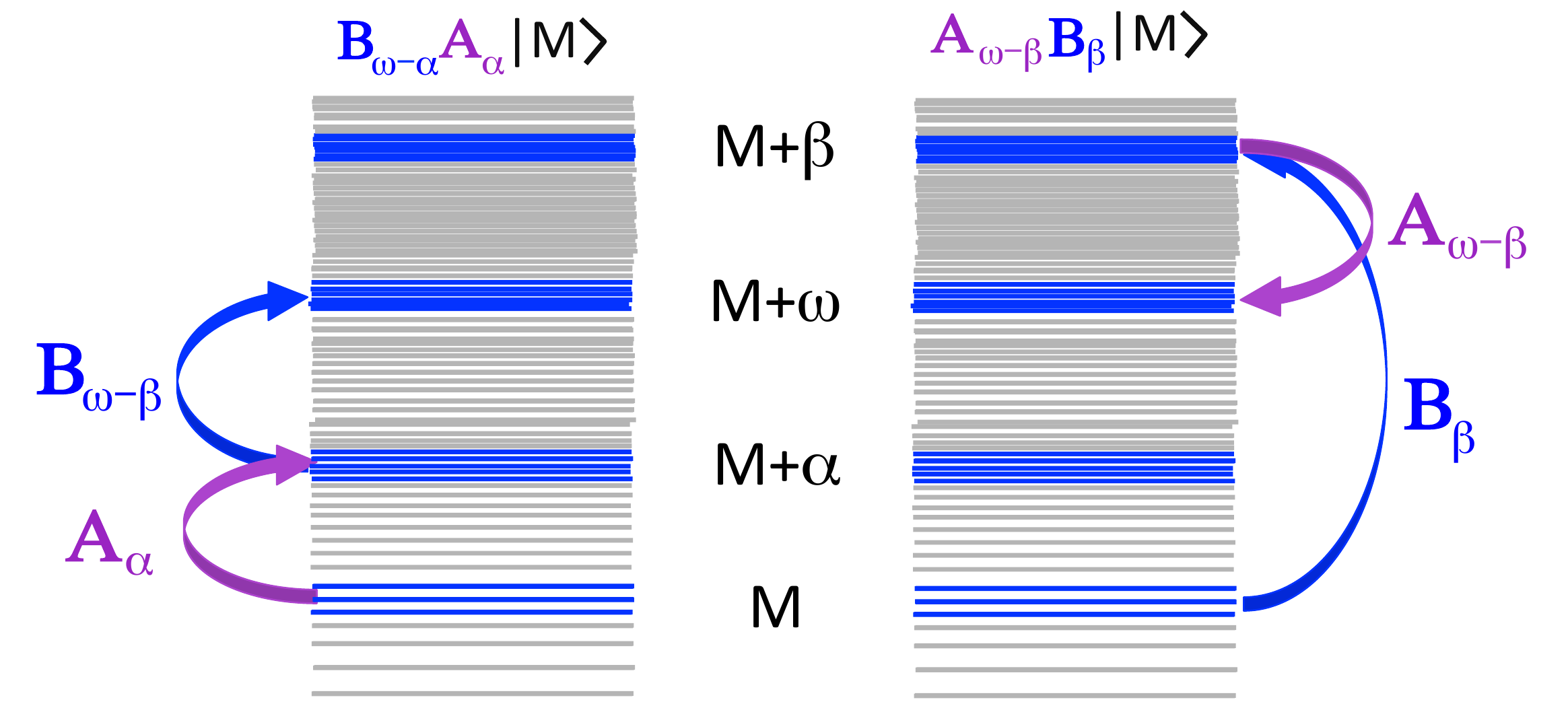}} 
\caption{The scrambling of  a signal (operator $A$)  due to the a perturbation (operator $B$)  at some earlier time $t_1<t_0$. An observer that measures the state can detect  signal $A$ only if $A$ acts on the state from the left. Passing A through B produces a new intermediate channel with energy $\beta$, which for $t_0-t_1> t_{\rm crit}$ exceeds $\omega$. Signal A becomes scrambled: its coherent phase information get washed out by the large entropy region of the spectrum near $M+\beta$. }
\end{center}
\vspace{-0.5cm}
\end{figure}  
  
Let us imagine the following scenario. 
Person ${\bf A}$ prepares a message A for person ${\bf C}$, destined to arrive at time $t_0$. Person ${\bf B}$ then perturbs the system by sending in some noise B at time $t_1$. Both the signal and the noise takes the form of a wave packet, localized in time and with some approximate frequency $\alpha$
\bea
\label{wavepackets}
\hat{A}_\alpha\spc e^{-i\alpha t_0}  \equiv \sum_{\nu \sim \alpha} \hat{A}_\nu \spc e^{-i\nu t_0},\quad & & \quad
\hat{B}_{\omega-\alpha} \spc e^{-i(\omega-\alpha) t_1} \equiv \sum_{\nu \sim \omega-\alpha} \hat{B}_\nu\spc e^{-i\nu t_1}.
\eea
We imagine that wave packet A carries some information encoded in the relative phases in this sum. 
Observer ${\bf C}$ has access to the state 
\bea
e^{-i\omega t_1- i\alpha(t_0-t_1)} \,\hat{B}_{\omega-\alpha} \hat{A}_\alpha \li M \ra. 
\eea
Let us suppose that ${\bf C}$ knows how to decode $\hat{A}$, but has no knowledge of what $\hat{B}$ looks like.  In other words, since ${\bf C}$ can only detect $\hat{A}$, she can act with $\hat{A}$ on the left of the state and decode whatever phase coherent signal she can extract from it. But since $\hat{A}$ is located on the right of $\hat{B}$, the signal is hidden behind the noise. No problem, so far: all ${\bf C}$ has to do is to interchange the order of $\hat{B}$ and $\hat{A}$, so that $\hat{A}$ acts from the left. So the exchange algebra becomes an intrinsic part of the story. 

The interchange proceeds via the R-operation $\hat{B}_{\omega-\alpha} \hat{A}_\alpha = {\cal R}_{\alpha\beta} \hat{A}_{\omega-\beta} \hat{B}_\beta$. Here ${\cal R}_{\alpha\beta}$ is a microscopic property of the CFT  that, according to our proposal, is well approximated by the quantum 6j-symbol (\ref{qsixj}). After applying R to the wave packets, we can use geometric optics to localize the sums over frequencies and deduce that observer ${\bf C}$ sees the  state  
\bea
e^{{\frac i \hbar S_{\alpha\beta}}-i\omega \tilde{t}_0 - i\beta(\tilde{t}_1-\tilde{t}_0)} 
\hat{A}_{\omega-\beta} \hat{B}_{\beta}\, 
\li M \ra
\eea
with $S_{\alpha\beta} = \hbar \log({\cal R}_{\alpha\beta})$ and  $\beta$ determined by the stationary phase condition. 
Given our result that $S_{\alpha\beta}$ matches the gravitational scattering phase, we recover the relation (\ref{relno}). We again find (but now from the CFT side!) that  $\beta$ quickly grows and soon becomes bigger than $\omega$. This is when, on the gravity side, signal A  has disappeared from view. 

The shifted  time variables $\tilde{t}_0$ and $\tilde{t}_1$ are defined via the saddle point equations (\ref{timeshift}). We  see that the equation for $\tilde{t}_0$ has no real solution for $\beta>\omega$. Instead it gives that 
\bea
\tilde{t}_0  - t_0 = - \frac 1\kappa\log\Bigl(\frac{\beta -\omega}{\alpha}\Bigr)\pm \frac{i\pi}{\kappa}.
\eea
The time variable $\tilde{t}_0$ has moved off to a different Riemann sheet. Picking the leading branch, we find that the semi-classical  state accessible to observer ${\rm C}$ takes the form (up to a phase)
\bea
\label{outstate}
\qquad \quad e^{\pi (\beta-\omega)/\kappa}   \hat{A}_{\omega- \beta} \hat{B}_{\beta}\, \li M\ra, \qquad \quad \beta>\omega.
\eea

This supercritical situation is depicted in fig 4. It shows that $M+\beta$ represents the intermediate channel in the transition from  the
initial state $|M\ra$ to the final state at energy $M+\omega$. In the space-time diagram in fig 1, the intermediate state represents the future black hole interior.
The mode $A_{\omega-\beta}$, that encodes the signal, is now a lowering operator that maps the excited state $M+\beta$ to the lower state $M+\omega$. 
Crucially, the entropy of the $M+\beta$ state is larger than that of the final state $M+\omega$,  in the sense that its neighborhood has the larger spectral density.

Observer ${\bf C}$ tries to measure the phase information encoded in the lowering mode $\hat{A}$. Can she do it?  If $\hat{A}$ had been a raising operator,
the answer would have been: Yes. In that case, it would map states from a  lower energy band with lower level density $\rho_{\rm low} = e^{S_{\rm low}}$, to a higher energy band with higher level density $\rho_{\rm high} = e^{S_{\rm high}}$. So states produced by $\hat{A}$ are relatively sparse within the final energy band. The excess entropy $\Delta S = S_{\rm high}-S_{\rm low}$ is the information that observer ${\bf C}$ has to her disposal to decode the signal. On the other hand, if $\hat{A}$ is a lowering operator, it maps a high energy band with entropy $S_{\rm high}$ to a lower energy band with entropy $S_{\rm low}$. In this case, there is no reliable way for observer ${\bf C}$ to distinguish the final state from a random state without any phase coherence. In this sense, signal $\hat{A}$ has indeed disappeared behind a black hole horizon.

 Now suppose that observer ${\bf C}$ does in fact know what mode $\hat{B}_\beta$ looks like. With this extra bit of knowledge, the entropy of the state $\hat{B}_\beta\li M\ra$ 
at energy $M + \beta$ is reduced to $S_{\rm Cardy}(M)$. This is smaller than the entropy $S_{\rm Cardy}(M+\omega)$ of the final state. So we're back in the first situation in which ${\bf C}$ is able to read the signal. From the dual  perspective, giving observer ${\bf C}$ the knowledge of mode $\hat{B}$ amounts to giving her  access to the black hole interior. From the interior, the lowering mode is a real particle and the state (\ref{outstate}) is obtained by acting with a creation operator 
\bea
\hat{A}^\dag_{\beta-\omega} \hat{B}_\beta \li M \ra \qquad {\rm with} \qquad \hat{A}^\dag_{\beta-\omega} \equiv e^{\pi(\beta-\omega)/\kappa} \hat{A}_{\omega-\beta}.
\eea
This formula for the interior creation operator, including the inverse of the square root of the Boltzman factor, is reminiscent of the recovery operator of a quantum error correcting code. By giving observer ${\bf C}$ access to the information of mode ${\bf B}$, she can perform the projection on a suitable code subspace of the Hilbert space, and recover the signal \cite{VV}.

\medskip

\subsection{Organization}

The organization of this paper is as follows. In section \ref{scatteringmat} we describe in some detail the calculation of the semi-classical scattering matrix $R_{\alpha\beta}$ from 2+1-D gravity. In sections \ref{bootstrap} and \ref{modulargeometry} we summarize the main results of the modular bootstrap and of Virasoro modular geometry and its connection with quantum Teichm{\ddotu}ller space. In section \ref{exchalgebra} we discuss the exchange algebra of Lorentzian 2D CFT and establish the match between CFT and gravity. We present our concluding remarks in section \ref{conclusion}. Some technical formulas  are collected in the Appendix. Most of the rest of this paper is review of known results. The main new observation is the application of these exact results to gravitational scattering in AdS/CFT.

The motivating question of our paper, how the effect of the shockwave interactions can be derived from the structure of CFT conformal blocks, 
has also been studied by Dan Roberts and Douglas Stanford  \cite{DanDouglas}. We are coordinating the submission of both our papers.


\noindent
\section{Gravitational Scattering Matrix}\label{scatteringmat}


In this section we present the computation of  ${\cal R}_{\alpha\beta}$ from 2+1-D gravity. On the bulk side of the holographic duality, the R-matrix defines the scattering amplitude between an outgoing mode $A$ and the ingoing mode $B$ in the background geometry of a BTZ black hole. Our definition of the scattering amplitude, and strategy for computing it, are as follows.

First we identify the classical phase space ${\cal P}_{0,4}$ of two particles inside the eternal BTZ black hole geometry.  If we fix the mass of the black hole and the particles,  the configuration space is parametrized by picking two points on the 2D hyperbolic cylinder. Dividing out by the global isometries leaves a two-dimensional  configuration space, that can be identified with the Teichm{\ddotu}ller space ${\cal T}_{0,4}$  of the sphere with four punctures -- or more accurately,  with two holes (the two asymptotic regions of the black hole) and two punctures (the two particles). 
If we would ignore backreaction, the phase space would simply be the tangent space to the configuration space. 
As we will see,  after including the backreaction, the phase space takes the form ${\cal P}_{0,4} = {\cal T}_{0,4} \times T_{0,4}$, the product of two copies of Teichm{\ddotu}ller space.  This space has a standard symplectic form, which follows from the 2+1-D Einstein action.

We then parametrize the initial and final kinematical state of the two particles, in terms of a convenient set of canonical variables on  phase space. The initial and final quantum state are then defined as eigenstates of a commuting subset of these variables. For the initial state, this subset includes the mass $M_\alpha$ of the initial black hole, while for the final state it includes the mass $M_\beta$ of the final black hole (see figs 1 and 5). The initial and final coordinate variables do not commute with each other. So in particular $[M_\alpha, M_\beta] \neq 0$.
The quantum scattering matrix ${\cal R}_{\alpha\beta}$ is then defined as the unitary operator that implements the associated canonical transformation.



\begin{figure}[t]
\begin{center}
\includegraphics[scale=.61]{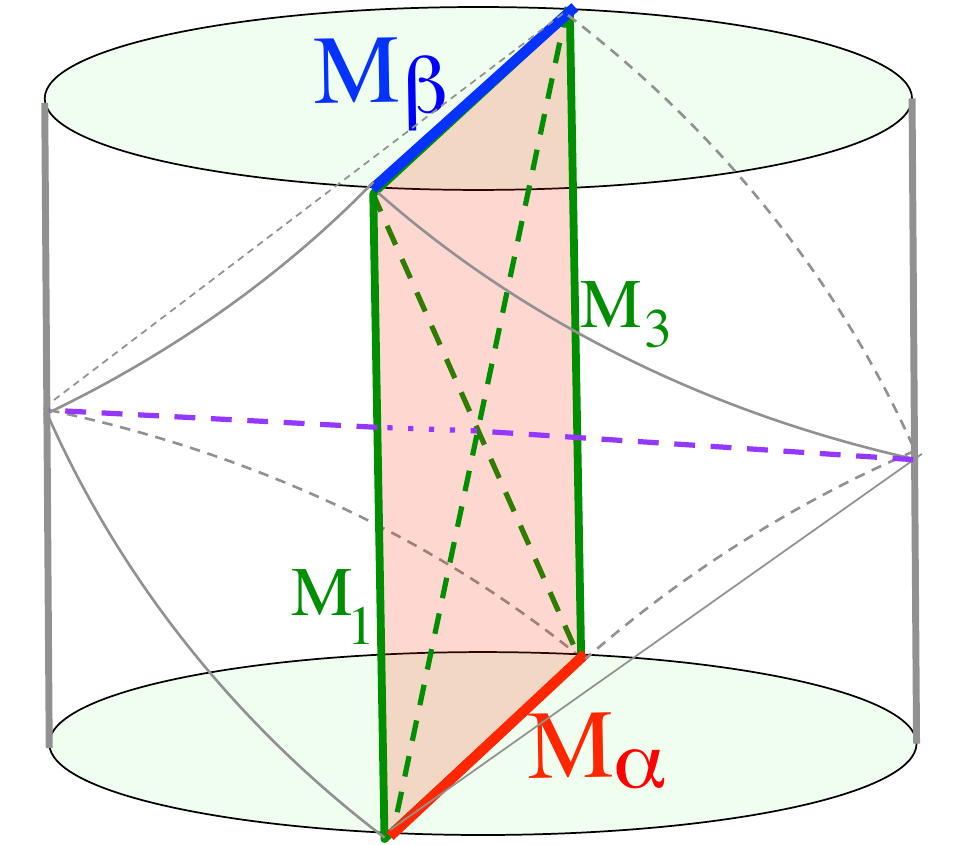}~~~~~~~~~~~~~~~~~~~~~~
\includegraphics[scale=.61]{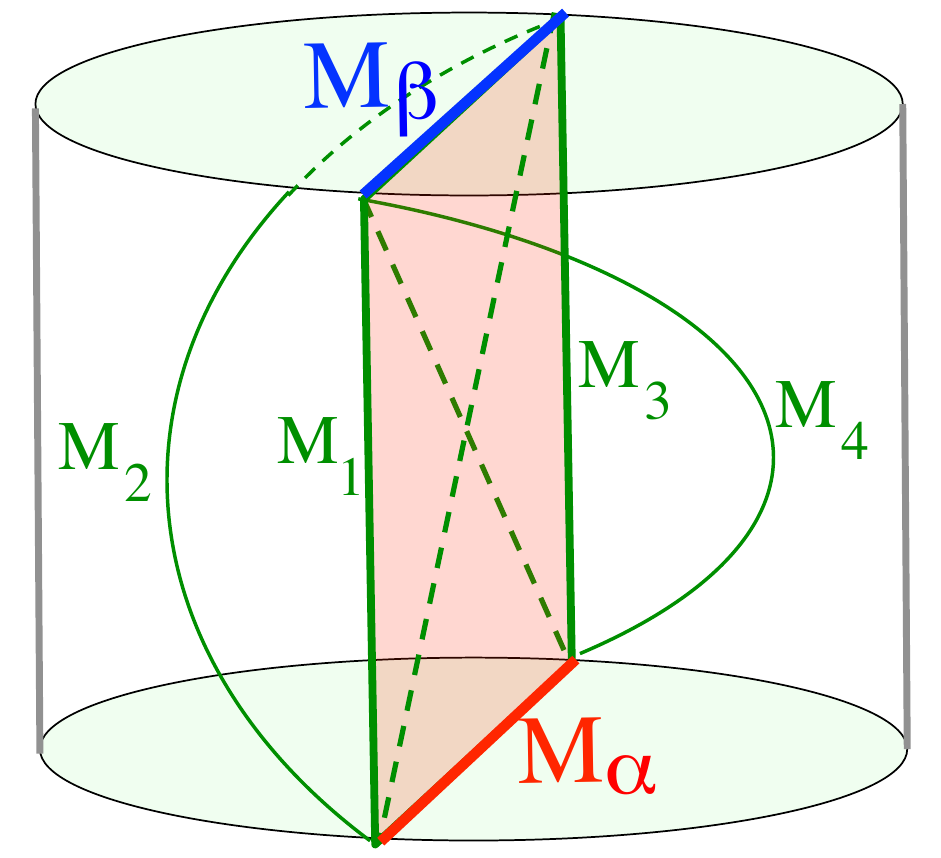}
\caption{The embedding of the Penrose diagram of fig 1 in the AdS${}_3$ cylinder. The left figure shows the two Poincar\'e wedges, connected at the corner points. The black hole horizon is indicated by the purple dashed line connecting the two corner points. The right
figure shows the two trajectories of the infalling and out going particle $B=M_2$ and $A=M_4$.}
\end{center}
\end{figure}

\subsection{Classical geometry}

The classical geometry of the BTZ black hole and of point particles in 2+1-d gravity is obtained by making global identifcations in AdS${}_3$. AdS${}_3$ is isomorphic to the $SL(2,\mathbb{R})$ group manifold; space-time points in AdS${}_3$ can thus be specified by group elements $g \in SL(2,\mathbb{R})$. The  AdS${}_3$ metric is given by the group invariant metric
\bea
\label{metric}
ds^2 \is - \ell^2 \tr(g^{-1} dg)^2
\eea
with $\ell$ the AdS curvature radius. The AdS${}_3$ isometry group $G = SL(2,\mathbb{R}) \times SL(2,\mathbb{R})$
 acts on the group element $g$ via left- and right multiplication $g \to h_+ g h_-.$

  To describe a non-rotating BTZ black hole of mass $M$, one useful coordinate choice is
\bea
\label{gparam}
g \is \left(\begin{array}{cc} \csc(\hat\rho) \spc e^{\hat\phi}  & \cot(\hat \rho) \spc e^{\hat \tau}  \\[2mm] \cot(\hat \rho) \spc e^{-\hat\tau}  & \csc(\hat \rho) \spc e^{-\hat \phi} \end{array}\right),\qquad \quad
\hat\rho = \frac{R\rho}{\ell},  \quad \hat \tau =\frac {R \tau}{\ell},  \quad \hat \phi = \frac{R \phi}{\ell},
\eea
with $\tau \in \mathbb{R}$, $\phi \in [0,  {2\pi }]$ and $\hat\rho \in [- \pi/2, \pi/2]$.
Here $R$ denotes the black hole Schwarzschild radius,  $R^2= 8M\ell^2$.
In this parametrization, the BTZ space-time takes the static form
\bea
\label{btzadstwo}
ds^2 \is R^2\, \Bigl( -  {\textstyle \cot^2({R}\spc \rho/\ell)}  dt^2\, +\, \frac{d\rho^2 + d\phi^2}{\cos^2({R}\spc \rho/\ell)} \, \Bigr).
\eea
The constant time slice is a 2D hyperbolic cylinder with two asymptotic regions connected by an ER bridge. The horizon is the geodesic at $\rho = 0$ with geodesic length $l=2\pi R$. 

The BTZ metric is invariant under shifts in $\phi$, but the group element  (\ref{gparam}) is not. In particular, a shift $\phi \to \phi+2\pi$ under a full period amounts  via  an isometry $g\to h_+g h_-$  with $h_+ = h_-=  e^{\pi R \sigma_2/\ell}$. Note that $R$ can take any positive real value.
The BTZ black  space-time is thus obtained by taking a quotient of the $SL(2,\mathbb{R})$ group manifold by the subgroup generated by the holonomy $h$.  
The space-time geometry of spinning black holes of mass $M$ and spin $J$ is obtained by taking  the quotient via a hyperbolic element $(h_+,h_-)\in G$
\bea
\label{quotient}
g \; \sim \; h_{+} g\spc  h_{-}, \qquad \qquad
h_\pm \is  e^{\pi (r_+ \pm r_-) \sigma_2/\ell}.
\eea
Here $r_+$ and $r_-$ denote the location of the outer and inner horizon, given by	
\bea
8\ell^2M\! \is \! r_+^2 + r_-^2, \qquad \ \ 
		4\ell J\, = \, r_+r_-\, .
\eea
The Bekenstein-Hawking entropy of the BTZ black hole is equal to 1/4 times the geodesic length of the black hole horizon $
S_{\rm BH} = {2\pi r_+}/{4}.$

The simplicity of the BTZ black hole space-time rests on two basic geometric facts: (i)~the Einstein equations in 2+1 dimensions enforce that the space-time outside any matter source is locally AdS${}_3$, (ii) in 2+1 dimensions, the space-time manifold ${\cal M}$ outside localized matter excitations  is non-simply connected. Every matter excitation creates a non-contractible loop $\gamma$ in ${\cal M}$, and thereby adds one extra element to the fundamental group $\Pi_1({\cal M})$. For every element of $\Pi_1 ({\cal M})$ one can associate a non-trivial space-time isometry $(g_+,g_-) \in G$, called the holonomy around the loop $\gamma$. The conjugacy class of the holonomy is determined by the mass and spin of the matter particles.

  In general, $SL(2,\mathbb{R})$ holonomies come in three types, depending on whether the conjugacy class of the group element is hyperbolic, parabolic, or elliptic. 
 For a black hole, both holonomies are in a hyperbolic conjugacy class; for localized point particles with mass $m$ in the subcritical regime $8\ell m < 1$, both elements $(g_+,g_-)$ are in the elliptic conjugacy class. For spinless particles $g_+=g_-$. The local geometry then has a simple conical singularity with angle $\theta$ related to the mass $m$ via  ${\theta}/{2\pi} = \sqrt{1-\ 8\ell m}.$

Classical 2+1-D space-times ${\cal M}$  with localized matter souces thus have a completely topological characterization in terms of a homomorphism
from the fundamental group $\Pi_1({\cal M})$ to the discrete subgroup ${\rm Hol}_*({\cal M}, G)$ of $G$ generated by the holonomies $(g_{+}^i, g_-^i)$ around all non-contractible loops $\gamma_i$. The holonomy is defined with respect to some arbitrary base-point, together with a choice of local frame. Changing the base point and local frame is a gauge symmetry that acts by overall conjugation of the holonomy elements. 
  
 This description can also be derived by considering the first order formulation of
classical 2+1-D gravity in terms of a triad $e^a$ and spin connection $\omega^a$. 
The linear combinations $
A^a_{\pm} = \omega^a \pm \frac 1 \ell e^a$
constitute two independent $SL(2,\mathbb{R})$ connections, in terms of which the Einstein action splits into a difference of two Chern-Simons
actions 
\bea
\label{csact}
S \is S_{CS}(A_+) - S_{CS}(A_-)\nonumber \\[-2mm]\\[-2mm] \nonumber
S_{CS}(A) \is \frac{\ell}{16\pi} \int  \tr\bigl(A \nspc \wedge \nspc d A + \frac{2}{3} A\nspc\wedge\nspc A\nspc \wedge \nspc A\bigl).\\[-6mm]\nonumber
\eea 
The  Einstein equation and torsion constraint take the form of flatness constraints:  $A_\pm$ are flat everywhere except at the localized matter sources. The group elements 
$g^i_\pm$ then coincide with the holonomies $g^i_\pm = P\exp \oint_{\gamma_i} A_\pm$ of the flat $SL(2,\mathbb{R})$ gauge fields around the closed loops $\gamma_i$ surrounding each matter source.
  
The classical phase space of a dynamical system is identical  the space of classical solutions.
We thus find that the phase space of 2+1-D gravity defined on ${\cal M}$ factorizes\footnote{\small \addtolength{\baselineskip}{.3mm} The holonomy group associated with the product group $G = SL(2,\mathbb{R}) \times SL(2,\mathbb{R})$ factorizes into the product of the holonomy groups of each $SL(2,\mathbb{R})$ factor.} into the product  ${\cal P} = {\cal T}_+ \times {\cal T}_-$ where  each factor ${\cal T}_\pm$ is given by the moduli space of flat $SL(2,\mathbb{R})$ connections over ${\cal M}$, with specified holonomies around each local matter source. The phase space ${\cal P}$ includes all positions and momenta of the matter particles. Unless the background space-time has non-trivial topology, we can identify the geometric phase space ${\cal P}$ with the phase space of the matter particles. This identification is supported by simple counting: each additional point particle adds one extra holonomy element $(g_+^i,g_-^i) \in G$. Of the six extra phase space variables per particle, four are the position and momentum, the other two are associated with the spin.

Given that all particle trajectories must pass through any given time slice, we can measure all holonomies of the 2+1-D space time within 
a constant time slice. Via the uniformization theorem the space of flat $SL(2,\mathbb{R})$ bundles on a 2D surface $\Sigma$ is isomorphic to Teichm{\ddotu}ller space, the space of constant negative curvature metrics on $\Sigma$. We will further clarify this identification in section 4, where we will use it to establish a precise connection between 2+1-D gravity and the modular geometry of conformal blocks in 2D CFT.

\begin{figure}[t]
\begin{center}
{\includegraphics[scale=.8]{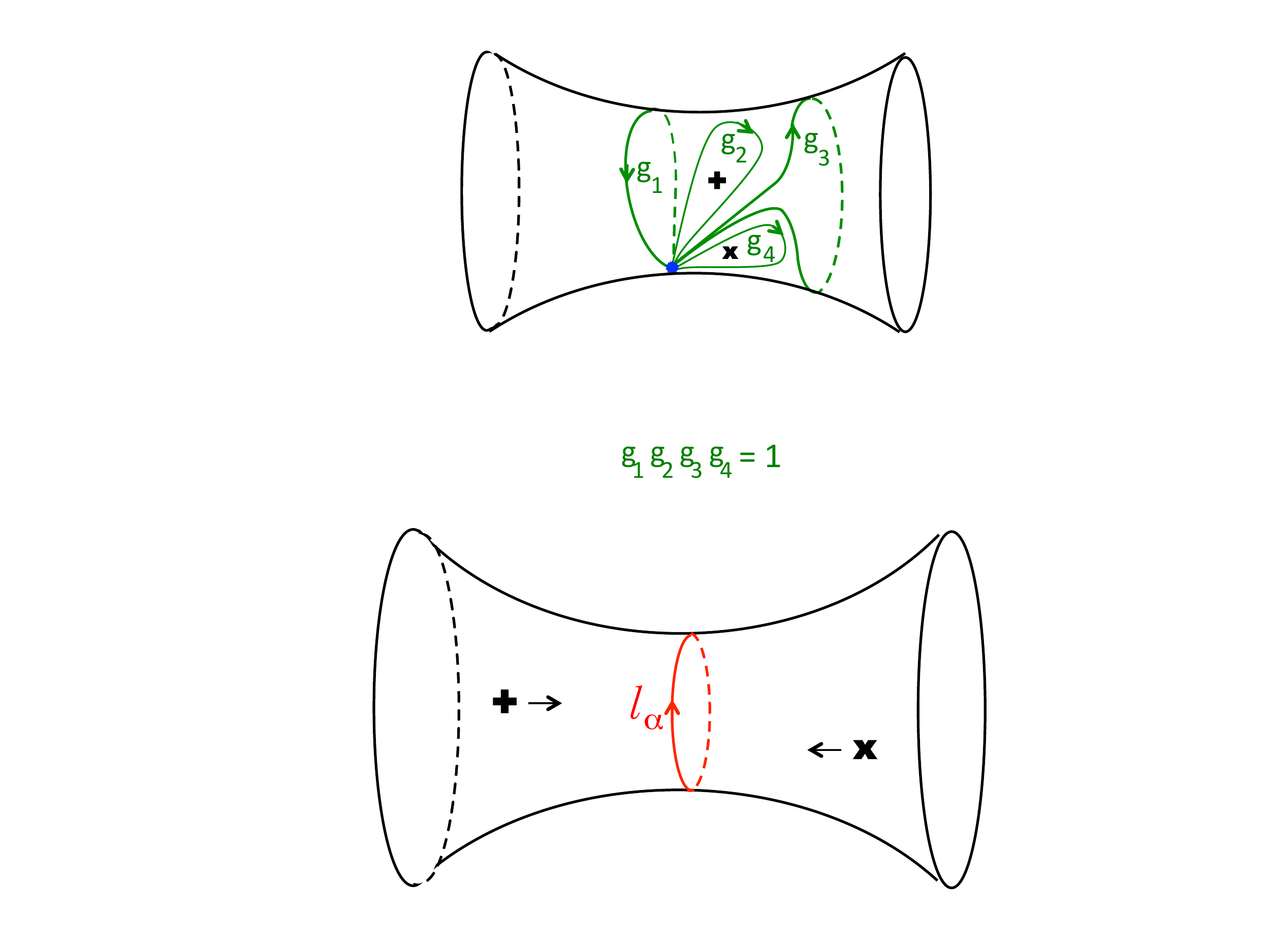}} 
\caption{The Teichm{\ddotu}ller space ${\cal T}_{0,4}$ of the hyperbolic cylinder (spatial section of a BTZ black hole) with two asymptotic regions and two punctures is parameterized by four holonomies $g_i \in SL(2,{\mathbb{R}})$ that satisfy the relation $g_1 g_2 g_3 g_4=1$, modulo overall conjugation.  }
\end{center}
\vspace{-.2cm}
\end{figure}  

Let us finally specialize to the case of interest. From the above general description, we can now confirm that the phase space ${\cal P}_{0,4}$ of two matter particles moving in the background of an eternal BTZ space-time is given by the product of two copies of the Teichm{\ddotu}ller space of the four-punctured sphere. For our purpose, the most convenient description of this space ${\cal T}_{0,4}$ is via its identification with the space of flat $SL(2,R)$ bundles
\bea
\label{teich}
{\cal T}_{0,4}  \is \Bigl\{ (g_1,g_2,g_3,g_4) \in SL(2,\mathbb{R})^4\, \Ri \; \tr(g_i) = L_i, \; g_1 g_2 g_3 g_4 =1\Bigr\}\mbox{\Large /} SL(2,\mathbb{R})\, .
\eea
Here the $SL(2,\mathbb{R})$ symmetry acts via simultaneous conjugation $g_i \to h g_i h^{-1}$. 
The relation $g_1 g_2 g_3 g_4 =1$ follows from the corresponding relation in $\Pi_1$.  For our application,  we must pick two of the  $SL(2,\mathbb{R})$ holonomies  to be hyperbolic, and the other two elliptic
\bea
L_1 \! \is \! \tr(g_1)  = 2 \cosh(l_1/2), \qquad \qquad 
L_2 = \tr(g_2)  = 2 \cos(\theta_2/2),
\nonumber\\[-2mm]\\[-2mm]
L_3 \! \is \! \tr(g_3) =  2 \cosh(l_3/2), \qquad \qquad
L_4 =\tr(g_4) =  2 \cos(\theta_4/2). \nonumber
\eea
To simplify our formulas somewhat, we will often restrict to the case of a non-rotating black hole and spinless particles with zero total angular momentum $j$. In this case, the conjugacy classes of the four holonomies are given in terms of the mass parameters (\ref{massp}) via
\bea 
  \frac{l_1}{2\pi} = \sqrt{8 M },\qquad  \frac{l_3}{2\pi} = \sqrt{8 (M \nspc+\nspc \omega)}, \qquad \frac{\theta_2}{2\pi} = \sqrt{1\nspc -\nspc 8\ell m_A}, 
 \qquad \frac{\theta_4}{2\pi} = \sqrt{1\nspc -\nspc 8 \ell m_B}.
\eea
Here $l_1$ and $l_2$ specify the horizon `area' (length) of BTZ black holes of mass $M$ and $M+\omega$, and $\theta_i$ are the conical angle associated with particle $A$ and $B$.

In the following we will keep all four mass parameters $M,\omega, m_A, m_B$ fixed.
The Teichm{\ddotu}ller space (\ref{teich}) is therefore $4 \times 3 - 4 - 3 -3 = 2$ dimensional.\footnote{\small \addtolength{\baselineskip}{.3mm} This is in accord with the familiar fact that the space of complex structures on the four-punctured sphere is parametrized by a single complex cross-ratio.}
The total phase space 
\bea
\label{ptt}
{\cal P}_{0,4} \is {\cal T}_{0,4} \times {\cal T}_{0,4}
\eea 
of the two particles is therefore 4-dimensional. This is four less than the usual $2\times 4 = 8$, since we have chosen to fix the total energy $\omega$ and angular momentum $j$ of the two particles. These are both first order constraints, that each reduce the phase space dimension by two.

\smallskip

\noindent
\subsection{Classical phase space}


As our next task, we need to determine the commutation relations between the physical phase space variables. Luckily, this work has already been done for us in \cite{Rosly}. We will now summarize their elegant calculation. Since the phase space takes the factorized form (\ref{ptt}), and because the Einstein action splits into a sum of two independent terms as in (\ref{csact}), we can focus on each of the two factors ${\cal T}_{0,4}$ separately, and combine them afterwards. 

The reader needs to be on guard, however, for a minor point of possible confusion. A point on Teichm{\ddotu}ller space  ${\cal T}_{0,4}$ is commonly specified by a 2D constant curvature metric on the hyperbolic cylinder with two punctures. It is tempting to identify this 2D constant curvature metric with the metric on a spatial slice of the 2+1-D geometry. However, this is in general {\it not} the correct identification: the 2D metric $g_{ij}$ on a spatial slice is obtained from the space components $e_i^\pm$ of the triad, which together with one of the spin connections $\omega^0$, also combines into a flat $SL(2,\mathbb{R})$ gauge field.\footnote{\small \addtolength{\baselineskip}{.3mm} The 2D torsion constraints $d e^\pm \mp [\omega_0, e^\pm] = 0$, and constant curvature condition $d\omega_0 = e^+ \wedge e^-$ combine into the $SL(2,\mathbb{R})$ flatness equation $F(A) =0$, with $A = (e^+,e^-,\omega)$.}
In contrast, the space components of the two  gauge fields $A_\pm = \omega^a \pm \frac 1 \ell e^a$ define {\it two} flat $SL(2,\mathbb{R})$ connections,
which are linear sums of the tetrad $e^a$ and $SO(2,1)$ spin connection $\omega^a$. We will try to steer clear of this possible confusion, by restricting ourselves to the case where the black holes and particles in the bulk all have zero spin and angular momentum. In this simplified situation, the quantum numbers are such that the spin connection $\omega$ effectively decouples, and all three Teichm{\ddotu}ller spaces can essentially be identified. It is completely straightforward, however, to extend the following analysis to the general case.

First we need to find some suitable coordinates ${\cal T}_{0,4}$.  Since the gauge invariant traces of single group elements $g_i$ are all held fixed, let us next look at the traces of products of two group elements. There are four combinations 
\bea
L_{\rmaa} \! \is \! \tr(g_1 g_2),\ \qquad \qquad \ \ L_{\sigma} \, = \,  \tr(g_1 g_3), \nonumber\\[-2.5mm]\\[-2.5mm]
L_{\rmbb} \! \is \! \tr(g_1 g_4),\ \qquad \qquad \ \ L_{\tau} \, = \, \tr(g_2 g_4) .\nonumber
\eea
These holonomy variables correspond to four  topologically distinct loops. However,
${\cal T}_{0,4}$ is two-dimensional, so there must be two relations among the four variables. These take the following form \cite{Rosly}
\bea
\label{relo}
L_\sigma + L_\tau\! \is L_1 L_3 + L_2 L_4 - L_{\rmaa} L_{\rmbb} \\[3mm]
L_\sigma  L_\tau + 4\! \is  L_{\rmaa}^2 + L^2_{\rmbb} +  L_1^2  + L_2^2 + L_3^2 + L_4^2  
\nonumber \\[-4mm]\label{relt} \\[-1mm]& & \ \ 
 - L_{\rmaa} (L_1L_2 \nspc+ \nspc L_3 L_4) - L_{\rmbb} (L_1L_4\nspc +\nspc L_2 L_3) \; +\; \nspc L_1L_2L_3L_4 . \nonumber
\eea
We will use these relations to solve for $L_\sigma$ and $L_\tau$. So our independent coordinates are $L_\alpha$ and $L_\beta$. They are related to the mass parameters $M_\alpha = M+\alpha$ and $M_\beta=M+\beta$  via
\bea
L_{\rmaa} \, = \, 2\cosh(l_{\rmaa}/2)\ \; \qquad &  & \qquad {l_\alpha}/{2\pi} = \sqrt{8 (M \nspc+\nspc \alpha)},
 \nonumber \\[-2mm]\\[-2mm]\nonumber
L_{\rmbb} \, = \, 2\cosh(l_{\rmbb}/2)\ \; \qquad &  & \qquad {l_\beta}/{2\pi} = \sqrt{8 (M \nspc+\nspc \beta)}.
\eea
Here $M_\alpha$, $M_\beta$, $l_\alpha$ and $l_\beta$ are the mass and horizon length of the initial and final black hole regions in figs 1 and 5.

The variables $L_\alpha$ and $L_\beta$ do not commute. Their commutator follows from the Einstein action. The first order form (\ref{csact}) is most convenient for this purpose.\footnote{\small \addtolength{\baselineskip}{.3mm} %
Here the CS form of the Einstein action is only used as a convenient short-cut towards the correct answer. In particular, we continue to require that the triad $e_\pm$ is always invertible. The results below can all be derived directly from the metric formulation of Einstein action.} The CS action can be rewritten in 2+1 notation as $
S_{CS} = \frac{\ell}{16\pi} \int \tr \bigl(A\wedge \partial_t A + A_t F(A)\bigr)$
with $F(A) = dA + A\wedge A$. Here $A$ and $\wedge$ denote 2D one-forms and wedge product. Applying the usual rules, one deduces \cite{WCS} that the symplectic form is given by (here $\hbar = 4\pi/\ell$ as before)
\bea
\label{wp}
\Omega = \frac{1}{2\hbar} \,\Omega_{\rm WP}, \quad & & \quad \Omega_{\rm WP} =  \int_{\cal M}\!\!\! \tr\bigl(\delta A \wedge \delta A\bigr).
\eea
restricted to the space of flat $SL(2,\mathbb{R})$ connections, $F(A)=0$, modulo gauge transformations. $\Omega_{\rm WP}$ is known as the Weil-Petersson symplectic form. It is the basic starting point for the quantum theory of Teichm{\ddotu}ller space. We will denote the associated Poisson bracket by $\{\ , \ \}_{\rm WP}$. In terms of the gauge potential, it takes the local form  
\bea
\label{acomm}
\{ A^a_i(x_1),A^b_j(x_2)\}_{{}_{\rm WP}} \is \epsilon^{ij} \delta^{ab}  \delta^2(x_{12}).
\eea
We can now compute the Poisson bracket between the holonomy variables $L_\alpha\nspc = \tr(\mbox{\small$P$} e^{ \oint_{\gamma_\alpha} A})$ and $L_\beta=\tr(\mbox{\small$P$}e^{ \oint_{\gamma_\beta} A})$. The two variables do not commute because (as seen in fig 5), the corresponding paths $\gamma_\alpha$ and $\gamma_\beta$ have an intersection point. From the  commutator  (\ref{acomm}) and the identity $\tr(gh^{-1}) + \tr(gh) = \tr(g) + \tr(h)$ for general $SL(2,\mathbb{R})$ matrices $g$ and $h$, one deduces the so-called skein relation
\bea
\label{aloopcomm}
\bigl\{\tr(P e^{ \oint_{\gamma_\alpha} A}), \tr(P e^{ \oint_{\gamma_\beta} A})\bigr\}_{{\!}_{\rm WP}} = \tr(P e^{ \oint_{\gamma_+} A})- \tr(P e^{ \oint_{\gamma_-} A}),\\[-2.5mm]\nonumber
\eea
where $\gamma_+$ and $\gamma_-$ are the two paths obtained by replacing the intersection via the anti-symmetric smoothing operation
\bea
\includegraphics[scale=.4]{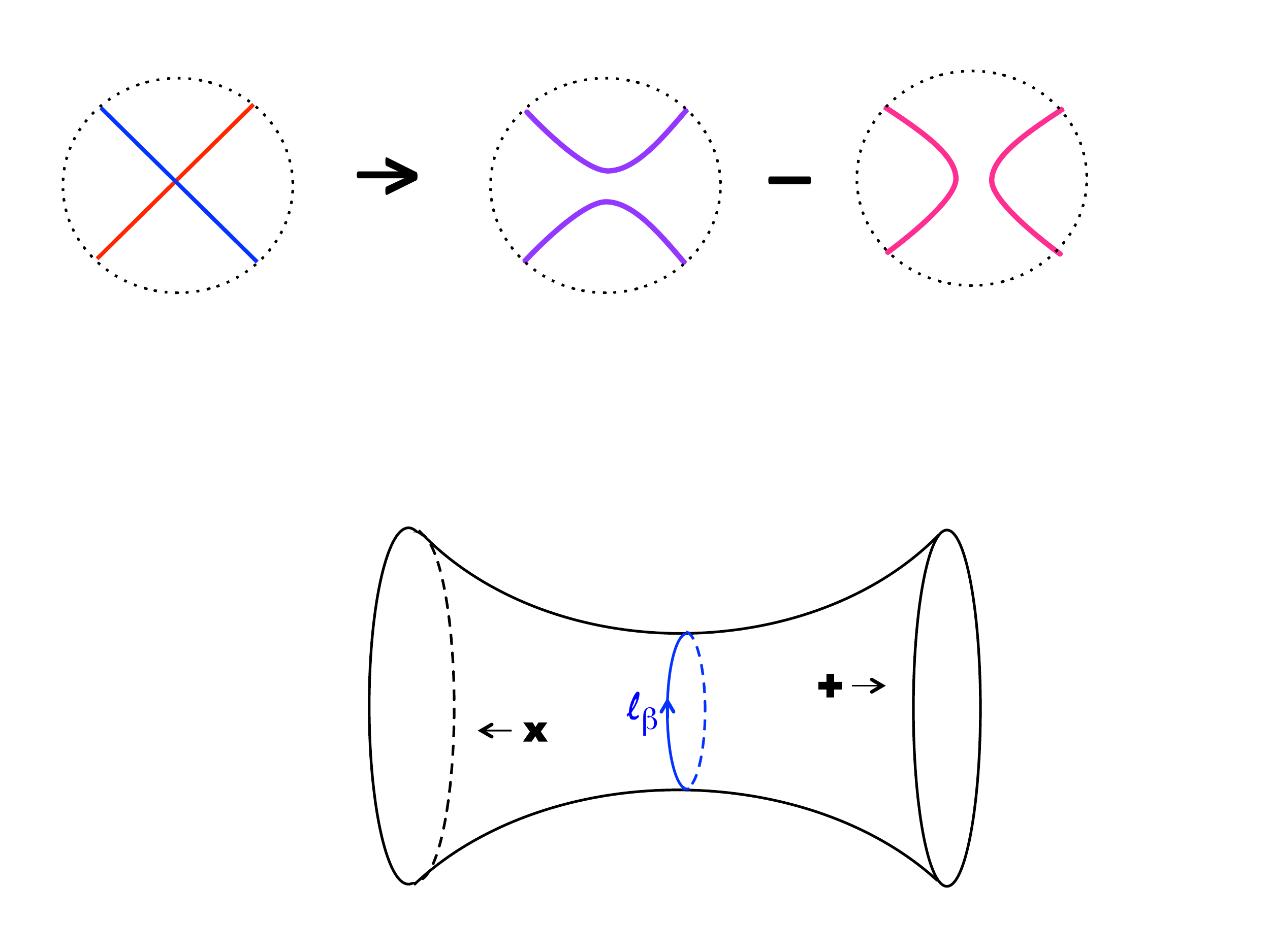} \nonumber
\eea
With a little bit of mental gymnastics, one recognizes that the two holonomy variables on the right-hand side of (\ref{aloopcomm})  are in fact equal to $L_\sigma$ and $L_\tau$.  We thus arrive at the following remarkably simple-looking result \cite{Goldman}
\bea
\label{goldman}
\bigl\{ L_{\rmaa}, L_{\rmbb} \bigr\}_{{\!}_{\rm WP}} \is L_\sigma - L_\tau.
\eea
This formula was first obtained by Goldman, without the use of the gauge theory formalism. Indeed, the variables $L_i$ represent the geodesic lengths of the corresponding curves, rather than traces of holonomies of a gauge field. 

\begin{figure}[t]
\begin{center}
\includegraphics[scale=.65]{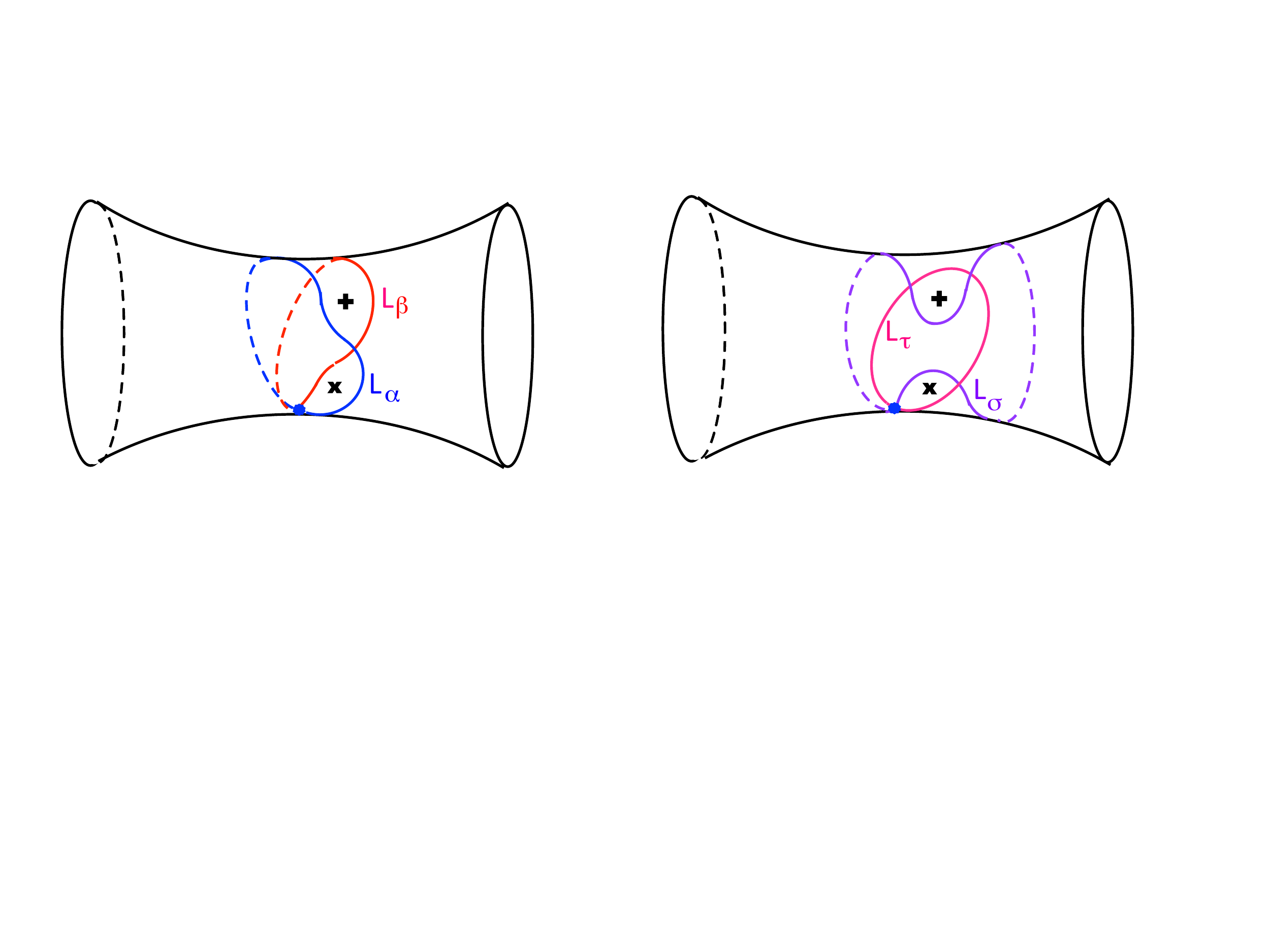}
\caption{The variables $L_\alpha$, $L_\beta$, $L_\sigma$ and $L_\tau$ are associated with four topologically distinct loops each of which surrounds a hole and puncture (left) or two punctures (right).}
\end{center}
\vspace{-0.2cm}
\end{figure} 

In spite of its apparent simplicity and geometrical beauty, eqn (\ref{goldman}) is not yet a suitable starting point for making a transition to the quantum theory.  The reason is clear: once we eliminate $L_\sigma$ and $L_\tau$ via eqns (\ref{relo}) and (\ref{relt}), the right-side becomes a complicated function of the phase space coordinates $L_\alpha$ and $L_\beta$. To make the quantization task more manageable, one typically looks for a set of Darboux coordinates, for which the symplectic form and Poisson bracket take a canonical form. 
A useful set of Darboux coordinates on Teichm{\ddotu}ller space \cite{Wolpert} are the so-called Fenchel-Nielsen coordinates, also known as `length and twists'. 

In our case, there are  two possible choices for the length coordinate: $l_\alpha$ or $l_\beta$. Each have their own canonical conjugate `twist' variable $\tau_\alpha$ and $\tau_\beta$ defined such that
\bea
\label{darboux}
& & \Omega_{\rm WP} = \spc dl_{\rmaa} \wedge d \tau_{\rmaa}\spc  =\spc  d l_{\rmbb} \wedge d\tau_{\rmbb}\nonumber \\[-2mm]\\[-2mm]\nonumber
& & \ \ \{ l_\alpha, \tau_\alpha\}_{{}_{\rm WP}} = \{ l_\beta, \tau_\beta\}_{{}_{\rm WP}} = 1
\eea
In terms of the complex geometry, a shift in the twist variable $\tau_\alpha \to \tau_\alpha + \delta$ acts by cutting the 2D surface open along the corresponding cycle $\gamma_\alpha$, rotating one side by an angle $\delta$, and gluing the two parts back again. The result that  the length and twists are Darboux variables was first shown by Wolpert \cite{Wolpert}. 
As we will see shortly, in terms of our scattering problem, $\tau_\alpha$ is a direct measure of the time-difference $t_0-t_1$ between the would-be arrival time $t_0$ of particle $A$ and the moment $t_1$ when particle $B$ is sent in.

\smallskip

\noindent
\subsection{Scattering matrix}

\smallskip

We are finally ready to define and compute the gravitational scattering amplitude between the outgoing particle $A$ and the infalling particle $B$ in the BTZ black hole background. First we specifiy the initial and final states.

The initial state decribes three objects: a BTZ black hole of mass $M$,  particle A that travels just outside of its horizon, and  particle  B that is falling in from asymptotic infinity. Particle A adds a finite amount of energy $\alpha$ to the black hole mass: from a distance, the geometry look like a single black hole of mass $M_\alpha=M+\alpha$. We take as a basis of initial states the eigenstates $\li \alpha\ra$ of the total mass operator that measures $M_\alpha$. So in particular
\bea
\hat{l}_\alpha \, \li \spc \alpha \spc \ra \is l_\alpha \li \spc \alpha \spc \ra.
\eea
Particle $B$ adds an additional amount of energy equal to $\omega-\alpha$. Again, we pick our basis states to be eigenstates of the  operator that measures the total energy $M_3 = M+\omega$. 

The final state also describes three objects: a BTZ black hole of mass $M$, particle B that has fallen in,  and particle A that has escaped to infinity. Particle B has added $\beta$ to the black hole mass, so from the outside,  it looks like a black hole of mass $M_\beta = M+\beta$. We choose as our basis of final states the eigen states $\li \spc \beta \spc\ra$ of $M_\beta$, satisfying
\bea
\hat{l}_\beta \, \li \spc \beta \spc \ra \is l_\beta \li \spc \beta \spc \ra\, .
\eea
Particle $A$ adds an additional amount of energy equal to $\omega-\beta$. As before, we pick our basis states to be eigen states of the total energy operator $M_3 = M+\omega$. 

The gravitational scattering matrix ${\cal R}_{\alpha\beta}$ is now simply defined as the overlap between an initial and a final basis state
\bea
\label{srmatrix}
{\cal R}_{\alpha\beta} \is \la \spc \beta \spc \ri \spc \alpha \spc \ra
\eea
From the previous discussion, we have learned that  the horizon lengths $\hat{l}_\alpha$ and $\hat{l}_\beta$  of the initial and final black holes do not commute with each other. The scattering matrix ${\cal R}_{\alpha\beta}$ should thus be thought of as the unitary operator that implements the canonical transformation between the Darboux coordinates $(l_\alpha,\tau_\alpha)$ associated with the initial state and the variables $(l_\beta,\tau_\beta)$ associated with the final state.
To construct this operator, we need to find the explicit relation between the two sets of  Darboux variables.
Luckily, also this calculation has already done for us at the semi-classical level in \cite{Rosly}, and at the full quantum level in \cite{TeschnerT,TeschnerR}. 
For now, we proceed with the semi-classical analysis. So we set 
\bea
 {\cal R}_{\alpha\beta} \is \exp\Bigl(\spc\frac {\raisebox{-2pt}{\small $i$}} {\raisebox{2pt}{\small $\hbar$}}\, S_{\alpha\beta}(l_\alpha,l_\beta)\spc \Bigr). 
\eea
Here $S_{\alpha\beta}(l_\alpha,l_\beta)$ is the generating function of the canonical transformation between the initial and final Darboux variables $(l_\alpha,\tau_\alpha)$ and $(l_\beta,\tau_\beta)$ 
\bea
\label{canons}
\tau_{\rmaa} \!\is \! \frac{\partial S
_{\rmab}
}{\partial l_{\rmaa}}, \qquad \qquad 
\tau_{\rmbb} \, = \,  -\frac{\partial S
_{\rmab}
}{\partial l_{\rmbb}}\, .
\eea

\begin{figure}[t]
\begin{center}
\raisebox{25mm}{\large $a)$}\ \; 
{\includegraphics[scale=.51]{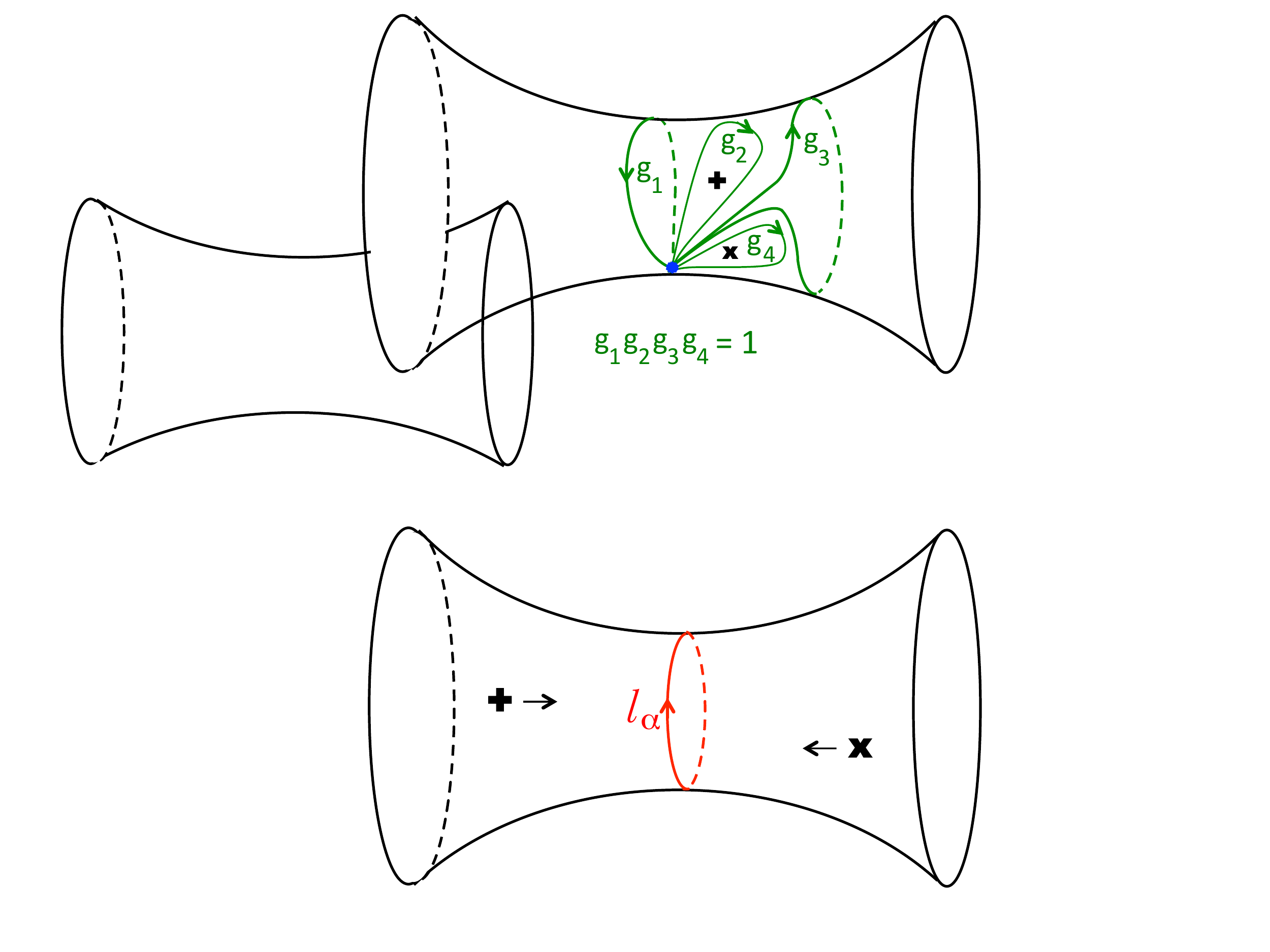}}\qquad \ \ \ \ \ 
\raisebox{25mm}{\large $b)$}\ \;  
{\includegraphics[scale=.51]{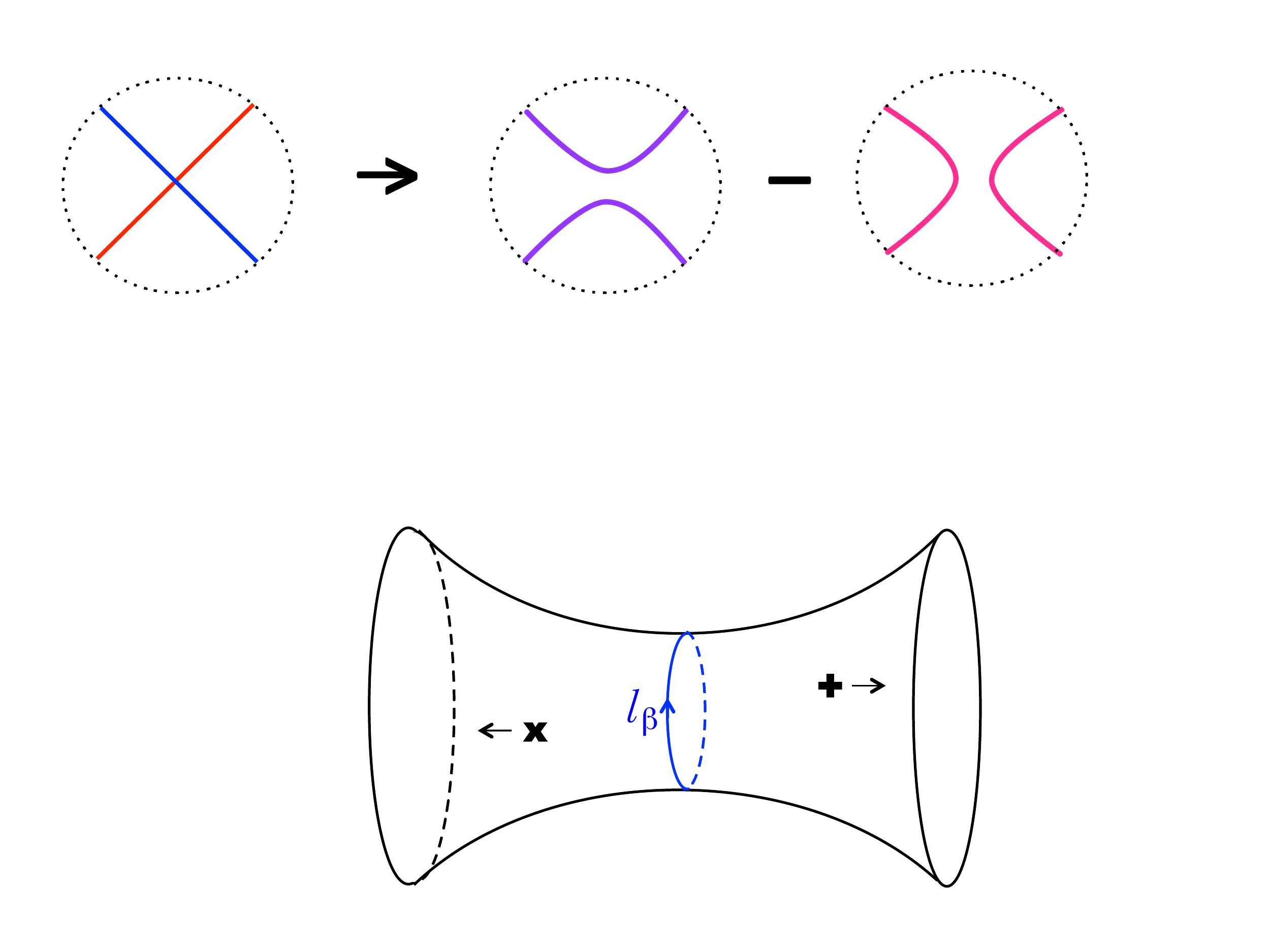}}

\caption{The gravitational scattering between outgoing particle $A$ (plus) and infalling particle B (cross). In the initial situation a), particle $A$ is inside a black hole of mass $M_\alpha$ and horizon length $l_\alpha$ and particle $B$ is outside.  In the final situation b), particle $A$ has escaped, and particle $B$ inside a black hole of mass $M_\beta$ with horizon length $l_\beta$. 
} 
\end{center}
\end{figure} 
The holonomy variable $L_\beta$ is expressed in terms of the initial Darboux coordinates as \cite{Rosly} 
\bea
\label{roslyrel}
L_{\rmbb}(L_{\rmaa}^2 \nspc - 4)\! \is\!  L_{\rmaa} (L_1 L_3\nspc + L_2 L_4)\nspc - 2 (L_2 L_3 \nspc + L_1 L_4) + 2 \cosh(\tau_{\rmaa}) \sqrt{c_{12}(L_{\rmaa}) c_{34}(L_{\rmaa})}\nonumber\\[-1mm] \\[-2mm]
& & \qquad c_{ij}(L_{\rmaa}) \, = \, L_{\rmaa}^2 + L_i^2 + L_j^2 - L_{\rmaa} L_i L_j - 4 \nonumber
\eea
This is a rather complicated looking relation. To understand its physical meaning, let us return to the discussion in Section 1.
Eqn (\ref{roslyrel}) should be compared with the kinematic relation (\ref{relno}) that expresses the final energy $\beta$ of the infalling particle B in terms of the initial energy of the outgoing particle A and the initial time difference $t_\alpha = t_0-t_1$. 

As a first check, let us compare the time coordinates, $t_\alpha$, which satisfies  the canonical relation $[ \alpha, t_\alpha] = i$ (see eqn (\ref{hj})) with the `twist' coordinate $\tau_\alpha$, defined via the canonical relation  $[ l_\alpha, \tau_\alpha] = i\hbar\spc \{ l_\alpha,\tau_\alpha\}_{{}_{\rm WP}} = i\hbar$ with $\hbar = 4\pi/\ell$. We can  confirm that the twist coordinate $\tau_\alpha$ can be identified with the time difference $t_\alpha$, via
\bea
\tau_\alpha \, = \, \kappa\, t_\alpha,& &\qquad\  \kappa\, = \,  \hbar\, \frac{\partial M_\alpha}{\partial l_\alpha} \, = \, \frac{R}{\ell^2},  \qquad \quad \ \hbar \, = \, \frac{4\pi}\ell,  
\eea
where we used that $l_\alpha = 2\pi \sqrt{8 (M+\alpha)} \simeq 2\pi \sqrt{8M} + \hbar\alpha/\kappa$. So (\ref{roslyrel}) and (\ref{relno}) exhibit the same exponential dependence on the time difference $t_\alpha$. To make the match complete, we should further specialize to the case where particles A and B are massless, and the regime where $\alpha$, $\beta$ and $\omega$ are small compared to $M$. In this regime we can make the replacements
\bea
\label{sim}
L_2 \spc \simeq \spc L_4\spc \simeq \spc 2, \quad \
\frac {L_3}{L_1} \, \simeq\, 1 +\,  \frac{\hbar \omega}{\kappa},
\   & &    \frac{L_\alpha}{L_1} \, \simeq\, 1 +   \, \frac{\hbar \alpha}\kappa ,\quad\  \frac{ L_\beta}{L_1}\, \simeq\, 1 + \, \frac{\hbar \beta}\kappa.\nonumber
\eea
A simple calculation now gives that (\ref{roslyrel}) takes the form
\bea
\beta \is \omega - \alpha +  2 \alpha \smpc (\omega-\alpha)
\spc e^{\kappa (t_\alpha-t_R)}.
\eea
with $\kappa t_R \simeq \log(\cosh(\pi R/\ell))$ + const. We see that, apart from a different off set of the time delay (which can't be fixed by the arguments presented here), there's a precise match with equation (\ref{relno}). This is not surprising, since both calculations are based on the same classical action. The lesson learned, however, is that our physical interpretation of eqn (\ref{roslyrel}) as describing the gravitational scattering process is confirmed, and that our schematic semi-classical analysis of the introduction has now been substantially refined.

Together with eqn (\ref{roslyrel}) and its time reversed partner, eqns (\ref{canons}) uniquely specify  the scattering phase $S_{\alpha\beta}$. 
Integrating (\ref{canons}) still looks like a substantial algebraic task. Given the above geometric formulation of the problem,  however, it is  not surprising that the solution can be found in terms of a natural geometric quantity. As announced in the introduction,
it turns out that the  $S_{\alpha\beta}(l_\alpha,l_\beta)$ can (up to a trivial phase) be identified with the volume of a hyperbolic tetrahedron,\footnote{\small \addtolength{\baselineskip}{.3mm} More accurately, $S_{\alpha\beta}$ denotes the excluded volume of the knot-compliment of a tetrahedron. This supports the interpretation that $S_{\alpha\beta}$ is equal to the Einstein action evaluated on the classical BTZ space-time, with two conical defects, that describes the scattering process.} with 
dihedral angles $\ell_i$, with $i=1,2,3,4, \alpha,\beta$.
\bea
\label{scatvol}
S_{\alpha\beta} =   {\rm Vol}\Bigl(\spc T\mbox{$\Bigl[\raisebox{2pt}{\fontsize{9.5pt}{.5pt}$\begin{array}{ccc} \! 1\! \! &\!\! {2} \! \!&\!\!\raisebox{1.25pt}{\fontsize{10pt}{.5pt}$\alpha$} \! \\[-1.5mm] \! 3\! \! &\!\! 4 \!\! &\! \! \raisebox{0pt}{\fontsize{10pt}{.5pt}$\!\beta$}
\!\! \end{array}$}\Bigr]$}\spc \Bigr)
\eea
 Here all geodesic lengths $l_i$'s are real, except for the two lengths associated with the conical defects:   $l_2 = i \theta_2$ and $l_4 = i \theta_4$.  The formula for $S_{\alpha\beta}$ is given in the Appendix.

This concludes the technical part of this section. 
It is straightforward to verify that the above formula indeed reduces to the solution to eqns (\ref{hj}) in the limiting regime (\ref{sim}).  
We end with a few comments about the physical interpretation of this result.

$\bullet$ As we will see in section 4, the semi-classical scattering matrix ${\cal R}_{\alpha\beta}$ lifts to a full unitary operator acting on the Hilbert space ${\cal H}_{0,4}$ spanned by all the energy eigen states $\li \spc \alpha \spc \ra$. 
This unitarity property follows from the identification between the volume of hyperbolic  tetrahedra and the quantum 6j-symbols of $U_q(\mathfrak{sl}(2,\mathbb{R}))$, the q-deformed universal enveloping algebra of $SL(2,\mathbb{R})$. We will describe this correspondence  in more detail in the next sections, where we will show that ${\cal R}_{\alpha\beta}$ is identical to the braid operator acting on the Hilbert space ${\cal H}_{0,4}$ spanned by the Virasoro conformal blocks on the four punctured sphere. 

$\bullet$ An important output of the unitary condition is that it prescribes the form of the inner product, or equivalently, the spectral density $\rho(\alpha)$ of the Hilbert space ${\cal H}_{0,4}$. Given the form of ${\cal R}_{\alpha\beta}$, we need to make sure that there exists an integration measure $\rho(\alpha)$ such that
\bea 
\label{runitarity}
\int d\beta \, \rho(\beta) \, {\cal R}_{\alpha_1\beta} \, {\cal R}^*_{\beta \alpha_2} \is ({\rho(\alpha_1)})^{-1} 
\; \delta(\alpha_1-\alpha_2).
\eea
This integration measure corresponds to a scalar product
\bea
\label{scalarp}
\la\spc \alpha_1 \spc \li \spc \alpha_2 \spc \ra \is  ({\rho(\alpha_1)})^{-1} \; \delta(\alpha_1-\alpha_2)
\eea 
On physical grounds, we should expect that the spectral density coincides with the Bekenstein-Hawking entropy  (recall that the Hawking temperature  equals $T_H = \kappa/2\pi$)
\bea
\rho(M+\alpha) \! \is\! \rho(M) \, e^{{2\pi\smpc \alpha}/\kappa}.
\eea
This match with the BH spectrum is crucial for the consistency of our story. Luckily, as we will see, this physical expecation is  confirmed by mathematics. 

$\bullet$ The integral  in (\ref{runitarity}) runs over both positive and negative values of $\beta$. Indeed, to obtain a unitary scattering matrix, it is necessary to include states $\li \alpha\ra$ with $\alpha< 0$. Since $\alpha$ measures the relative energy between two black hole states $\li M \ra$ and $\li M+\alpha\ra$, states with negative values of $\alpha$ are black hole states that are lighter than $\li M\ra$, or states in which the trajectory of particle A or B is (shifted to) behind the (black or white hole) horizon.

$\bullet$ There is one unusual aspect of our scattering set up which needs some further discussion. We have presented our computation as that of an ordinary $S$-matrix between asymptotic particle states. However, both the initial and final states contain at least one and possibly two particles
that end up in the interior of the black hole. These particles have no asymptotic states. So how can we even talk about a scattering matrix?

Our formalism  incorporates the gravitational backreaction. It thus can distinguish initial and final black hole states before and after a particle has escaped or fallen in. Moreover, it is designed to give a global perspective on the geometry, including the black hole interior. 
The scattering matrix should therefore be thought of as describing the elastic scattering 
$$
\li M + A\ra \, \li B\ra\; \to\; \li M + B\ra\, \li A\ra,
$$
or depending on whether A also gets absorbed by the black hole, as the fusion process
$$
\li M + A \ra \, \li B \ra \; \to \; \li M+ A + B\ra.
$$
Note that in this case the final state has higher entropy than the the initial state, in the sense that its Hilbert space neighborhood has a larger spectral density than the initial state.

\begin{figure}[t]
\begin{center}
${}$~~~~\includegraphics[scale=.58]{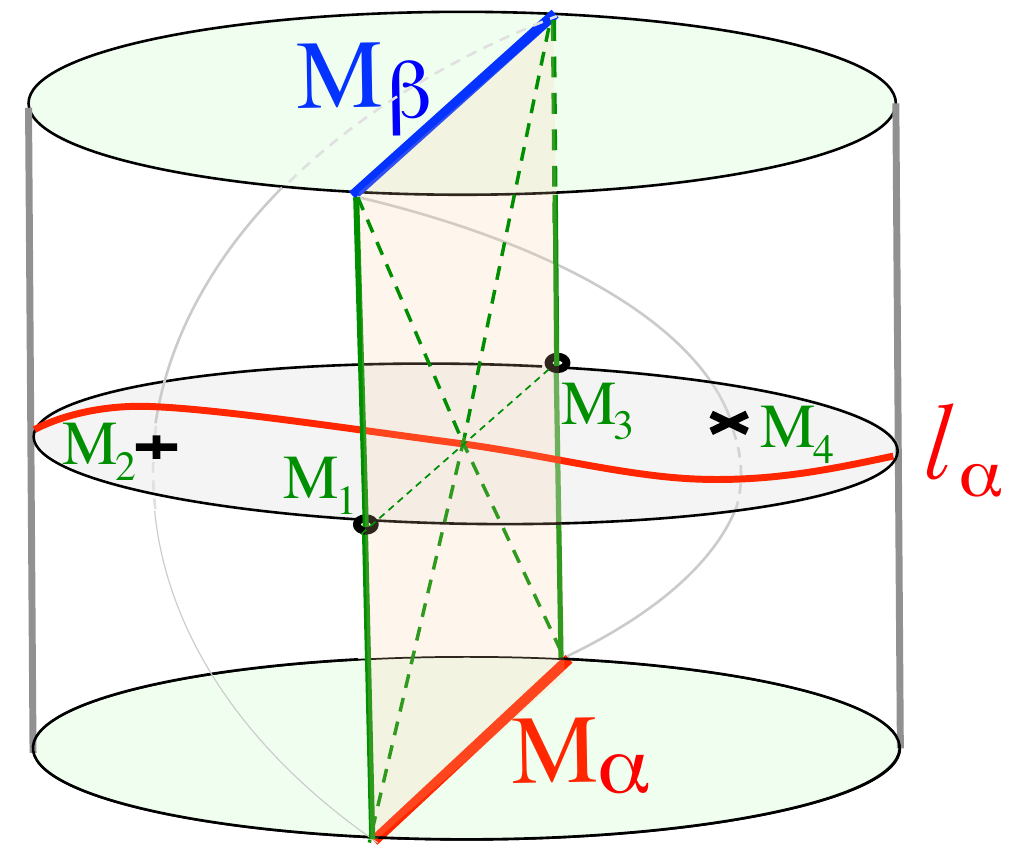}~~~~~~~~~~~~~~~~~~~~
\includegraphics[scale=.58]{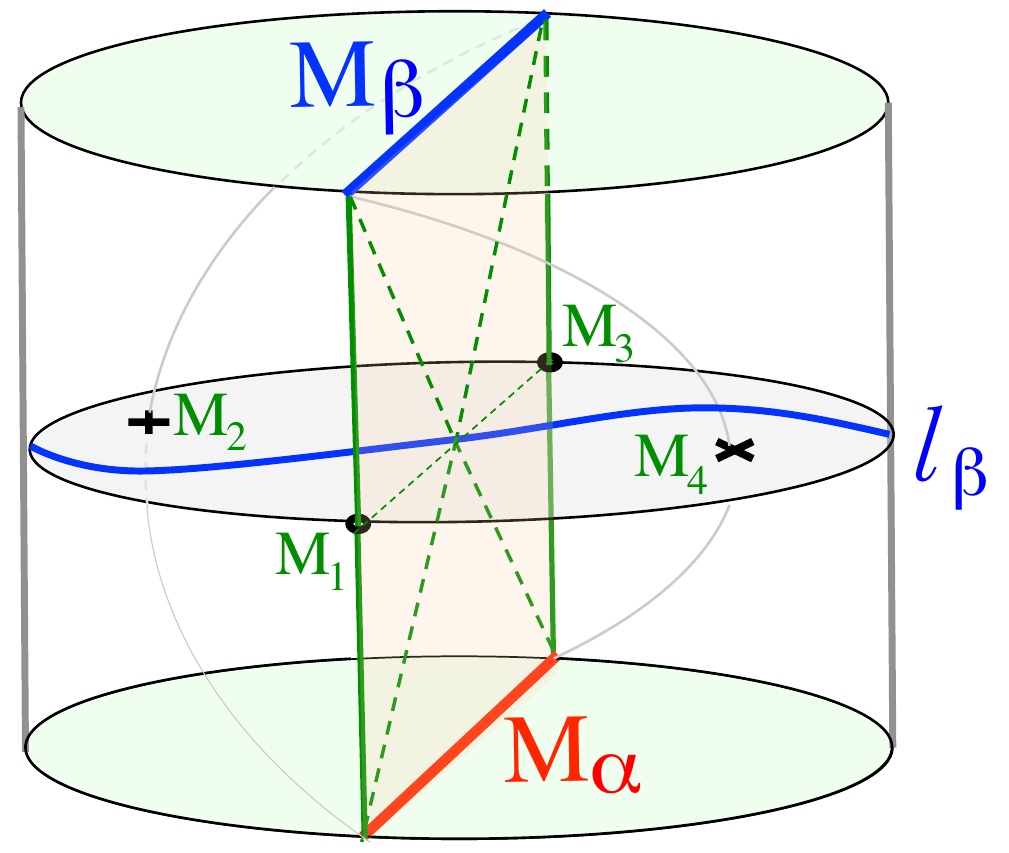}
\caption{A global 2+1-D perspective of the two geodesics $\gamma_\alpha$  with length $l_\alpha$ (left) and $\gamma_\beta$ with length $l_\beta$ (right), both drawn on the same constant time slice inside the AdS${}_3$ cylinder. The geodesics intersect and as a result, the length operators $\hat{l}_\alpha$ and $\hat{l}_\beta$ do not commute.}
\end{center}
\vspace{-5mm}
\end{figure}

$\bullet$ We represent the in- and out-states as two particles traveling on a 2D hyperbolic geometry with two asymptotic regions, corresponding to the spatial section of an eternal black hole space time. It looks therefore as if the Hilbert space should factorize into two sectors, one associated with each asymptotic region.  (See also fig. 9.) Our interest, however, is to apply and relate these results to the holographic map between a single 2D CFT and a one-sided black hole, with only one asymptotic region. So how can we talk about the global geometry of an eternal black hole, and what is the meaning of the other asymptotic region?

This question is  closely related to the firewall issue \cite{AMPS}, and a complete physical answer would go beyond the scope of this paper. All we can say here is that the phase space, whose quantization gives us the Hilbert space in which  the R-matrix ${\cal R}_{\alpha\beta}$ acts as a unitary operator, looks like that of particles moving on the two-sided black hole geometry. Moreover, as we will see in the next section,
this same matrix ${\cal R}_{\alpha\beta}$ will appear as the R-matrix that describes the braiding properties of Virasoro conformal blocks, and appears in the exchange algebra between local operators of a single 2D CFT. This is of course a strong hint that the CFT in fact knows more about the interior black hole geometry than one might have thought.

\renewcommand{\rmaa}{{\rm a}}
\renewcommand{\rmbb}{{\rm b}} %
\renewcommand{\rmcc}{{\rm c}} %
\renewcommand{\rmdd}{{\rm d}} %
\renewcommand{\rmee}{{\rm e}} %
\renewcommand{\rmff}{{\rm f}} %
\renewcommand{\rmab}{{\rm ab}} %
\renewcommand{\rmac}{{\rm ac}} %
\renewcommand{\rmbc}{{\rm bc}} %


\noindent
\section{Modular Bootstrap}\label{bootstrap}


We will consider a general 2D CFT with a large central charge $c\gg 1$, and with no other exact continuous symmetries than conformal invariance. We thus assume that the chiral algebra consists of only the Virasoro algebra, generated by the $L_n$'s and $\bar{L}_n$'s. The class of such theories is usually referred to as  irrational CFTs. In contrast to rational CFTs which are charactized by the presence of a large chiral symmetry algebra with only a finite number of representations, irrational CFTs have an infinite spectrum of primary fields, with a large asymptotic level density. More precisely, the primary fields ${\cal O}_a(x)$ divide into light fields, with conformal dimension $\Delta_a< c/12$, and heavy fields, with conformal dimension $\Delta_a> c/12$ \cite{Tom}. We will assume that the theory
has a small number of light primary fields and a large number of heavy states, with level density $\rho(\Delta,\bar{\Delta}) = e^{S(\Delta,\bar{\Delta})}$ that saturates the Cardy entropy formula \cite{Cardy}
\bea
\label{cardy}
S(\Delta,\bar{\Delta}) \! \is\! 
2\pi \sqrt{\frac c 6 \spc \Delta_c} + 2\pi \sqrt{\frac c 6 \spc \bar\Delta_c}\; 
\, = \, S_L(\Delta) + S_R(\bar{\Delta}),
\eea
with $\Delta_c = \Delta-\mbox{\large  $\frac c {24}$}$.
Our restriction to irrational CFT's with a sparse light spectrum  is aimed at ensuring that the dual theory, at low energies, will be accurately described by Einstein gravity coupled to a few light matter fields \cite{Tom}.  The heavy operators then describe BTZ black hole states with Bekenstein-Hawking entropy  (\ref{cardy}).
 
For a generic irrational CFT, we expect the spectrum of conformal dimensions to be non-degenerate. The Cardy formula (\ref{cardy}) does therefore not count the exact degeneracy of a given level $(\Delta,\bar\Delta)$, but the average inverse level spacing in the neighborhood of the level. Note further that for $c\gg 1$, the Virasoro algebra contributes only a small fraction of the entropy: most of the entropy is supplied by the dense spectrum of primary states. 

The fact that (\ref{cardy}) looks like the sum of a left- and a right-moving entropy reflects the dynamical decoupling of the two chiral sectors. 
As we will see, however, it is somewhat misleading to view the two chiral sectors as completely decoupled: the two sectors are intimately linked by the non-trivial and restrictive requirements of locality and crossing symmetry. As a result,
the allowed pairs  $(\Delta_{\rmaa}, \bar{\Delta}_{\rm \bar{a}})$ of left and right conformal dimensions are highly correlated. This entanglement between the two sectors is instrumental for the conformal bootstrap program for 2D CFTs.\footnote{\small \addtolength{\baselineskip}{.3mm}
In the following we will make use of this left-right correlation to simplify our notation and make an identification a = $\rm\bar{a}$ between left and right moving labels. }


\noindent
\subsection{Chiral vertex operators}

To set notation, we start with a lightning review of the spectrum of states and operators of a general 2D CFT on a space-time with topology of $S^1 \times \mathbb{R}$. The periodic space coordinate $x \in [\spc 0,\spc 2 \pi\smpc\rangle$ and  time coordinate $t$ combine into light-cone coordinates via $u = t + x$, $v = x-t$.
For any 2D CFT, we can find a basis of Hilbert states that factorize into left- and right-movers.
  We call the Hilbert subspace  ${\cal H}_{\alph}$ 
  spanned by all descendent states of a given chiral primary state $\li\spc \rm a \spc \ra$ a {\it chiral sector}.
The CFT Hilbert space decomposes into a direct sum of factorized sectors
\bea
\label{hfact}
{\cal H}_\CFT \is \raisebox{-5pt}{\Large $\Oplus\atop\raisebox{1pt}{\footnotesize$({\rmaa},\nspc\rm\bar{a})$}$}\; {\cal H}_\alph\, \sotimes \, {\cal H}{}_{\mbox{\footnotesize $\rm \bar a$}}
\eea
The pairing between the left- and right-moving labels $\rmaa$ and $\rm \bar a$ is constrained by locality. In particular, the momentum  ~$\Delta_\alph\!- \! \bar\Delta_{\bar\alph}$ of every state  must be an integer.

It will often be convenient to apply a Wick rotation, replacing $t$ by $\tau = it$, and to map the Lorentzian cylindrical space-time to the complex plane via $(z,\bar{z}) =(e^{i x + \tau} , e^{-i x + \tau})$ with $\tau = i t$
The infinite past thus gets mapped to the origin $z=0$, and the infinite future  to $z=\infty$.  The inverse mapping from $\mathbb{C}$ to $S^1\times \mathbb{R}$ is  radial quantization.  The operator state correspondence
\bea
\label{opstate}
\li\spc {\rm a}\spc \ra \is {\cal O}_{\rm a}(0) \li\spc 0\spc \ra,
\eea
provides a one-to-one map between primary states and primary operators. 
In eqn (\ref{opstate}) $\li\spc {\rm a}\spc \ra$ and ${\cal O}_{\rm a}$ are short-hand for $\li\spc {\rm a,\bar{a}}\spc \ra$
and ${\cal O}_{\rm a,\bar{a}}$, see footnote 10.

\def\vv{\mbox{\fontsize{10pt}{0.5pt}${\rm V}$}}
\def\nn{\mbox{\fontsize{11pt}{0.5pt}${\rm N}$}}


The left-right factorization (\ref{hfact}) of the Hilbert space, combined with the operator state correspondence (\ref{opstate}), suggests that  local operators can also be factorized into chiral components. Indeed,  the matrix element of a primay operator ${\cal O}_c(z,\bar{z})$ between two basis states factorizes as
\bea
\label{threepoint}
\la\smpc {\rmaa}, n \spc \ri  \spc {\cal O}_{\rmcc}(z,\bar{z})\smpc \ri \spc{\rm  b}, m   \ra \is \, f^{\rmcc}_{\rm \; ab}\; C_{nm}\; z^{\Delta_{\rmaa} - \Delta_{\rmbb} - \Delta_{\rmcc}+ n - m}\,  {\overline{z}}^{\bar{\Delta}_{\rmaa} - \bar{\Delta}_{\rmbb} - \bar{\Delta}_{\rmcc}+ \bar{n} - \bar{m}} \, .
\eea
The overlap coefficients $C_{nm}$ between the different descendant states are universal and uniquely determined by the Virasoro algebra and the conformal weights of the three primary sectors. They are normalized so that $C_{00} =1$.
The operator product coefficients 
\bea
f^{\rmcc}{}_{\! \rm ab} = \la\spc  {\rmaa} \spc \ri {\cal O}_{\rmcc}(1) \li {\rmbb} \ra
\eea
are real numbers, that depend on the specific CFT.

A general matrix element $\la \phi\smpc \ri  {\cal O}_{\rmcc}(z,\bar{z}) \li\smpc \psi \ra$
between two arbitrary states decomposes as a sum of factorized terms
$\la \phi\smpc \ri  {\cal O}_{\rmcc}(z,\bar{z}) \li\smpc \psi \ra = \sum_{\rm a,b} \spc \phi^*_{{\rmaa},n} \spc \psi_{{\rmbb},m}\spc \la {\rmaa},n \ri {\cal O}_{\rmcc} (z,\bar{z}) \ri {\rmbb},m\ra $.
Correspondingly, the left-right decomposition of a local operator $ {\cal O}_{\rmcc}(z,\bar{z})$ takes the  form
\bea
\label{lrdeco}
{\cal O}_{\mbox{\footnotesize c}}(z,\zbar) \is \sum_{\rm{a,b}} 
\;\, \psi^\rmcc_{\rmaa\rmbb} 
(z)\spc \otimes\spc
\; \bar\psi^\rmcc_{\rmaa\rmbb} 
(\zbar).
\eea
The chiral component $\psi^\rmcc_{\rmaa\rmbb} %
(z)$ is called a {\it  chiral vertex operator}. It associates to the chiral primary state $\li\spc  {\rmcc}\spc \ra$ a linear intertwiner map between two chiral sectors
\bea
\label{psimap}
\psi^\rmcc_{\rmaa\rmbb} 
(z) \!&\! : & {\cal H}_{\rmbb} \ \to \ {\cal H}_{\rmaa}.
\eea
The intertwiner property of the chiral vertex operator is equivalent to the statement that it satisfies the  conformal Ward identity.
Chiral vertex operators are invariant tensors, or Clebsch-Gordan coefficients, of the conformal group.\footnote{\small \label{foottwo} \addtolength{\baselineskip}{.3mm}The operator product expansion, as encoded in the chiral vertex operators, supplies a rule for taking tensor products of the chiral sectors ${\cal H}_{\rmaa}$ \cite{BPZ,EV,GregNati}. In other words, the CVOs prescribe the decomposition of the tensor product ${\cal H}_{\rmaa} \otimes {\cal H}_{\rmbb}$ into a direct sum
\bea
{\cal H}_{\rmaa} \otimes {\cal H}_{\rmbb} \is {{\raisebox{-8pt}{$\Oplus$}}\atop{\raisebox{0pt}{\scriptsize $\rmcc$}}}\; \vv^{\spc \rmcc}_{\! \rm a \rm b} \otimes\, {\cal H}_{\rmc}.
\eea
Here $\vv^{\rmcc}_{\! \rmab}$ denotes the multiplicity space, with dimension given by the fusion number 
$\nn^{\rmcc}_{\rmab}$ 
that counts the number of invariant tensors (CVOs) for a given fusion channel $
{\rm dim} \spc \vv^{\rmcc}_{\! \rmab} = \nn^{\spc \rmcc}_{\rmab}$
We will typically assume that $\nn^{\rmcc}_{\rmab}$ is either $0$ or $1$. 
It is natural to generalize the definition of the CVOs as elements of a triple tensor product of chiral sectors 
$$\Psi^\rmcc_{\rmab}(z) \, \in\, {\cal H}_{\rmaa} \spc \sotimes \,  {\cal H}_{\rmbb}\spc \sotimes\, {\cal H}_{\rmcc}
$$
This definition treats the three chiral sectors symmetrically and is therefore most suitable  starting point for braid operations that permute the different sectors.\\[-3mm]} 

Their matrix elements between basis states are holomorphic monomials in $z$
\bea
\label{cvo}
\la\smpc {\rmaa}, n\spc \ri  \spc \psi^{\rmcc}_{\rmaa\rmbb} 
(z)\smpc \ri \spc {\rmbb}, m\spc  \ra \is \lambda^{\rmcc}{\!}_{\rmab}\; \gamma_{nm} \; z^{\Delta_\rmaa - \Delta_\rmbb - \Delta_\rmcc + n-m}.\\[-2mm] \nonumber
\eea
Here we introduced the chiral square root  $\lambda^c{\!}_{\rmab}$ of the OPE coefficients $f^{\rmcc}{}_{\! \rmab}$ via $
f^{\rmcc} {\!\!}_{\rmab} = \lambda^{\rmcc}{\!}_{\rmab}\, \bar{\lambda}{}^{\rmcc}{\!}_{\rmab}$.
Similarly, $\gamma_{nm}$ is the chiral half of the universal overlap coefficient $C_{nm} = \gamma_{nm} \bar{\gamma}_{\bar{n}\bar{m}}$ that appears in the non-chiral three point function  (\ref{threepoint}). 
Note that in our normalization, the chiral vertex operators include the chiral operator product coefficients $\lambda_{\rm abc}$ as a factor. So with our definition, the CVOs carry detailed dynamical information.

\def\fs#1{\mbox{\scriptsize $#1$}}
\def\p{\partial}
\def\mat#1#2#3#4{\bigl[\hbox{\footnotesize$\begin{matrix}
 \spc #1\!\! &\!\!\nspc #2\spc \\[-1.75mm]
\spc #3\!\! &\!\!\nspc #4\spc
\end{matrix}$}\bigr]}


\noindent
\subsection{Conformal  bootstrap}


The chiral decomposition (\ref{lrdeco}) of local operators is a key step of the 2D conformal bootstrap \cite{BPZ,EV,GregNati}. Solving a CFT is tantamount to finding the spectrum  and absolute normalization of all chiral vertex operators, which are the elementary building blocks of the CFT.
 Every chiral CFT correlation function, or {\it conformal block}, can be obtained,
via a  {\it pants decomposition}, by gluing together a collection of CVOs.

To visualize this procedure, it will be helpful to graphically represent  the chiral vertex operators as  a cubic vertex
connected to three external double lines
\bea
\psi^\rmcc_{\rmaa\rmbb} 
(z) \is \raisebox{-3mm}{\includegraphics[scale=.23]{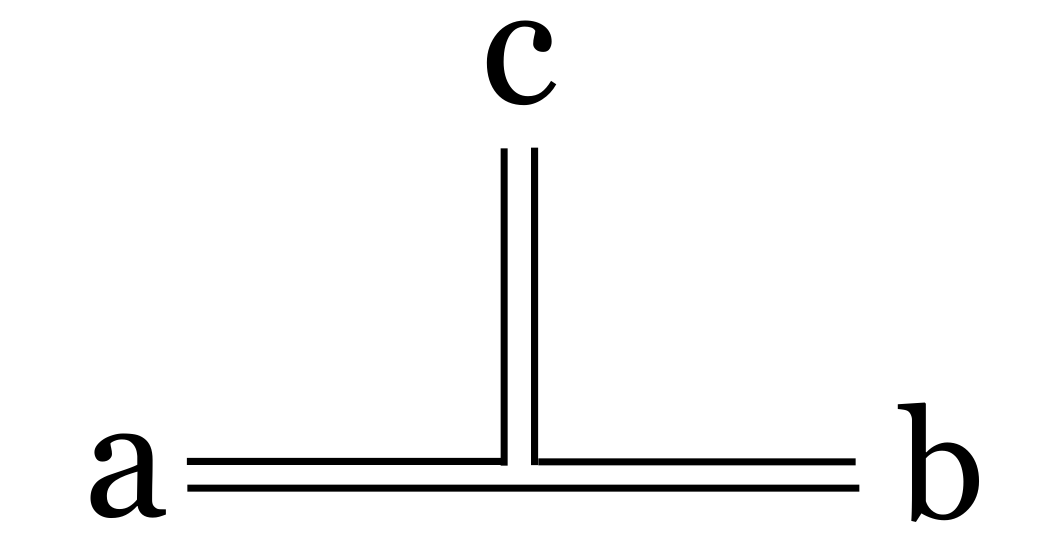}}\\[-1mm]\nonumber
\eea
The labels indicate the three chiral sectors that are linked by chiral vertex operator.\footnote{\small \addtolength{\baselineskip}{.3mm}The double line notation is helpful for visualizing the multi-valuedness of the chiral vertex operators. We can consider the  R operation, which acts as the square root of the full rotation and implements an interchange of two of the external labels. This R-operation is most naturally defined on the extended chiral vertex operator defined in footnote \ref{foottwo} 
\bea
{R} \,  \Psi^\rmcc_{\rmab}(z) \, \equiv \,
\Psi^\rmbb_{\rmaa\rmcc}(e^{i\pi} z) \! \is \!
 e^{i\pi (\Delta_a - \Delta_b - \Delta_c)} \Psi^\rmcc_{\rmab}(z) .
\eea
\small
This relation has a simple pictorial derivation as follows$$
\raisebox{-16pt}{
\includegraphics[scale=.16]{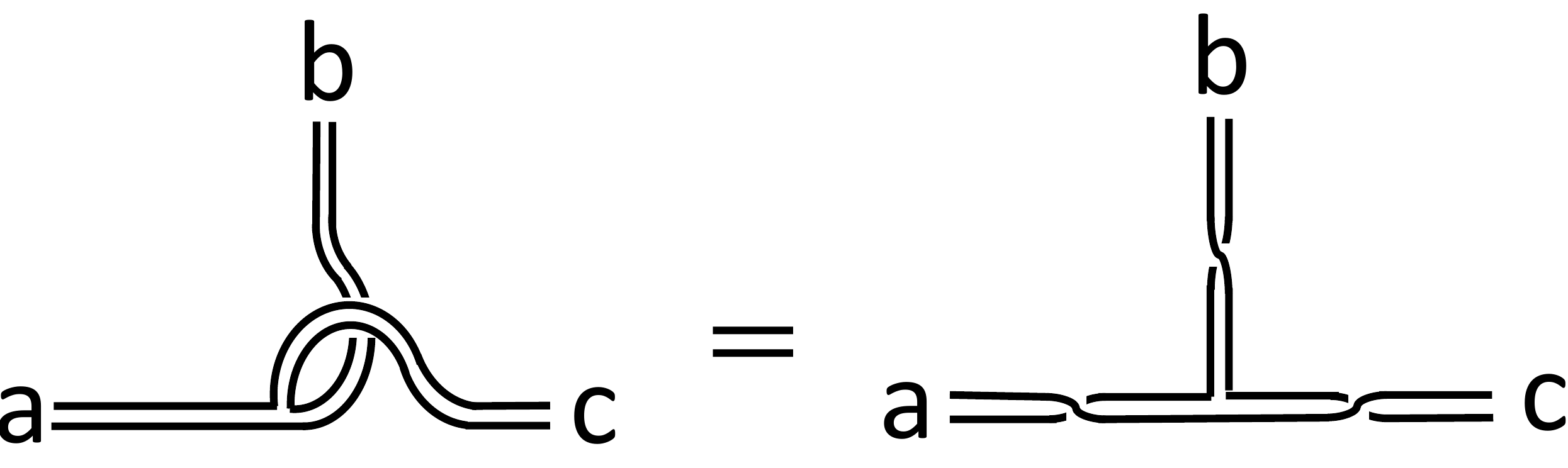}}$$
The l.h.s. depicts the CVO with the b and c end-points switched, and $z$ replaced by $e^{i\pi} z$.  The double lines indicate the framing, the choice of local coordinate at each end-point. The r.h.s. is obtained by straightening the three external legs, while keeping the frame at each end point fixed. The $\pi$-rotation of the local frames leads to the phase~factor $ e^{i\pi (\Delta_a - \Delta_b - \Delta_c)} $.}

Consider the four point function between four primary operators
\bea
 \la \spc 1 \spc \ri \, {\cal O}_{2}(w,\bar w) \, {\cal O}_3 (z,\bar z) \, \li\spc 4\spc \ra \is \La {\cal O}_1(\infty) \spc {\cal O}_2(w,\bar{w}) \spc {\cal O}_3(z,\zbar) \spc {\cal O}_4(0) \Ra.
\eea
To specify a conformal block decomposition, we insert a complete set of intermediate states  $
\mathbb{1} \, = \, \sum_\alpha P_{\rmaa}$ with  ${\cal P}_{\rmaa}\, =\!\! \sum_{\alpha \in {\cal H}_{\rmaa} \otimes \bar{H}_{\rmaa}} \li\alpha\ra\la \alpha \ri$
the projection operator on a given intermediate channel. The contribution for a given channel factorizes into the absolute value square of holomorphic conformal blocks ${\cal F}_{\nspc \rmaa}(z,w)$. Usually, the conformal blocks are normalized so that when $z$ and $w$ approach 0 and $\infty$, respectively, ${\cal F}_{\nspc \rmaa}(z,w) \simeq z^{\Delta_\rmaa - \Delta_3 - \Delta_4} w^{\Delta_1 + \Delta_2 - \Delta_\rmaa}$, with unit prefactor.
With this canonical normalization, the chiral factorization of the four point function is expressed as
\bea
\label{bpz}
 \la \spc 1 \spc \ri \, {\cal O}_{2}(w,\bar w) \, {\cal O}_3 (z,\bar z) \, \li\spc 4\spc \ra \is \sum_{\rmaa} \;f_{12\rmaa}\, f_{\rmaa 34}\; \bigr| \spc 
{\cal F}_{\nspc \rmaa}
(w,z) \spc \bigr|^2 ,
\eea
with $f_{\rm abc}$ the OPE coefficients. 
The canonically normalized conformal blocks ${\cal F}_{\rmaa}$ are defined independently of the specific CFT in question and eqn (\ref{bpz}) explicitly involves the OPE coefficients of the CFT.

The convential form of the 2D conformal bootstrap equation \cite{BPZ} (here we set $w=1$) 
\bea
\label{bpzbootstrap}
\sum_{\rmaa} \;f_{12{\rmaa}}\, f_{{\rmaa}34}\; \bigr| \spc {\cal F}_{\rmaa}  \bigl[\!\! \begin{array}{cc} \mbox{\scriptsize 2} \!\!&\!\! \mbox{\scriptsize 3}\\[-2.25mm] \raisebox{1pt}{\scriptsize 1} \!\! &\!\! \raisebox{1pt}{\scriptsize 4} \end{array} \!\!\bigr]\spc 
(z) \bigr|^2 \is  \sum_{\rmbb} \;f_{23{\rmbb}}\, f_{{\rmbb}14}\; \bigr| \spc {\cal F}_{\rmbb}  \bigl[\!\! \begin{array}{cc} \mbox{\scriptsize 3} \!\!&\!\! \mbox{\scriptsize 4}\\[-2.25mm] \raisebox{1pt}{\scriptsize 2} \!\! &\!\! \raisebox{1pt}{\scriptsize 1} \end{array} \!\!\bigr]\spc
(1-z) \spc \bigr|^2 
\eea
employs the canonically normalized conformal blocks ${\cal F}_{\rmaa}(z)$. This choice has an obvious rationale. Since the analytic form of the conformal blocks ${\cal F}_{\rmaa}(z)$ is known, the above equation amounts to an explicit  and highly non-trivial condition on the OPE coefficients $f_{\rm abc}$. Analogous equations have recently been used to great success in higher dimensions than 2 \cite{Rychkov}.

For reasons that will soon become more evident, we prefer to absorb the OPE coefficients into the definition of the conformal blocks\\[-6mm]
\bea
{\Psi}_{\nspc\rmaa}
(w,z) \is \la \spc 1 \spc \ri \,\psi^2_{1\rmaa} 
(w) \spc \psi^3_{\rmaa 4} 
(z)\spc \li\spc 4\spc \ra \;\;\, = \, \   \raisebox{-7mm}{\includegraphics[scale=.38]{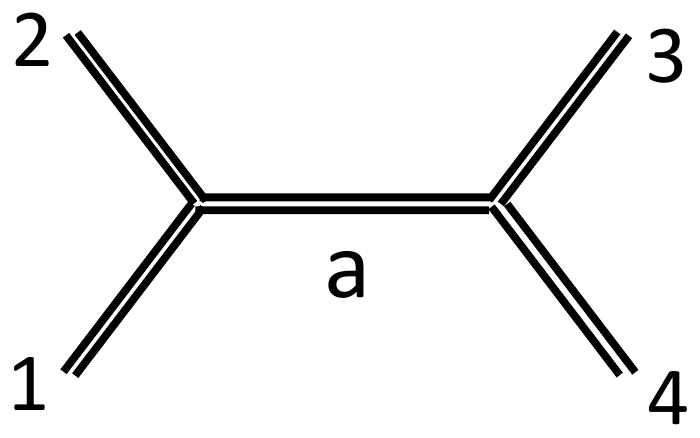}}.
\eea
The blocks ${\Psi}_{\nspc\rmaa}(w,z)$ encode the information of the OPE coefficients, and are thus specific to the individual CFT.
This seems a step backwards.
The pay-off, however, is that the relation between the full non-chiral and chiral four-point functions now takes a more canonical form
\bea
\label{unitaryfact}
 \la \spc 1 \spc \ri \, {\cal O}_{2}(w,\bar w) \, {\cal O}_3 (z,\bar z) \, \li\spc 4\spc \ra \is \sum_{\rmaa} \; \bigr| \spc 
{\Psi}_{\nspc \rmaa}
(w,z) \spc \bigr|^2 .
\eea
Here the sum over ${\rmaa}$ is over the full spectrum $({\rmaa},{\rm \bar \rmaa})$ of primaries of the CFT. 

\def\CW{{\cal W}}
\def\CF{{\cal F}}

Now let us rewrite these same equations into the unitary normalization of the conformal blocks. In this notation, the conformal bootstrap equation (\ref{bpzbootstrap})  takes a much simpler form
\bea
\label{unitarybootstrap}
\sum_{\rmaa} \; \bigr| \spc {\Psi}_{\rmaa}  \bigl[\!\! \begin{array}{cc} \mbox{\scriptsize 2} \!\!&\!\! \mbox{\scriptsize 3}\\[-2.25mm] \raisebox{1pt}{\scriptsize 1} \!\! &\!\! \raisebox{1pt}{\scriptsize 4} \end{array} \!\!\bigr]\spc (z) \bigr|^2 \is  \sum_\rmbb \; \bigr| \spc {\Psi}_\rmbb  \bigl[\!\! \begin{array}{cc} \mbox{\scriptsize 3} \!\!&\!\! \mbox{\scriptsize 4}\\[-2.25mm] \raisebox{1pt}{\scriptsize 2} \!\! &\!\! \raisebox{1pt}{\scriptsize 1} \end{array} \!\!\bigr]\spc
(1-z) \spc \bigr|^2 
\eea
In the graphical notation introduced above, we may represent this equation as
\bea
\raisebox{-20pt}{
\mbox{$\quad$}  \includegraphics[scale=.35]{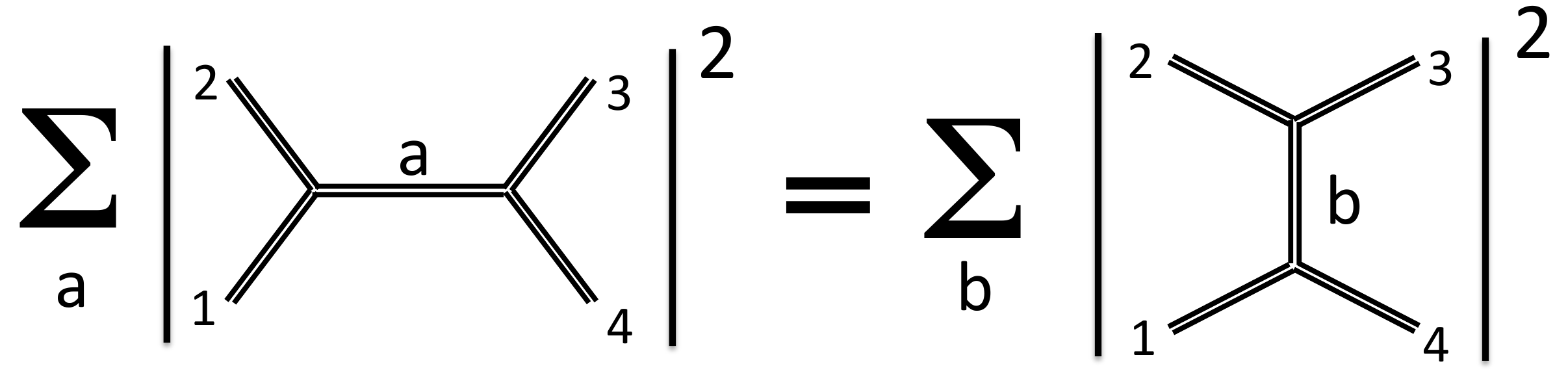}}
\eea
The unitary bootstrap equation (\ref{unitarybootstrap}) seems much simpler, and thus a priori much less instructive, than the original bootstrap equation (\ref{bpzbootstrap}). The two equations, however, are completely equivalent.

We will refer to our definition $\Psi_{\rmaa}(w,z)$ as the {\it unitary normalization} of 2D conformal blocks.
The factorization formula (\ref{unitaryfact}) is indeed suggestive of an interpretation of the blocks ${\Psi}_{\nspc \rmaa}
(w,z)$ as orthonormal  states in some suitable Hilbert space with inner product
\bea
\label{psinner}
\la \Psi_{\rmaa_1}\li \Psi_{\rmaa_2}\ra = \delta_{\rmaa_1\rmaa_2}.
\eea   
This interpretation of conformal blocks as Hilbert states will be a central theme of our story.


\noindent
\subsection{Fusion and braiding}

In view of our proposed interpretation of the conformal blocks $\Psi_{\rmaa}$  
as orthonormal basis states in some Hilbert space ${\cal H}$, it is natural to look for unitary operators that act on ${\cal H}$. Two examples of such a unitary operators are the fusion matrix ${\rm F}_{\rmab}$ and braid matrix, or R matrix ${\rm R}_{\rmab}$. These 
 implement the two basic crossing operations 
 \bea
\raisebox{-20pt}{
\includegraphics[scale=.36]{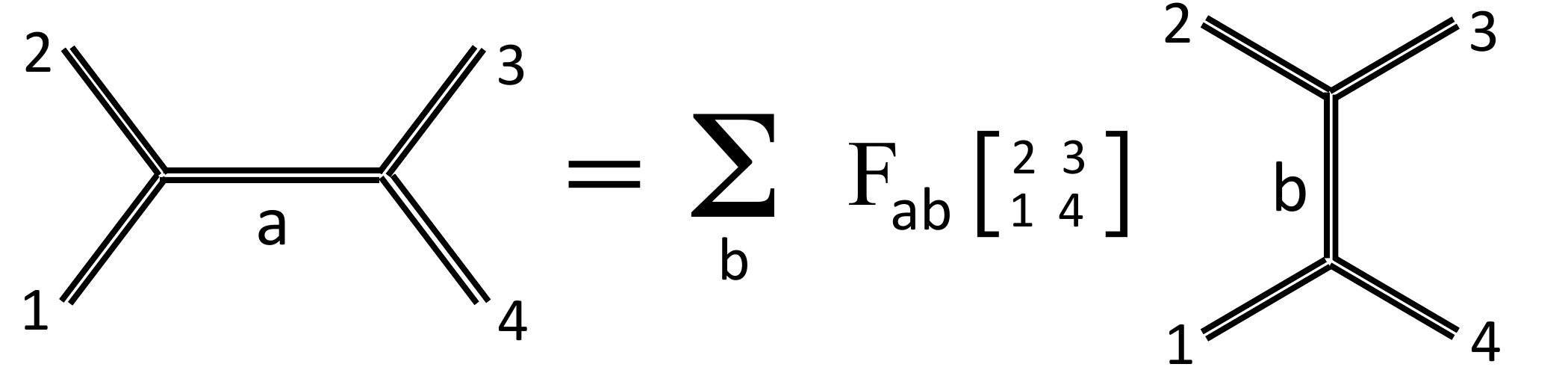}}\qquad \\[3mm]
\mbox{${}$}\qquad \raisebox{-20pt}{\includegraphics[scale=.40]{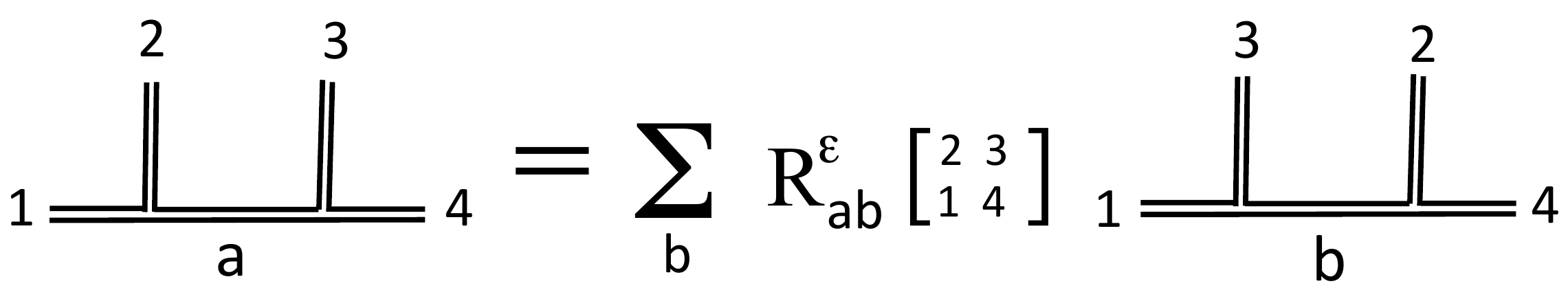}}
\eea
 or in equations  (here $\epsilon = \pm 1$ encodes the orientation of the braiding move)
 \bea
 \label{fcross}
 {\Psi}_{\rmaa}  \bigl[\!\! \begin{array}{cc} \mbox{\scriptsize 2} \!\!&\!\! \mbox{\scriptsize 3}\\[-2.25mm] \raisebox{1pt}{\scriptsize 1} \!\! &\!\! \raisebox{1pt}{\scriptsize 4} \end{array} \!\!\bigr]\spc (z) \is \sum_{\rm b} \; {\rm F}_{\rmab}\bigl[\!\! \begin{array}{cc} \mbox{\scriptsize 2} \!\!&\!\! \mbox{\scriptsize 3}\\[-2.25mm] \raisebox{1pt}{\scriptsize 1} \!\! &\!\! \raisebox{1pt}{\scriptsize 4} \end{array} \!\!\bigr]\;\spc {\Psi}_{\rmbb}  \bigl[\!\! \begin{array}{cc} \mbox{\scriptsize 3} \!\!&\!\! \mbox{\scriptsize 4}\\[-2.25mm] \raisebox{1pt}{\scriptsize 2} \!\! &\!\! \raisebox{1pt}{\scriptsize 1} \end{array} \!\!\bigr]\spc
(1-z)\\[2mm]
\label{bcross}
 {\Psi}_{\rmaa}  \bigl[\!\! \begin{array}{cc} \mbox{\scriptsize 2} \!\!&\!\! \mbox{\scriptsize 3}\\[-2.25mm] \raisebox{1pt}{\scriptsize 1} \!\! &\!\! \raisebox{1pt}{\scriptsize 4} \end{array} \!\!\bigr]\spc (z) \is \sum_b \; {\rm R}^\epsilon_{\rmab}\bigl[\!\! \begin{array}{cc} \mbox{\scriptsize 2} \!\!&\!\! \mbox{\scriptsize 3}\\[-2.25mm] \raisebox{1pt}{\scriptsize 1} \!\! &\!\! \raisebox{1pt}{\scriptsize 4} \end{array} \!\!\bigr]\;\spc {\Psi}_{\rmbb}  \bigl[\!\! \begin{array}{cc} \mbox{\scriptsize 3} \!\!&\!\! \mbox{\scriptsize 2}\\[-2.25mm] \raisebox{1pt}{\scriptsize 1} \!\! &\!\! \raisebox{1pt}{\scriptsize 4} \end{array} \!\!\bigr]\spc
(1/z)
\eea 
The fusion and R-matrix  ${\rm F}_{\rmab}
$ and ${\rm R}_{\rm ab}$ are constant matrices independent of the coordinate $z$.
They satisfy the following relation \cite{GregNati} \begin{equation}
 {\rm F}_{\rmab}\fourj{2}{3}{1}{4}
=e^{-\epsilon i\pi (\Delta_{1}+\Delta_{3}-\Delta_{a}-\Delta_{b})} {\rm R}^\epsilon_{\rmab}\fourj{2}{4}{1}{3}
\end{equation}
Crossing symmetry is equivalent to the condition that crossing operations (\ref{fcross}) and (\ref{bcross})  expresses a unitary change of basis. This unitarity condition
\bea
\label{unitaryc}
 \sum_{\rmcc}\, {\rm F}_{\rmac}
 \, {\rm F}^*_{\rm bc}
\, = \, \delta_{\rmab} \qquad & & \qquad
 \sum_{\rmcc} \; {\rm R}_{\rmac}
 \,  {\rm R}^*_{\rm bc}
 \, = \, \delta_{\rmab}
\eea 
is one of several defining relations of the modular bootstrap.
The modular group generated by the braid and crossing operations is characterized by a set of relations, which in turn amount to a set of non-trivial polynomial identities among  ${\rm F}_{\rmab}$ and ${\rm R}_{\rmab}$.
The most well known  polynomial equations are the so-called pentagon identity and the Yang-Baxter equation. A set of unitary matrices ${\rm F}_{\rmab}$ and ${\rm R}_{\rmab}$ that constitute a solution to these polynomial equations \cite{GregNati} defines a so-called modular tensor category, or in more physical language, a solution to the conformal bootstrap -- and thereby a local,  unitary 2D CFT with an associative operator algebra and crossing symmetric correlation functions.

For rational CFTs with only a finite number of conformal blocks, the fusion matrices ${\rm F}_{\rmab}\bigl[\!\! \begin{array}{cc} \mbox{\scriptsize $\rmcc$} \!\!&\!\! \mbox{\scriptsize $\rmdd$}\\[-2.25mm] \raisebox{1pt}{\scriptsize $\rmee$} \!\! &\!\! \raisebox{1pt}{\scriptsize $\rmff$} \end{array} \!\!\bigr]$ are finite
dimensional. The fusion matrices of WZW models  have an elegant representation as the expectation value of a collection of six open Wilson lines glued together in the shape of a tetrahedron. The physical derivation of this result  is summarized via
\bea
\raisebox{-20pt}{\includegraphics[scale=.41]{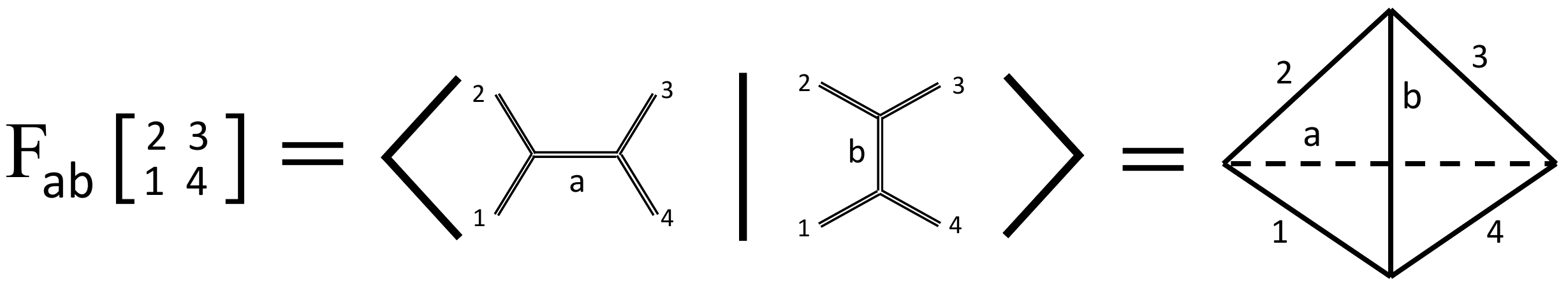}}
\eea
\noindent
First one makes an identification of conformal blocks as states of 2+1-D Chern-Simons theory, created by a Wilson line configuration in the shape of the corresponding trivalent graph. The representation of each open Wilson line matches  with the label of the chiral sectors. The lines are attached at the cubic vertices via an invariant tensor, or Clebsch-Gordan coefficient. The remaining open ends of the Wilson lines correspond to the location of the four local CFT operators ${\cal O}_i(z)$. Taking the inner product between the two conformal blocks amounts to gluing together the open ends. The resulting configuration is a tetrahedron, with six edges labeled by the six representations.

This graphical representation makes clear that the fusion matrices ${\rm F}_{\rmab}\fourj{2}{3}{1}{4}$  enjoy a tetrahedral symmetry acting on the six labels $\rmaa,\rmbb, 1,2,3,4,$. Tetrahedral symmetry is a hallmark of Wigner 6j-symbols. This is not coincidental:
if we view the chiral sectors of the CFT as representation spaces $V_{\rmaa}$ of some group, the definition of the fusion matrices becomes identical to that of the 6j-symbols. In mathematical terms, both are given by matrix coefficients of the associator isomorphism between tensor product representations $(V_{\rmaa}\otimes V_{\rmbb}) \otimes V_{\rmcc} \simeq V_{\rmaa} \otimes (V_{\rmbb} \otimes V_{\rmcc})$.
This representation theoretic interpretation of the fusion matrix is quite evident for WZW models. 
The tensor algebra of Kac-Moody representations is known to be isomorphic to that of the corresponding quantum group $U_q(\mathcal{G})$. The fusion matrices are quantum 6j-symbols
\bea
{\rm F}_{\rmab}\bigl[\!\! \begin{array}{cc} \mbox{\scriptsize $\rmcc$} \!\!&\!\! \mbox{\scriptsize $\rmdd$}\\[-2.25mm] \raisebox{1pt}{\scriptsize $\rmee$} \!\! &\!\! \raisebox{1pt}{\scriptsize $\rmff$} \end{array} \!\!\bigr] \is \left\{ \begin{array}{ccc} \!\! \mbox{\small $\rmaa$} \! &\! \mbox{\small $\rmcc$} \! &\!  \mbox{\small $\rmee$} \! \\[-1mm]\! \! \mbox{\small $\rmbb$} \! &\! \mbox{\small $\rmdd$} \! &\!  \mbox{\small $\rmff$} \! \end{array}\right\}{\!}_{\strut q}.
\eea
The subscript $q$ indicates the quantum deformation. The classical limit sends $q\!\to\! 1$.

\noindent
\section{Virasoro Modular Geometry}\label{modulargeometry}

The generalization of this algebraic perspective  to all rational CFTs is now well understood. It is less clear, perhaps, how much of it applies to generic irrational CFTs.  Nonetheless, this is the perspective that we would like to take and explore. Our basic assumption will be that the whole structure persists, except that all  sums over chiral sectors are replaced by infinite sums. The other idea that we will take guidance from is that the space of 2D conformal blocks 
has a dual interpretation as a Hilbert space which arises from quantization of a suitable geometric space
\cite{HV,TeschnerT,Ponsot,TZ,Kashaev-Fock,Kashaev-V,Tudor,Rosly,TeschnerV}.


The generalization of this idea to irrational CFTs with only Virasoro symmetry was first explored in \cite{HV}. The conformal blocks  span a linear vector space, given by the space of solutions to the conformal Ward identity 
\bea \label{virid}
\La T(z) \, \cO_1(z_1) \, ...\, \cO_4(z_4)\Ra \is  \hat{T}(x) \; \La  \cO_1(z_1) \, ...\, \cO_4(z_4)\Ra
\eea
\vspace{-8mm}
\bea
\label{virido}
\hat{T}(z) \is \sum_{i=1}^4 \left(\frac{\Delta_i}{(z-z_i )^2} + \frac{1}{(z-z_i)} \pdd{}{z_i}\right) 
\eea
In \cite{HV}\cite{Freidel}  it was shown that, when we view the conformal blocks as functionals of the 2D geometry, this Ward identity is equivalent to the physical state conditions (the Wheeler-De Witt equation and the diffeomorphism constraint)  of 2+1 gravity in asymptotic AdS${}_3$. We now briefly summarize the most incisive version of this result \cite{TZ,TeschnerT}.

 
 \noindent
 \subsection{Conformal blocks as wave functions}
 \smallskip

In section 2, we have seen that the phase space of four particles is isomorphic to two copies of the Teichm{\ddotu}ller space ${\cal T}_{0,4}$. We will now show that the left- and right conformal blocks of a CFT  combine into the wave functions obtained by quantization of the phase space ${\cal T}_{0,4} \times {\cal T}_{0,4}$. Consider a {\it classical} stress tensor  $T(z)$ of the form
\bea
\label{tzz}
T(z) \is \sum_{i=1}^{4}\Bigl( \frac{\Delta_i}{(z-z_i)^2} \, + \, \frac{\gamma_i}{z-z_i}\Bigr)
\eea
Here $\gamma_i$ are $c$-number coefficients that are usually referred  to as {\it accessory parameters}.
Regularity of $T(z)$ at $z\to \infty$ imposes the three conditions
\bea
\label{cstraints}
\sum_{i=1}^{n} \gamma_i \, = \, 0, \  \qquad \ \ 
\sum_{i=1}^{n} \bigl(\Delta_i + z_i \gamma_i\bigr)\!\is \! 0,\qquad \ \ 
\sum_{i=1}^{n} \bigl(z_i \Delta_i + z^2_i \gamma_i\bigr) \, = \, 0\, .
\eea
After removing the redundancy under global $SL(2,\mathbb{C})$ transformations $z \to (az+b)/(cz+d)$, the space of stress tensors of the form (\ref{tzz}) constitutes an 8 $- \; 2 \times 3$ = 2-dimensional complex space, which can be identified with the Teichm{\ddotu}ller space ${\cal T}_{0,4}$. This is shown as follows.

$T(z)$  in (\ref{tzz}) is the classical limit of the operator $\hat{T}(z)$ in the conformal Ward identity (\ref{virid})-(\ref{virido}). Indeed, our goal is to obtain $\hat{T}(z)$ as  the quantum version of $T(z)$.
 Alternatively, we can think of $T(z)$ as the  expectation value $
\la T(z)\ra $ 
 in a CFT with central charge $c = 6/b^2 \gg 1$.
   This expectation value transforms under conformal transformations via 
\bea
 T(z) \! & \to & \! T(w) \Bigl(\frac{dw}{dz}\Bigr)^2+ \frac{1}{2b^2}\bigl\{\spc w,\spc z\spc \bigr\} ,
\eea
with $
\bigl\{\spc w, \spc z\spc \bigr\} = \bigl(\frac{w''}{w'}\bigr)' - \frac  1 2 \bigl(\frac{w''}{w'}\bigr)^2$
 the Schwarzian derivative. Mathematically, such an object is called projective connection, or an {\it oper}. To an oper we can associate constant curvature metric $ds^2  =\spc e^\phi{(z,\zbar)} dz d\zbar$ 
via
\bea
\label{tphi}
b^2 T(z)\, = - \frac 1 2 (\partial \phi)^2 + \partial^2 \phi, \qquad \qquad
\partial \overline\partial \phi\! \is\! e^{\phi}\, .
\eea
Via the uniformization theorem, solutions to (\ref{tphi}) are of the form
\bea
 ds^2 = e^\phi dz d\zbar\! \is\! \frac{4 d Z\bar{d}\bar{Z}}{\ (1-Z\bar{Z})^2},
\eea
where  the uniformization coordinate $Z(z)$ is a holomorphic function of $z$, with non-trivial $SL(2,\mathbb{R})$ monodromy $Z \to (a_iZ+b_i)/(c_iZ+d_i)$ around the four punctures $z_i$. 

Near a local puncture $z \to z_i$, the conformal factor behaves as $\phi(z) \simeq -2 \alpha_i \log|z-z_i|$  with $2b^2\Delta_i =  \alpha_i (2-\alpha_i)$.  This shows that the puncture is a conical singularity with deficit angle $\theta_i = 2\pi (1-\alpha_i).$ 
Since  $\alpha_i < 1$ for a local puncture, we deduce that local punctures correspond to the light regime $\Delta_i < 1/2b^2 =c/12$. 
When one of the double poles has $\Delta_i = (k^2+1)/2 b^2>c/12$, the associated monodromy $Z(e^{2\pi i}z) = e^{2\pi k} Z(z)$ is in the hyperbolic conjugacy class of $SL(2,\mathbb{R})$. This monodromy fixes the local uniformization coordinates to $Z= z^{ik}, \bar{Z} = \zbar^{ik}$. The corresponding 2D metric \cite{Nati}
\bea
\label{natimetric}
ds^2 = \frac{\spc k^2 dz d\zbar}{z\bar z \sin^2\bigl(\frac k 2\log(z\bar{z})\bigr)},
\eea
is identical to the spatial metric in eqn (\ref{btzadstwo}): it describes the metric of the Einstein-Rosen bridge of a BTZ black hole.
This is not a coincidence.

Teichm{\ddotu}ller space has a K{\ddota}hler form, given by  the Weil-Peterson form $\Omega_{\rm WP}$. To quantize 
${\cal T}_{0,4}$, we start by equipping it with a symplectic form
\bea
\Omega  \! \is\! \sum_{i=1}^{4} \, d\gamma_i \wedge dz_i \, = \, \frac{1}{2\pi b^2}\; \Omega_{\rm WP}
\eea
Applying the standard rules of quantum mechanics shows that the accessory parameters $\gamma_i$ and the positions $z_j$ are canonically conjugate variables 
\bea
\bigl[ \gamma_i, z_j \bigr]\is  \spc \delta_{ij}.
\eea

What does the corresponding Hilbert space look like? Wave functions must be functions of a commuting set of coordinates. It is natural to choose  wave functions of the form 
\bea
\label{psiz}
\Psi(z_i)
\eea
subject to the constraints (\ref{cstraints}), which through the replacement $\gamma_i =$ {\large $\frac{\partial\ }{\partial z_i}$} impose invariance under the $SL(2,\mathbb{C})$ global conformal symmetry.
This same replacement also turns the classical stress tensor (\ref{tzz}) into the quantized stress tensor $\hat{T}(z)$ that appears in the conformal Ward identity (\ref{virid})-(\ref{virido}). The geometric meaning of this identification is as follows.

The  variables $(z_i,\gamma_i)$ provide a complex coordinate system on ${\cal T}_{0,4}$.  In other words,  ${\cal T}_{0,4}$ inherits its complex structure from that of the four puntured sphere $\Sigma_{0,4}$.
Choosing wave functions to be of the form (\ref{psiz}) thus defines a complex polarization.  This complex polarization is not unique: it  depends on choice of complex structure on the four punctured sphere. So there exists a 2-parameter family of complex polarizations,
and correspondingly, we should view the wave functions (\ref{psiz}) as sections of a suitable bundle, defined over the space of all possible polarizations. An important step in geometric quantization is the construction of a hermitian connection on this bundle.
 
 A general 2D metric looks like $ds^2 = e^{\varphi}|dz + \mu d\bar{z}|^2$. It picks a special complex coordinate system $z$ in which it  reduces to a diagonal metric $ds^2 = e^{\varphi} dz d\bar{z}$ with $\mu=0$.  Varying $\mu$ thus amounts to varying the complex polarization.  The connection that prescribes how the wave functions (\ref{psiz}) vary with the complex structure is given by 
\bea
\label{virado}
 \frac{\delta\  }{\delta \mu (z)\!\!}\;\Psi\bigl(z_i; \mu \bigr)_{\li \mu = 0}  \is\, \hat{T}(z)\Psi\bigl(z_i\bigr),
\eea
with $\hat{T}(z)$ the quantum stress tensor given in (\ref{virido}).
This linear equation on wave functionals $\Psi(z_i,\mu)$ is  the Virasoro Ward identity. 
The space of all Virasoro conformal blocks is defined as the linear space of all independent solutions to   (\ref{virado}). Via the above reasoning, we learn that this linear space is the Hilbert space obtained by quantizing the Teichm{\ddotu}ller space ${\cal T}_{0,4}$.
\smallskip

 \noindent
 \subsection{Virasoro modular geometry  and the quantum 6j-symbol}
 \smallskip
 
The quantum theory of Teichm{\ddotu}ller space  has been developed in great detail \cite{TeschnerT,Ponsot,TZ,Kashaev-Fock,Kashaev-V,Tudor,Rosly,TeschnerV}, resulting in important new  insight into the modular geometry of Virasoro conformal blocks  and of quantum Liouville theory \cite{Zamolodchikov,TeschnerRevisited,Ponsot,TeschnerT}. In this section, we give a short summary of the main results.
We refer to the original literature for a more complete treatment.

As explained in section 3, solving the conformal bootstrap is equivalent to finding a unitary representation of the fusion and braiding operations  on the Hilbert space of conformal blocks, specified by the spectrum of primary operators. The OPE coefficients then follow from factorizing the  conformal block (defined in the unitary normalization) into 3-point functions. Virasoro modular geometry is the program of finding a unitary representation of fusion and braiding on the space of {\it all} Virasoro conformal blocks. The CFT associated which this continuum bootstrap program is called Liouville CFT. Liouville CFT can be viewed as a limit of irrational CFT, in which the dense discrete spectrum of primary operators is replaced by a continuum. 

We now recall a few basic facts about Virasoro modular geometry and its relation with Liouville theory.  In essence, Liouville theory provides the mapping betwen the fusion ring of Virasoro algebra representations in irrational CFT to the representation ring of the universal enveloping algebra $U_q\bigl(\mathfrak{sl}(2,\mathbb{R}) \times \mathfrak{sl}(2,\mathbb{R})\bigl)$. The mapping starts with a relabeling of the central charge $c$ and conformal dimensions $\Delta$ in terms of the deformation parameter $b$ (related to $q$ via $q = e^{i\pi b^2}$) and $\mathfrak{sl}(2,\mathbb{R})$ representation labels $j$ via
\bea
\label{ltrans}
c = 1+ 6Q^2, \quad \quad Q\! \is \! b+ b^{-1}, \qquad \qquad \Delta = j (Q-j).
\eea
We will mostly focus on the heavy states with conformal weight $\Delta > Q^2/4$. These correspond to continuous series representations, labeled by
\bea
j \!\is \!\textstyle \frac 1 2 Q + i p, \qquad \qquad \Delta = \frac 1 4  {Q^2} +  p^2.
\eea
The $\mathfrak{sl}(2,\mathbb{R})$  representation labels $j$ are often referred to as Liouville momenta.\footnote{\small\addtolength{\baselineskip}{.4mm} The identification of $j$ as a `momentum'  is somewhat misleading and has given rise to some longstanding misconceptions about Liouville theory. The term momentum is suggestive of translation invariance, and the spectral density of momenta is typically  uniform. For this reason it is often said that Liouville theory does not have the enough degrees of freedom to produce a Hilbert space with level density given by the Cardy formula. This conclusion is wrong: Liouville CFT (with its continuum spectrum) has {\it more states} than any other irrational CFT with (a discrete spectrum and) the same central charge, not~less.} Eqn (\ref{ltrans})  indeed designates  the central charge $c$ of the  Liouville stress tensor $T(z) =\frac 1 2 (\partial\phi)^2 + Q\partial^2 \phi$ and conformal weight of the vertex operators $V =e^{j\phi}$ with specified Liouville momentum~$j$. For our purpose, Liouville CFT is defined as the geometric theory whose quantization gives the  modular tensor category of Virasoro conformal blocks \cite{TeschnerT}.

To a non-compact Lie groups like $SL(2,\mathbb{R})$ one can associate measure on the space of continuous series of representations, called the Plancherel measure, equal to the weight of each representation in the Peter-Weyl decomposition of the space of functions on the group. 
The Plancherel measure $d\mu(p)$ on the space of representations of $U_q(\mathfrak{sl}(2,\mathbb{R}))$ is given  by
\footnote{\small \addtolength{\baselineskip}{.4mm} The Plancherel measure is the closest counterpart of the quantum dimension \cite{TeschnerR}. As pointed out in \cite{McGough}, applying the formula for topological entanglement entropy $S_{\rm top}(j) = \log S_0^j$ with $S_j^k$ the Virasoro modular S-matrix, identifies  (\ref{plancherel}) as the natural measure for the number of states as a function of conformal dimension. The argument of \cite{McGough} provides a new derivation of the Bekenstein-Hawking formula, and a new connection between microscopic and entanglement entropy. }
\bea\label{plancherel} & & d\mu(j) = \rho(j)\, d\smpc j \nonumber\\[-2mm]\\[-2mm] \nonumber
 \rho(j)\! & \! \equiv\! &  \! 4\sinh(2\pi b\smpc p)\sinh(2\pi b^{-1} p) .  \eea
One of the main results of the Virasoro modular bootstrap program is that eqn (\ref{plancherel}) specifies the spectral density of Liouville CFT. This density grows exponentially with $p$. Plugging in the redefinitions (\ref{ltrans}), we see that (for small $b$ = large $c$), it agrees with the Cardy formula
\bea
d\mu(\Delta) \, \simeq\, 
d\Delta \, e^{2\pi \sqrt{\frac{c}6 (\Delta- \frac{c}{24})}}.\\[-5mm]\nonumber
\eea
This result is central to our proposal that Liouville theory captures the universal behavior of irrational CFTs in the regime of large $c$ and large conformal dimension $\Delta>c/12$.

The non-chiral correlation function of Liouville theory is given by an integral, with Plancherel measure, over all representations $j$. The integrand for given $j$ factorizes into chiral conformal blocks
\bea
\label{lsumsq}
\la {\cal O}_1 \spc {\cal O}_2 \spc {\cal O}_3 \spc {\cal O}_4 \ra \is
\int\!\nspc d\mu(j_{\alpha}) \,\spc  \mbox{\LARGE $\Bigl|$} \spc \raisebox{-15pt}{\includegraphics[scale=.28]{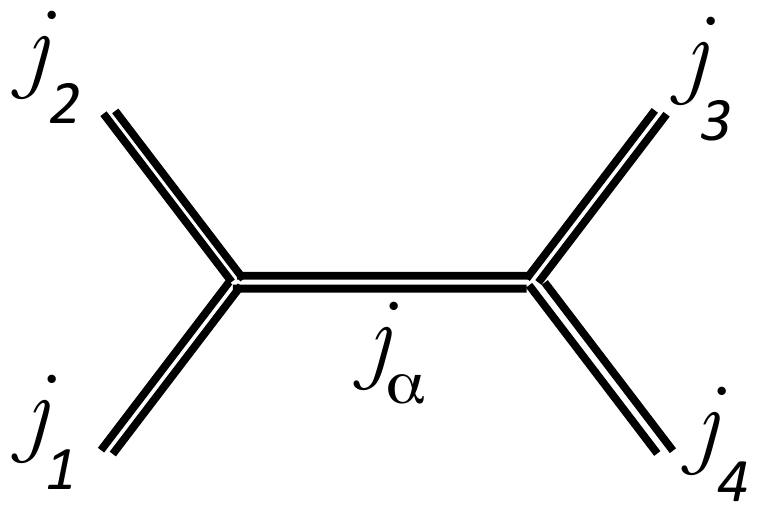}\spc}\mbox{\LARGE $\Bigr|$}^2.
\eea
We interpret this equation as the continuum limit of the formula (\ref{unitaryfact}) that describes the left-right factorization of  correlation function of an irrational CFT in terms of conformal blocks in the unitary normalization. Just like before, the OPE coefficients of Liouville CFT
(given by the  DOZZ formula) are absorbed into the unitary basis of conformal blocks.  We can recover them by factorizing the four point function into three point functions.

The chiral three point functions define chiral vertex operators and can be viewed as invariant tensors of the Virasoro symmetry.
Via the isomorphism between the tensor algebra of the Virasoro algebra and $U_q(\mathfrak{sl}(2,\mathbb{R}))$, one can show \cite{Ponsot,TeschnerT} that duality transformation of the Virasoro conformal blocks are specified by the quantum 6j-symbols
\bea
 \label{ncross}
\raisebox{-17pt}{\includegraphics[scale=.32]{ablock.pdf}}
 \is \int\!\nspc d\mu(j_{\beta}) \; \sixj{j_1}{j_2}{j_\alpha}{j_3}{j_4}{j_\beta}_q\; 
\raisebox{-24pt}{\includegraphics[scale=.34]{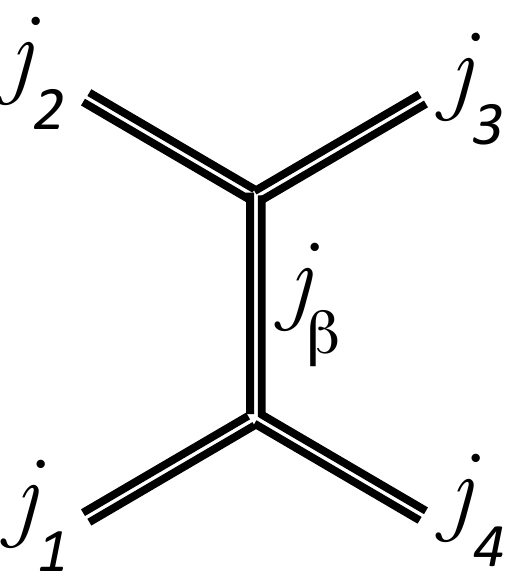}}
\eea
Crossing symmetry is the condition that (\ref{ncross})  expresses a unitary change of basis, when the space of conformal blocks is equipped with a Hilbert space structure 
\bea
\la \Psi_{j_1} \li \Psi_{j_2}\ra \is \frac{1}{\rho(j_1) 
} \, \delta(j_1\nspc - j_2)
\eea
This condition is fulfilled:  the quantum $6j$ symbol for fixed $j_i$, $i=1,..,4$ 
defines a unitary matrix 
\bea
\int
\! d\mu(p)\,
\sixj{j_1}{j_2}{j_p}{j_3}{j_4}{j_\alpha}_q^*
\sixj{j_1}{j_2}{j_p}{j_3}{j_4}{j_\beta}_q
\is \frac 1 {\rho(j_\alpha)
} \, \delta(j_\alpha \nspc - j_\beta).
\eea
The above three equations are the continuum version of the discrete equations (\ref{fcross}), (\ref{psinner}) and (\ref{unitaryc}).

Liouville CFT captures the universal structure of the modular bootstrap equations of a general irrational CFT in the supercritical regime $c\gg1$ and $\Delta> c/12$. The quantum 6j-symbols of $U_q(\mathfrak{sl}(2,\mathbb{R}))$ indeed satisfy all the required polynomial equations and define a proper modular tensor category.
Given the non-triviality of these modular consistency equations and the match of the Plancherel measure with the Cardy density of states, it seems more than plausible that these same 6j-symbols also capture the course grained quantitative behavior of the fusion matrices of general irrational CFTs in the entropy dominated regime, where the dual bulk physics is expected to be dominated by gravity.

We can perform a precise quantitative check of this proposed match with the gravity dominated regime of the holographic dual.
For this, we need the explicit form of the quantum 6j-symbols. This formula is given in the Appendix. It is expressed in terms of so-called double Sine functions, which are closely related to the quantum dilogarithm function. 
The explicit expression  (\ref{qsixjay}) looks a bit difficult to work with. However, as mentioned in the introduction, it has a concrete geometrical interpretation as the quantum volume of a hyperbolic tetrahedron. Using the limiting behavior of the  quantum dilogarithm function for small $b$, one can indeed recognize that 
\bea
\label{sixjvol}
\sixj{\alphaj_1}{\alphaj_2}{\alphaj_\alpha}{\alphaj_3}{\alphaj_4}{\alphaj_\beta}_q \! & \simeq & \! \exp\Bigl\{\frac{i}{2\pi b^2} {\rm Vol\bigl( T 
\bigr)}\Bigr\}
\eea
where $T$ is the hyperbolic tetrahedron with six dihedral angles specified by $\nu_i = 2\pi b^2 j_i$, see eqn (\ref{tvolume}). The details of this calculation are found in \cite{TeschnerV}.

The exact match between  (\ref{scatvol}) and (\ref{sixjvol}) establishes that the modular matrices of Liouville CFT provide the exact quantum solution to the 2+1-D gravitational scattering problem described in sections 1 and 2, and vice versa.  In the next section, we will show that the modular martices dictate the form of the exchange algebra between local vertex operators in the Lorentzian CFT. This will complete the identification. 

\bigskip
\bigskip

\noindent
\subsection{Verlinde operators as geodesic lengths}


\def\IR{{\mathbb{R}}}

The match between Virasoro CFT and gravitational scattering  can be made more evident by considering the Verlinde loop operators \cite{EV} of Liouville CFT.
In rational CFT, Verlinde loop operators are hermitian operators associated with closed loops $\gamma$ that act on the space of conformal blocks.  For a given loop $\gamma$, they generate the commutative and associative algebra, isomorphic to the fusion algebra of the CFT. The Verlinde operators for two different loops $\gamma_1$ and $\gamma_2$ commute only if the loops don't intersect. A basis of conformal blocks is provided by the simultaneous eigenvectors of a maximally commuting set of Verlinde operators, associated to all  dividing cycles of a pant decomposition of the 2D surface. The modular transformations represent canonical transformations between the eigenbases of different maximal sets of commuting loop operators. 
 
There exist analogous loop operators in Virasoro CFT \cite{HV,Alday}. They correspond to a quantized  version of the operators $L(\gamma) = 2\cosh( l(\gamma)/2)$ that measure the geodesic length of the corresponding loop. The idea put forward in \cite{HV,Alday} is that Virasoro modular geometry is organized by the algebra of these quantized loop operators, in the same manner that RCFT is codified by the Verlinde loop operators. Their relevance to our story is that they supply the inverse map from Virasoro CFT to the quantum geometry employed in section 2.

The simplest  loop operator makes use of the degenerate primary field $\phi_{2,1}$. It satisfies
\bea
\label{nullc}
(L_{-1}^2 + \spc b^2 L_{-2})
\li\phi_{2,1}\ra = 0
\eea 
The degenerate field $\phi_{2,1}$ can be
viewed as the operator with Liouville momentum equal to $j_{2,1}= -b/2$. 
The fusion algebra of the degenerate field is $[\, - b /2 ] \, \times \, [ \, j \, ] = [ \, j- {b}/{2} ] + [ \, j+ {b}/{2} ]\, .$

\newcommand{\PPsi}{{\Psi}} 

\smallskip

\noindent
The Verlinde monodromy operators $L(\gamma)$ associated with $\phi_{2,1}$ is defined  via the 
recipe~\cite{EV}:

\vspace{-2.5mm}

\begin{enumerate}
\item Insert the identity operator $\mathbf{1}$ inside a Virasoro conformal block.
\vspace{-2.5mm}
\item Replace  $\mathbf{1}$  by the short-distance OPE beween two chiral  degenerate fields $\phi_{2,1}$.
\vspace{-2.5mm}
\item Transport one of the operators $\phi_{2,1}$ along a closed (non-self-intersecting) path $\gamma$.
\vspace{-2.5mm}
\item Re-fuse  the two degenerate fields together into identity $\mathbf{1}$, via their OPE.
\end{enumerate}

\vspace{-3.5mm}

\noindent
This procedure defines a linear map on the space of Virasoro conformal blocks.
The loop operators are tolopogical:
 they only depend on the topology of the loops  and not on the position of operators or on the shape of the 2D surface.

\def\CT{{\cal T}}

The identification of the Verlinde loops with quantized geodesic lengths $L(\gamma)$ can be understood as follows.
Consider the situation right after step 2 in the above procedure: a Virasoro conformal block with the added insertion of a pair of degenerate fields $\phi_{2,1}$. The null state relation (\ref{nullc}) implies that this correlator satisfies the differential equation 
\bea 
\label{degendiff}
\bigr( \partial_z^2+ b^2 \hat{T}(z)\bigr)\La \phi_{2,1}(z) \phi_{2,1}(0) {\cal O}_1(z_1) \cdots {\cal O}_4(z_4)\Ra = 0,
\eea
with $\hat{T}(z)$ the differential operator defined in (\ref{virido}). This equation can be formally rewritten as a parallel transport via a flat, operator valued $SL(2,R)$ gauge field $\hat{A}= \tau_- +\,  b^2 \, \hat{T} \, \tau_+$ in the spin 1/2 representation. (The two components of the spin 1/2 representation correspond to the two fusion channels of the degerate field.)  Hence eqn (\ref{degendiff})  tells us that moving $\phi_{2,1}$ along the loop $\gamma$ produces the holonomy of the flat connection 
\bea
L(\gamma) \is \tr_{\frac 12}\Bigl( P \exp\oint_\gamma  \hat{A}\Bigl).
\eea
The trace arises because of the splitting and merging at the first and last step in the Verlinde loop recipe.
In the limit $b\to 0$, we can replace $\hat{T}(z)$ by its classical version $T(z)$. This loop operator then coincides with the $SL(2,\mathbb{R})$ holonomy operators introduced in section~2.

A more precise check on the identification is to simply evaluate the Verlinde operators, by using the known formulas for the modular matrices. Consider the a Virasoro conformal block $\Psi_\alpha\fourj{2}{3}{1}{4}$, with a specified intermediate channel $j_\alpha$ along the A-cycle. From the above definition, it is clear that this  conformal block is an
eigen state $|\Psi_\alpha\rangle$ of the Verlinde operator $\hat{L}_A$, associated to the A-cycle. The dual loop $\hat{L}_B$, on the other hand, acts non-trivially on the representation label. A straightforward calculation
\footnote{\small \addtolength{\baselineskip}{.3mm} This calculation was performed in \cite{Alday} in the context of the AGT correspondence. The Wilson and `t Hooft loops in ${\cal N}=2$ gauge theory correspond to the A-cycle and B-cycle Verlinde operators,  and the fusion matrix $F_{\alpha\beta}$ decribed the S-duality map.} gives \cite{Alday}
\bea
\hat{L}_{A}\spc | \Psi_{\rm a} \rangle \is 2\cosh(2\pi b \smpc p_{\rm a}) \, |\Psi_{\rm a}\rangle\, \qquad\qquad\ 
j_{\rm a} =\textstyle \frac{Q}{2} + ip_{\rm a}, \\[3.5mm]
\hat{L}_B \spc | \Psi_a \rangle\is  L_\beta^0(a) \spc |\Psi_a\rangle\spc +\spc  L_\beta^+(a)\spc | \Psi_{a+ b}\rangle \spc +\spc L_\beta^-(a)\spc |\Psi_{a- b}\rangle\, .
\eea
The formula for $\hat{L}_A$ should be compared with the classical formula $L_\alpha = 2\cosh( l_\alpha/2)$ with $l_\alpha/2\pi = \sqrt{8M}$. The functions
$L_\beta^{0,\pm}$ are  quantum deformations of the three terms in the $e^{\tau_\alpha}$ expansion of $L_\beta$ given in eqn (\ref{roslyrel}).

\smallskip

\noindent
\section{Exchange Algebra of 2D CFT}\label{exchalgebra}


Crossing symmetry is the requirement that, in the Euclidean regime,  the modular fusion and braiding matrices F and R are invisible from the point of view of the full non-chiral CFT. A useful analogy is to think of chiral conformal blocks  as invariant tensors that carry a non-trivial charge under a `quantum statistics group', the group that governs the fractional statistics of the chiral vertex operators. For rational CFT this statistics group is some finite quantum group, for Liouville theory it is given by the non-compact quantum group  $U_q(\mathfrak{sl}(2,\mathbb{R}))$, and for a general irrational CFT, it is some different yet unknown quantum group. The modular matrices F and R are unitary basis transformation that relate different  bases of  invariant tensors. Locality and crossing symmetry then takes the form of a gauge constraint: it imposes the condition that non-chiral correlation functions and non-chiral local operators of the Euclidean CFT are 
singlets under this quantum group symmetry. They are neutral bound states between left- and right chiral blocks or chiral vertex operators, in a similar way that mesons are color singlets. We could characterize this by saying that the quantum statistics group is `confined' in Euclidean CFT. This confinement ensures the single-valuedness of all non-chiral correlation functions.

\subsection{Locality}

In Lorentzian signature, one needs to pay special attention to what happens on the light-cone. Most interestingly, as pointed long ago in \cite{Rehren}, when operators are time-like separated, the monodromy properties of the chiral correlation functions no longer cancel out, but the add up and become fully visible. The  quantum statistics group deconfines in the time-like domain, and manifests itself via the operator algebra between any pair of time-like separated local operators. As we will now explain,
this operator algebra  must take the form of an exchange algebra, specified by the unitary R-matrix that prescribes the braid relations of holomorphic conformal blocks.

We write local operators  in terms of the light-cone coordinates $u = t+x$ and $v=x-t$. The  locality conditions that non-chiral operators have single valued correlation functions and  commute when
space-like separated
\bea
\label{ocomm}
\qquad \bigl[ \spc \hat{B}(u,v), \spc \hat{A}(0)\spc \bigr] \is 0, \qquad \qquad {\rm for} \ \  uv> 0
\eea
become non-trivial requirements once we factorize the operators into chiral vertex operators
\bea
\hat{A}(0,0) = \sum_\alpha \bar{A}_\alpha(0) A_\alpha(0), \quad & & \quad \hat{B}(u,v) = \sum_\beta \bar{B}_\beta(v) B_\beta(u).
\eea
At given frequency, the operators factorize. $A_\alpha(0)$ is a chiral vertex operator that maps the chiral sector with energy  $M$ and to the chiral sector  of energy $M+\omega$. 

Let us first look at the condition (\ref{ocomm}) from the point of view of the factorization of correlation functions into chiral conformal blocks
\bea
\label{nca}
\la M\!+\nspc\omega\spc | \spc \hat{B}(u,v)\spc \hat{A}(0,0)\spc \li M\ra \is \sum_{\alpha}\, \bar{\Psi}_{\nspc\alpha}
(v)\, {\Psi}_{\nspc\alpha}
(u)\; \equiv \; \bigl(\spc \bar{\Psi},\spc \Psi\spc \bigr),\\[1.5mm]
\label{ncb}
\la M\!+\nspc\omega\spc |\spc \hat{A}(0,0) \spc \hat{B}(u,v)\spc \li M\ra \is \sum_{\beta} \, \bar{\Phi}_{\beta}
(v)\, {\Phi}_{\beta}
(u)\; \equiv \; \bigl(\spc \bar{\Phi}\smpc,\spc \Phi\spc \bigr),
\eea
\vspace{-7mm}

\noindent
with
\vspace{-10mm}
\bea
\Psi_\alpha(u) \is
\la M\!+\!\omega | \spc B_{\omega-\alpha}(u)\spc A_\alpha(0)\spc \li M\ra ,\\[3.5mm]
\Phi_\beta(u) \is
\la M\!+\!\omega | \spc A_{\omega-\beta}(0)\spc B_\beta(u)\spc \li M\ra .
\eea
The locality requirement (\ref{ocomm}) tells us that, in the Euclidean regime when $uv>0$, the two 2-point functions (\ref{nca}) and (\ref{ncb}) should be the same. 

The conformal blocks $\Psi_\alpha(u)$ and $\Phi_\beta(u)$ are related via a braiding operation, which takes the form of a unitary basis transformation executed by the left- and right-moving R-matrices. We abbreviate this relation as
\bea
\label{euclidr}
\Psi_\alpha(u)\, = \, \sum_\beta \, {\cal R}_{\alpha\beta}\, \Phi_\beta(u), \quad & & \quad
\bar\Psi_\alpha(v)\, =\, \sum_\beta\, \bar{\cal R}_{\alpha\beta} \,\bar\Phi_\beta(v),
\eea
where ${\cal R}$ is short-hand for the R-matrix.
Using the reality condition $\bar{\cal R}^T = {\cal R}^\dag$, we are ensured that (\ref{nca}) and (\ref{ncb}) are indeed identical 
\bea
(\spc \bar{\Psi} ,\spc\Psi\spc) \, = \, (\spc \bar\Phi \spc \rm {\cal R}^\dag,\spc {\cal R}\Phi\spc ) \, = \, 
(\spc \bar{\Phi} ,\spc\Phi\spc) .
\eea
provided we impose the unitarity condition ${\cal R}^\dag {\cal R} =1$.

This R-matrix is the linchpin of our story. It is specified by six arguments. In the context of our CFT description of the gravitational scattering problem,  the six arguments are 
the conformal dimension of the two light primary fields that create the two particles A and B, and  of the four heavy primary fields associated with the four sectors of the BTZ black hole space-time of fig~1, with masses $M, M\nspc +\nspc \alpha, M\nspc +\nspc \beta$ and $M\nspc +\nspc \omega$.
\bea
\label{rfour}
{\cal R}_{\alpha\beta} \is {\rm R}_{\sca\scb}\fourj{\scm}{\omega}{\alpha}{\beta}
\,\; = \;\;\;
\raisebox{-25pt}{\includegraphics[scale=.56]{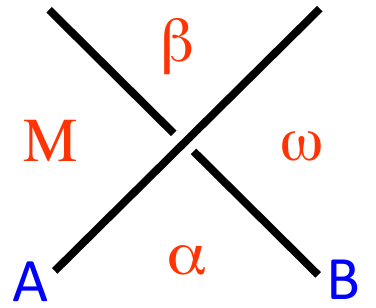}}
\eea
The R-matrix satisfies the unitarity condition  
 \bea
 \label{unir}
\sum_\alpha  \;\, {\rm R}^*_{\sca\scb}\fourj{\scm}{\omega}{\alpha}{\beta}\, {\rm R}_{\sca\scb}\fourj{\scm}{\omega}{\alpha} {\delta} \spc  \is\, \delta_{\beta\delta} .
 \eea

 
 \noindent
 \subsection{Exchange algebra}
 

The unitarity condition (\ref{unir}) ensures that space-like separated operators commute. However, it of course does not imply that time-like separated operators also commute. 

\begin{figure}[t]
\begin{center}
\includegraphics[scale=.34]{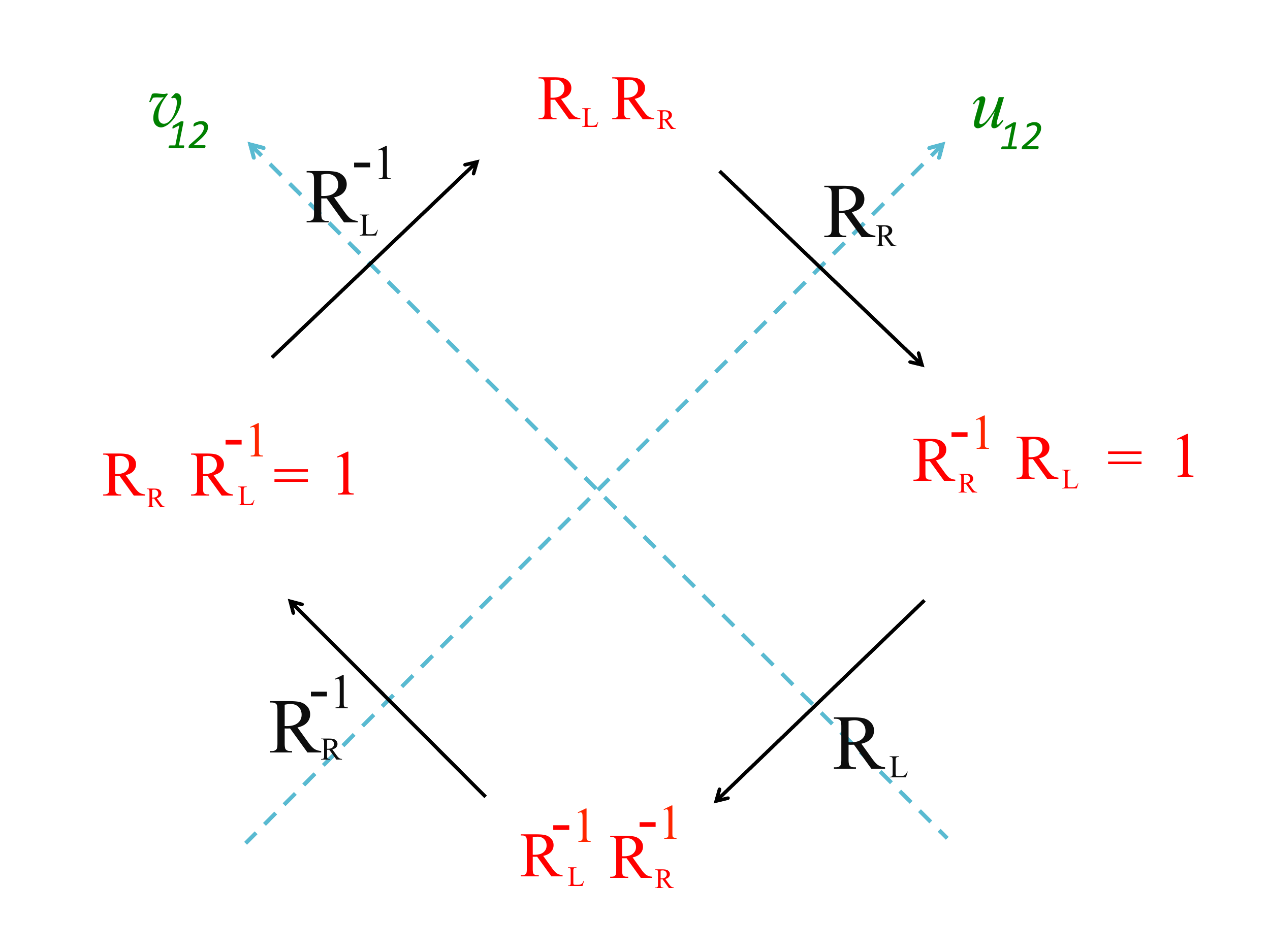}
\caption{This diagram indicates (in black) the R-matrix associated with flipping the sign (via the appropriate analytic continuation) of each light-cone coordinate $u$ and $v$. It also shows (in red) the full exchange matrix associated with the combined analytic continuation from $(u,v)$ to $(-u,-v)$. }
\end{center}
\vspace{-0.5cm}
\end{figure} 
As it stands, eqn (\ref{euclidr}) is not complete: we in fact implicitly picked a particular sign for the light cone coordinates, let's say $u>0$ and $v>0$.  Why would this sign be relevant? The braiding operation comes with an orientation. To see how this orientation relates to the sign of $u$, it is convenient to specialize to the case that $\hat{A} = \hat{B}$ are the same operator.\footnote{\small \addtolength{\baselineskip}{.3mm} This following reasoning closely follows the argumentation presented in \cite{Rehren}.} In this case, the conformal block  $
 {\Psi}_{\alpha}  (u)$ on the left-hand side of (\ref{euclidr}) is by definition the analytic continuation of  the conformal block  ${\Phi}_\beta(u)$ on the right-hand side. The analytic continuation sends $u$ to $-u$. The standard $i\epsilon$ prescription dictates that the new value of $u$ is approached  via the upper complex half plane. Thus for $u>0$, the coordinate $u$ circumnavigates the origin $u=0$ with an anti-clock wise orientation. On the other hand, if $u<0$, the $i\epsilon$ prescription dictates that the orientation should be reversed, and the R-matrix should thus be replaced by its inverse. Although we derived this conclusion for the case that two operators $\hat{A}$ and $\hat{B}$ are the same local operator, it also holds for the general case.

The braid relation (\ref{euclidr}) generalizes to arbitrary n-point functions. The R-matrix looks the same for all cases: R only depends on the labels of the local operators that take part in the braiding operation and on the adjacent intermediate channels. So the braiding relation of the conformal blocks lifts to a property of the local operators $\hat{A}$ and $\hat{B}$ themselves. This argumentation shows that the left and right chiral vertex operators in 2D CFT satisfy the  exchange algebra
\bea
\label{exchl}
  B_{\omega-\alpha}(u) \,  A_{\alpha}(0) \,\is \sum_{\beta} \; {\cal R}^{\spc {\rm \epsilon}(u)}_{\alpha\beta}
  \; A_{\omega-\beta}(0) \,  B_{\beta }(u) \\[2mm]
  \bar{B}_{\omega-\alpha}(v) \,  \bar{A}_{\alpha}(0) \,\is \sum_{\delta} \; \bar{\cal R}^{\spc {\rm \epsilon}(v)}_{\alpha\smpc \delta}
  \; \bar{A}_{\omega-\delta}(0) \,  \bar{B}_{\delta }(v) ,\label{exchr}
\eea
where $\epsilon(u)$ is the sign function: $\epsilon(u) = 1$ for $u>0$ and $\epsilon(u) = -1$ for $u<0$, and same for $\epsilon(v)$. The appearance of this sign function  has important consequences for the Lorentzian non-chiral CFT, as we will see in a moment.

As a helpful warm-up exercise, let us  briefly pause and take a look at a simple example: the $c=1$ CFT  of a free massless boson. It has non-chiral local vertex operators of the form 
$$
V_p(u,v)  = e^{ip \phi(u,v)}.
$$ 
These operators are mutually local: they commute at space-like separation.  Using the CBH formula $$e^{ip \phi(u,v)} e^{iq \phi(0,0)} = e^{iq \phi(0,0)} e^{ip \phi(u,v)} e^{-pq[\phi(u,v),\phi(0,0)]}$$
and the basic commutator $[\spc \phi(u,v),\spc \phi(0,0)] = 2\pi i\smpc \epsilon(u,v)$ with
\bea
\label{epsdef}
\epsilon(u,v) \! \is\! \frac 1 2 \bigl(\epsilon(u)-\epsilon(v)\bigr) = \left\{\begin{array}{c}{0 \quad {\rm if} \quad uv>0 \qquad \quad  \ \ \; \mbox{(space-like)}\ }\\{\! 1\quad {\rm if} \quad u>0,v<0 \qquad \ ({\rm future}) \ \ \ }\\{\! -1\quad {\rm if} \quad u<0,v>0\qquad\ \; ({\rm past})\qquad}\end{array}\right.
\eea
we deduce  that the gaussian  vertex operators  satisfy a non-trivial exchange algebra 
\bea V_p(u,v)  V_q(0,0)= V_q(0,0) V_p(u,v) e^{-2\pi i\smpc pq\smpc \epsilon(u,v)}
\eea
The phase factor is non-zero inside the past and future light-cone. We see that the anyon statistics of the chiral theory is still relevant for describing the operator algebra of the non-chiral Lorentzian CFT. 
This is the simplest example of the deconfinement phenomenon of the quantum statistics group in the time-like domain. 


  This simple calculation has some unexpected immediate consequences. In Euclidean CFT, one is used to the statement that primary states transform irreducibly under the conformal group. This is no longer the case in Lorentzian signature. Special conformal transformations $u\to \tilde{u} = u/(1+u)$ can map space-like separations into time-like separations. If it would act irreducibly on all operators, then space-time locality would forbid any non-trivial commutators. In other words, if we want to define the CFT Hilbert space  such that  the conformal group acts irreducibly, we need to decouple the left and right-moving sectors. We will return this point momentarily.

Let us now return to the general case. 
The product of two local operators decomposes into chiral components as
\bea
B(u,v)  A(0) \is 
\sum_{\omega,\alpha} \,
  \bar{B}_{\omega-\alpha}(v)  \spc  B_{\omega-\alpha}(u) \;  \bar{A}_{\alpha}(0) A_{\alpha}(0).
\eea
Applying the exchange algebra (\ref{exchl})-(\ref{exchr}) for each chiral half, we find that\footnote{Here ${\cal R}_{\beta\smpc\delta}$ in fact depends on $\omega$ as indicated in eqns (\ref{rfour}) and (\ref{reight}).}
\bea
\label{ooexch}
B(u,v) \, A(0)  \is \sum_{\rm \omega,\beta,\delta}\, {\cal R}^{2\epsilon(u,v)}_{\beta\smpc \delta} \;  \bar{A}_{\omega-\beta}(0) A_{\rm \omega-\delta}(0) \;  \bar{B}_{\beta}(v) B_\delta(u),
\eea
with $\epsilon(u,v)$ as given in eqn (\ref{epsdef}), and where (taking (\ref{ooexch}) to act on $| M\rangle$)
\vspace{-1mm}
\bea
\label{reight}
{\cal R}^{2\epsilon}_{\beta\smpc \delta} \is \sum_\alpha\, {\rm R}^\epsilon_{\sca\scb}\fourj{\scm}{\omega}{\beta}{\alpha}\, {\rm R}^\epsilon_{\sca\scb}\fourj{\scm}{\omega}{\alpha}{\delta},
\eea
with $\epsilon = \pm 1$.  Here we used the reality relation $\bar{\cal R}^T = {\cal R}^\dag$ (which follows from the fact that $u$ and $v$ are real, and our assumption that $\hat{A}$ and $\hat{B}$ are hermitian) and the unitarity relation  (\ref{unir}).
Since  ${\cal R}^{0} = \mathbb{1}$ we recover  the result that $\hat{A}(0,0)$ and $\hat{B}(u,v)$  commute for $uv>0$. 

Confirming locality  is not the most exciting lesson. The more interesting conclusion is that we find an explicit formula for the  operator algebra in the time like region
\bea
\label{ooexcht}
B(u,v)\,  A(0)  \is\, \sum_{\rm \omega,\beta,\delta}\,  {\cal R}^{2\epsilon}_{\beta\delta} \;  \bar{A}_{\omega-\beta}(0) A_{\rm \omega-\delta}(0) \;  \bar{B}_{\beta}(v) B_\delta(u),
\eea
where $\epsilon\! =\!1$ inside the future  light-cone and $\epsilon\! =\! -1$ inside the past light-cone. This remarkable result, that the operator algebra
of any 2D CFT can be characterized via an exchange algebra of this form, was first pointed out by Rehren in \cite{Rehren}.  
It has its origin in  two basic characteristics of two dimensional physics, namely that the statistics operation (of interchanging the position of two operators) is a simple re-ordering, and secondly, that the statistics operator (the matrix that expresses the effect of the interchange) does not need to be a simple sign or a phase, but can take the form of a unitary matrix ${\cal R}$. The exchange algebra and the R-matrix comprise detailed dynamical information about the CFT.

Associativity of the exchange algebra implies that ${\cal R}$ must satisfy the Yang-Baxter equation. Let us briefly describe how this comes about. Suppose we add a third operator $C$ that acts in-between operator $A$ and $B$. We can then perform the exchange operation (here we suppress the positions in our notation)
\bea
B_{\omega-\gamma} \spc C_{\gamma-\alpha}\spc A_\alpha \li M\ra  = \sum_{\beta,\delta} \;  {\cal B}
\!\!\nspc\raisebox{1pt}{$\begin{array}{cc}\mbox{\scriptsize\spc$\beta$} \!\!&\!\!\! \mbox{\scriptsize$\delta$}\\[-2.25mm] \raisebox{1.25pt}{\scriptsize{$\alpha$}} \!\! &\!\!\!\! \raisebox{1pt}{\scriptsize$\gamma$} \end{array} $}\!  A_{\omega-\delta} \spc C_{\delta-\beta}\spc B_{\beta} \li M\ra.
\eea
The reordering process of going from the left-hand side to the right-hand side involves three elementary R-operations. However, these three operations can be performed in two different sequences
\bea
\includegraphics[scale=.55]{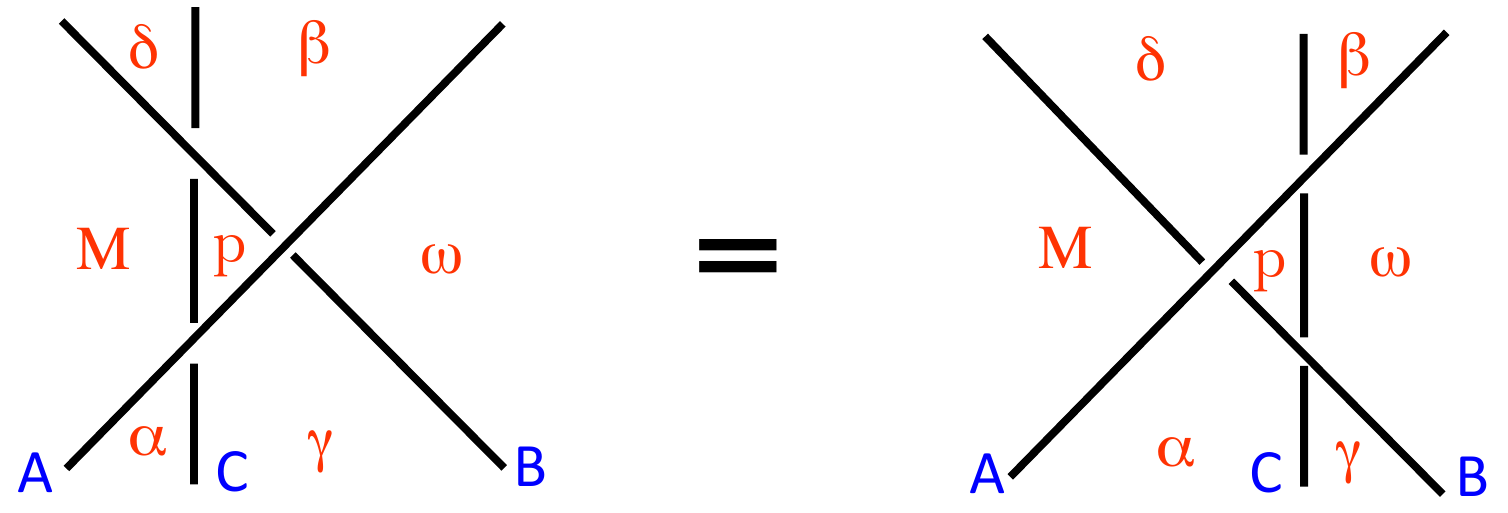} \nonumber
\eea
Hence the matrix ${\cal B}\!\!\nspc\raisebox{1pt}{$\begin{array}{cc}\mbox{\scriptsize\spc $\gamma$} \!\!&\!\!\! \mbox{\scriptsize$\delta$}\\[-2.25mm] \raisebox{1.25pt}{\scriptsize{$\alpha$}} \!\! &\!\!\!\! \raisebox{1pt}{\scriptsize$\beta$} \end{array} $}$ can be decomposed in two ways into a triple product of R-matrices.  The condition that  the two are equal gives the
familiar Yang-Baxter equation
\bea
\label{yb}
\sum_p \;  {\rm R}^\epsilon_{\sca\scc}\fourj{\scm}{\gamma}{\alpha}{p}\,
 {\rm R}^\epsilon_{\sca\scb}\fourj{p}{\omega}{\gamma}{\delta} \,
 {\rm R}^\epsilon_{\scb\scc}\fourj{\gamma} {\scm}{p}{\beta}  \is
\sum_p\; {\rm R}^\epsilon_{\scb\scc}\fourj{\omega}{\alpha}{\gamma}{p} \,
 {\rm R}^\epsilon_{\sca\scb}\fourj{\scm}{p}{\alpha}{\beta} \,
 {\rm R}^\epsilon_{\sca\scc}\fourj{\beta}{\omega}{p}{\gamma} .
\eea
This Yang-Baxter equation is a close cousin of the conformal bootstrap equations. It puts equally strong constraints on the spectrum and OPE coefficients of the CFT. But it is also equally hard to solve, especially for irrational CFTs without additional symmetries besides the Virasoro algebra.

Via the same arguments as before, we will make the reasonable assumption that for correlation functions between heavy states, we can approximate the discrete sums over states by continuum integrals, with a measure set by the Cardy entropy. Since the Cardy formula is universal, the resulting consistency equations for R, 
i.e. the  unitarity condition (\ref{unir})  and the Y-B equation (\ref{yb}), also take a universal form. Via this physical argument, we have mapped the mathematical problem of finding the R-matrix in some specific irrational CFT  to the more tractable problem of finding the R-matrix for Liouville theory. The latter problem has been solved.
Using the standard relation between the R-matrix and the fusion matrix F, we find that in this continuum approximation
\bea
{\rm R}^\epsilon_{\sca\scb}\fourj{\scm}{\omega}{\alpha}{\beta} \is e^{\epsilon i\pi\ell (\omega - \alpha -\beta)} 
\sixj{j_{\mbox{\scriptsize \sc a}}}{j_{\scm}}{j_\alpha}{j_{\mbox{\scriptsize \sc b}}}{j_\omega}{j_\beta}.
\eea
The expression for the quantum 6j-symbol is given in Appendix B.

To conclude, we see that the R-matrix has direct relevance for the  operator algebra between time-like separated operators in the full non-chiral CFT. 
The quantum group symmetry that characterizes the non-local operator algebra of the chiral CFT, that seems hidden in the non-chiral Euclidean CFT, becomes liberated and fully visible in the Lorentzian regime.

\newcommand{\Omegs}{{\mbox{\tiny $\Omega$}}}
\newcommand{\Omeg}{{\mbox{\footnotesize $\Omega$}}}

\section{Concluding Remarks}\label{conclusion}

Let us summarize our results. We have exhibited universal behavior of 2D CFT correlation functions and operators in the regime in which dual gravitational interactions dominate. We focussed on  the matrix element of two light fields $A(u_0,v_0)$ and $B(u_1,v_1)$ between two heavy states $\li M\ra$ and $\ri M+\omega\ra$. In the bulk this describes the scattering of two particles A and B inside the BTZ black hole geometry of mass $M$. We argued that if the CFT is irrational,  the conformal bootstrap equations in this regime approach that of Liouville CFT. This observation is sufficient to establish a detailed quantitative match between the  algebra of CFT operators and scattering amplitudes computed from 2+1 gravity. On both sides, the interaction between the particles is captured by an exchange algebra specified by an R-matrix.
Moreover,  the mathemetical problem of finding the R-matrix is soluble and identical on both sides of the holographic duality. The explicit answer is expressed as a 6j-symbol of the quantum group $U_q(\mathfrak{sl}(2,\mathbb{R}))$. 

\smallskip

\begin{figure}[hbtp]
\begin{center}
{\includegraphics[scale=.28]{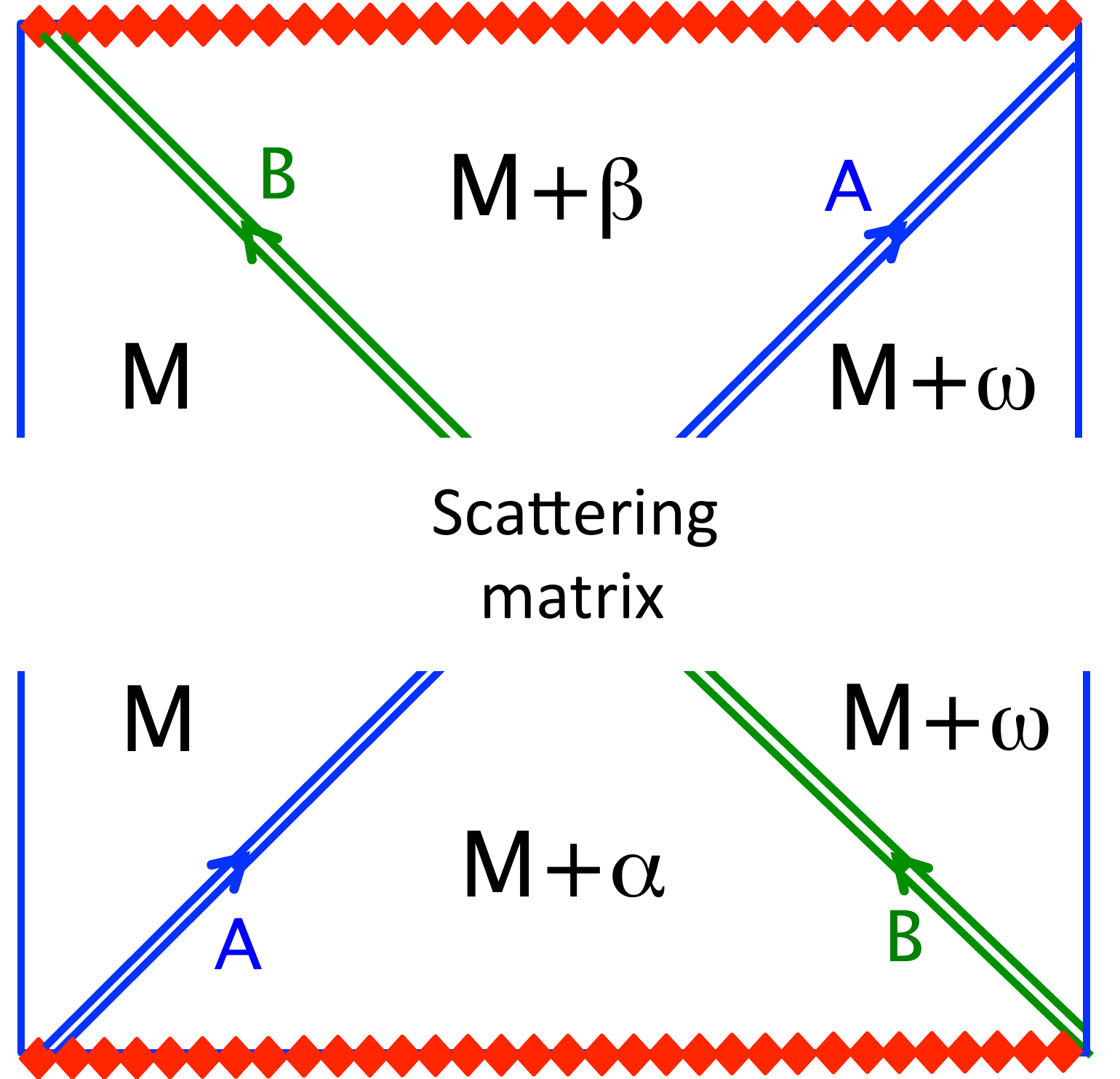}}\qquad \ \
\raisebox{20mm}{\Large $\leftrightarrow$}\qquad\ \
{\includegraphics[scale=.38]{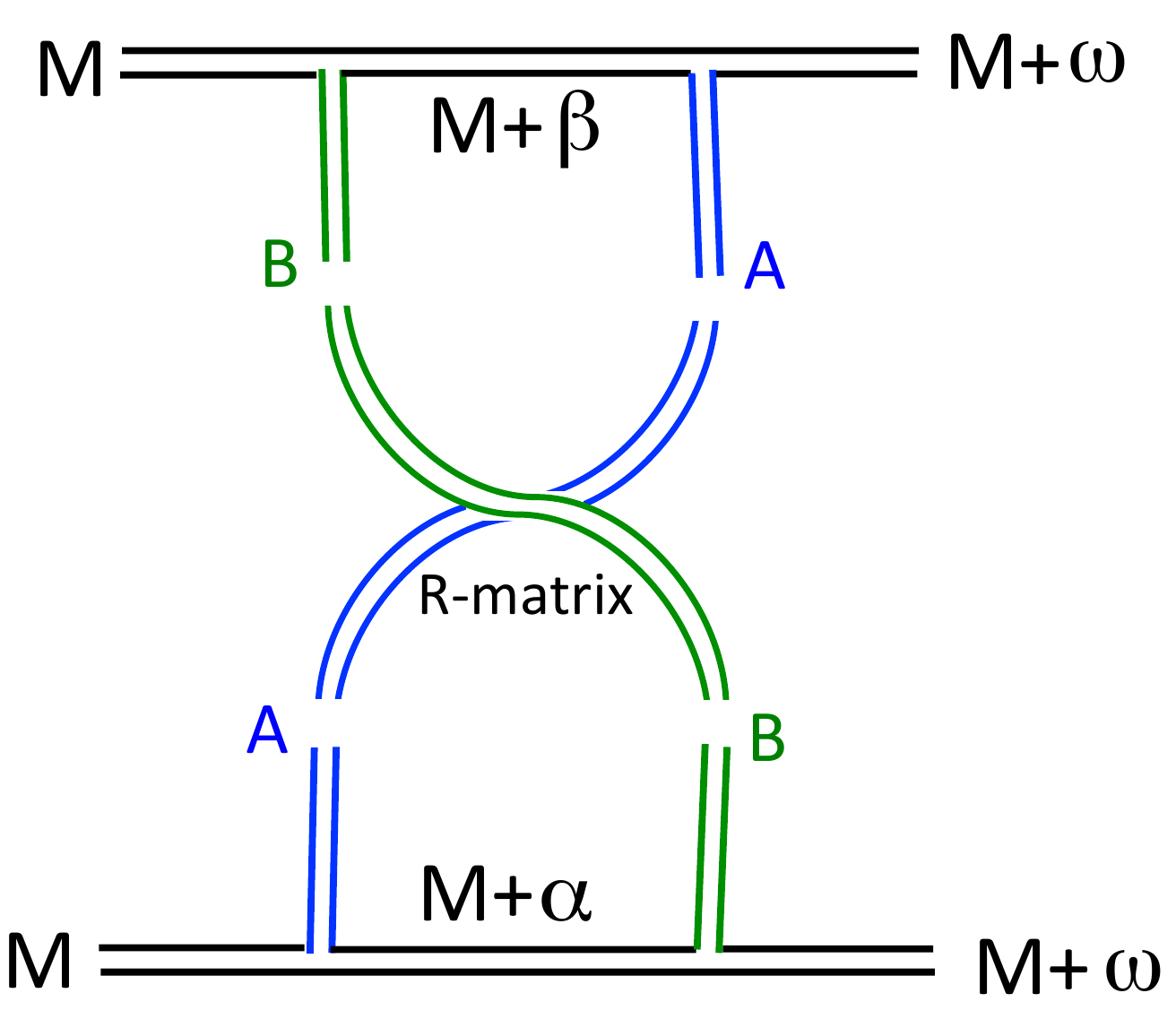}}

\caption{Summary of our calculation and identification. The gravitational scattering matrix (left) is described by the R-operation on conformal blocks (right). 
} 
\end{center}
\vspace{-0.5cm}
\end{figure} 
We end with listing some possible general lessons, open questions and future directions.
\begin{center}
{\it Possible generalizations}
\end{center} 

\vspace{-2mm}
We have  focused on  Virasoro conformal blocks and gravitational interactions. However, it is in principle straightforward to include gauge interactions by considering CFTs with additional continuous symmetries and extended chiral algebras. This would correspond to adding massless gauge and/or higher spin fields in the bulk.

A more interesting question is how to include the backreaction and interaction due to massive quantum fields in the dual bulk. This requires quantifying the difference between a given irrational CFT  and Liouville theory, by isolating the imprint on the conformal bootstrap equations due to the presence of a given light primary field.  
Our detailed dictionary for pure gravity suggest that a gravity+matter system in the bulk should somehow be describable as a Liouville CFT + matter system on the boundary, even if, perhaps paradoxically, an irrational CFT with a discrete spectrum  has fewer states than Liouville theory with a continuum spectrum. 

In our study we made critical use of the special properties of 2D CFT. So it may seem unlikely that any of our findings can be generalized to higher dimensions. 
On the gravity side, on the other hand, the argument goes through without much change. Exponential redshifts  and the resulting appearance of gravitational shockwaves as dominant long distance effects are a common property of black holes in any dimensions. As explained in  \cite{Kiem}, their effect can be taken into account by means of an exchange algebra between infalling and outgoing fields. In 3+1-dimension, it takes the form
\bea
\label{exchange}
\phi_{out} ( u , \Omeg_1 ) \phi_{in} ( v , \Omeg_2 )
\is e^{ i  f(\Omegs_{12} ) e^{\kappa(u - v)}
\partial_v \partial_u}
\phi_{in} (v , \Omeg_2 ) \phi_{out} ( u , \Omeg_1 )
\eea
where $u$ and $v$ now denote longitudinal  light-cone coordinates $u,v = r\pm t$ in the bulks, $\Omeg = (\theta,\varphi)$ are sphere coordinates,  and $f(\Omeg_{12})$ denotes a suitable Green function on the horizon, and $\kappa = 1/4M$ is the surface gravity. The shockwave interaction dominates over a sizable kinematic range, i.e. as long as the infalling and outgoing particle do not come too close to each other to produce excited string states or create their own black hole, and as long as the collision energy remains sufficiently smaller than $M$. So we can now make a plausible conjecture: for large $N$ CFTs with a gravity dual, there should exist a universal regime, similar to the one studied here, where the  algebra of CFT operators takes the form of an exchange algebra (\ref{exchange}).

\vspace{-1mm}

\begin{center}
{\it Left-right (dis)entanglement}
\end{center} 

\vspace{-2mm}

 Both for rational CFT and irrational CFTs with only conformal symmetry, there is in general a unique pairing (up to automorphisms of the fusion algebra) between left- and right-moving chiral sectors. This coupling between left and right-movers plays a central role in our argumentation. Its presence is dictated, and its form is constrained,  by the conditions of locality and modular invariance. However, it introduces some a rather subtle puzzles.
 
The left-right coupling can be quantified via the entanglement that is injected into a small space interval ${\cal I}$, by acting with a local operator $\hat{A}$ inside ${\cal I}$. The chiral degrees of freedom in $\hat{A}$ move in opposite directions. So one can measure their mutual information by computing the increase in the entanglement entropy of ${\cal I}$ when a chiral signal passes the boundary $\partial{\cal I}$. This computation was performed for rational CFTs in \cite{Takayanagi}. It was found that the left-right entanglement of the local field ${\hat A}$ is equal to the log of the  quantum dimension of the associated chiral sector  $[ A]$
\bea
S_{\hat{A}} \! \is \! - \log\bigl(F_{00}(A)\bigr) = 2 \log S^A_0 = S_{\rm top}(A),
\eea
with $F_{00}(A) = F_{00}\fourj{A}{A}{A}{A}$.
For irrational CFT this calculation appears to be somewhat more involved \cite{TomQuench} and the notion of a quantum dimension does not have an obvious generalization. However, for Liouville CFT there is a close analog. One can define $F_{00}(A)= \lim_{\beta\to 0} \beta^2 F_{\beta\beta}(A)$, and use the identity $F_{00}(A)  = 1/(S_{0}^A)^2$ with $S_0^A = \sinh(i\pi b p_A) \sinh(i\pi p_Ab^{-1})$ the Plancherel measure of $U_q(\mathfrak{sl}(2,\mathbb{R}))$. Using this formula for the left-right entropy
for the case that $\hat{A}$ is a heavy field with scale dimension $\Delta_A>c/12\gg1$, we obtain the result that the left-right entanglement entropy saturates the Cardy entropy
\bea
S_{\hat{A}} \is S_{\rm Cardy}(\Delta_A).
\eea
This result seems surprising at first, but it is physically reasonable. For irrational  CFTs with a random spectrum of primary  conformal weights, without any integer gaps,\footnote{A CFT primary spectrum with integer gaps would indicate the presence of exact (continuous) symmetry.}
 the condition that $\Delta_{\rm A} -\bar{\Delta}_{\bar A} \in \mathbb{Z}$  in fact boils down to the requirement that $\Delta_{\rm A} =\bar{\Delta}_{\bar A}$. So the mutual information between the left and right sectors then indeed equals the Cardy entropy. 
 
There is some tension between this conclusion and the standard view that the Cardy entropy counts the microscopic entropy of the actual CFT spectrum. 
The Cardy formula (\ref{cardy}) looks like a sum of left and right-moving entropy, which suggests that the two sectors are completely decoupled. How can one reconcile this with the fact that there should also be a comparably large entanglement entropy between the two chiral sectors?

A pragmatic point of view, that leaves both results intact, is to consider the CFT Hilbert space on an infinite real line, or equivalently, on a line segment. In this case, the left and right-movers are truly decoupled
and the Cardy formula correctly quantifies the microscopic spectral density of the Hilbert space. Non-chiral local operators, however, should still be left-right entangled --  even if the chiral components each start living their own life.
 
\vspace{0mm}

\begin{center}
{\it Scrambling and Thermalization}
\end{center} 

\vspace{-2mm}

It would clearly be very interesting if our techniques could shed some light on scrambling and thermalization in the 2D CFT \cite{ShSt}. Our set up is designed so that, on the bulk side, gravitational backreaction effects are dominant (near the black hole horizon), which means that thermalization effects dominate on the CFT side (in the regime with a large spectral density). We have given an intuitive description of the underlying mechanism in section 1.4. Our explanation of how bulk dynamics is encoded in the CFT may point to a possible proof that irrational CFTs with a large central charge are indeed fast scramblers.

As a  more modest first step, it would be worthwhile to work out how the exhange algebra structure harmonizes with the eigen state thermalization hypothesis (ETH) \cite{ETH}, i.e. the expectation that general expectation values
$\Psi_{{\!}_{1\spc ...\spc n}}(t_1, \ldots, t_n) = \la M\ri {\cal O}_1(t_1)\,\ldots \, {\cal O}_n(t_n) \li M \ra$  between sufficiently excited  energy eigenstates $\li M\ra$ behave, to a good approximation, like thermal expectation values. Since the R-matrix tells us about how to re-order operators, it is natural to try to test the ETH by writing a set of TBA like equations 
of the form 
\bea
\label{exchagain}
\Psi_{{\!}_{1\smpc ..\smpc ij\smpc..\smpc n}}(t_1, \spc ..\spc , t_i,t_j, \spc ..\spc, t_n) \is {\cal R}_{ij}\,
\Psi_{{\!}_{1\smpc ..\smpc ji\smpc ..\smpc n}}(t_1, \spc ..\spc , t_j,t_i, \spc ..\spc, t_n),\\[3mm]
\Psi_{{\!}_{1\spc ..\spc n-1,n}}(t_1, \spc ..\spc ,t_{n-1}, t_n) \! & \! \raisebox{4pt}{$\mbox{\small $?$}\atop\mbox{$\simeq$}$}& 
\Psi_{{\!}_{n\smpc 1..n-1}}(t_n\nspc +\nspc i\beta, t_1, \spc ..\spc , t_{n-1} ),
\label{kms}
\eea
and ask if the approximate  KMS condition (\ref{kms}) is   compatible with the exchange property (\ref{exchagain}). It seems likely that the interplay between thermodynamics and the Virasoro modular bootstrap will be informative.

\vspace{-0mm}

\begin{center}
{\it Going into the bulk}
\end{center} 

\vspace{-2mm}

The main motivation for our study  is to gain insight into the holographic reconstruction of bulk physics from CFT data. As illustrated by the firewall debate \cite{AMPS}, there are deep open questions about how classical bulk geometry arises and whether it persists into the black hole interior. The basic conceptual issues are already present for AdS${}_3$/CFT${}_2$. Given our explicit map between gravitational particle dynamics in AdS${}_3$ and  CFT${}_2$ data, it seems that there is room for progress on this front. 

In the old connection between 2D CFT and 2+1-D topological QFT, there are two main elements to the dictionary. For a topological theory defined on a
2+1-D space-time with a boundary (like AdS${}_3$) the CFT describes the edge state dynamics. This relation follows the usual holographic dictionary: the Hilbert space of the 2D and 3D theories can be identified, and the CFT really lives on the boundary. However, there is also a second component of the dictionary:  the wavefunctions of the 2+1-D TQFT are given by conformal blocks of the 2D CFT. This is the relation that relates the braid statistics of the 2+1-D theory with the exchange algebra of the CFT.
In this paper we have made use of both types of identifications.

In AdS/CFT, the bulk theory is not topological. Nonetheless, if it has a mass gap (i.e. if the CFT has no additional continuous symmetries), there is a universal long distance regime governed by a TQFT, given by pure 2+1 Einstein gravity. The Hilbert states of 2+1 gravity are obtained by quantizing the physical phase space, or as wavefunctionals $\Psi[g]$ of the spatial metric. In the latter case, the wave functionals must satisfy physical state conditions:  reparametrization invariance and the Wheeler-DeWitt equation. It has been known for some time \cite{HV}\cite{Freidel} that there is a precise map between the correlation functions of the CFT,  $\Psi_{\rm CFT} [\smpc A_\pm,z_i] = \la \, {\cal O}_1(z_1,\zbar_1)\ldots \,{\cal O}_1(z_n,\zbar_n) \ra$, viewed as functionals of the 2D metric $g^{\rm CFT}_{\mu\nu} = A^+_{(\mu} A^-_{\nu)}$, and solutions to the 2+1-D WDW equation.
Using that $\Psi_{\rm CFT}$ satisfies the Virasoro Ward identity and Weyl anomaly equation, one can explicitly show that the WDW wavefunction  $\Psi_{\rm WDW}[g,z_i]$ is obtained via simple Legendre transform \cite{Freidel}
\bea
\label{wdwmap}
\Psi_{\rm WDW}[\smpc e_\pm,z_i] \is \int[dA_\pm] \, e^{\frac 1 {\hbar}\nspc \int (A_\pplus\nspc -\spc e_\pplus)\wedge(A_\mmin\nspc -\spc e_\mmin)} 
\, \Psi_{\rm CFT} [\smpc A_\pm ,z_i].
\eea
This transform implements a change in polarization from the two $SL(2,\mathbb{R})$ connections $A^a_\pm = \omega^a \pm \frac 1 \ell e^a$ to the spatial metric variable $g_{\mu\nu} = \eta_{ab} e^a_{(\mu} e^b_{\nu)}$ used in the ADM canonical formalism. While it seems evident that this result must play a central role in the holographic reconstruction of the bulk, it is not yet clear to us how it connects with the standard GKPW dictionary \cite{AdSCFT}.

\vspace{0mm}

\begin{center}
{\it Passing through a horizon}
\end{center} 

\vspace{-2mm}

Consider the state $\li M \ra  = {\cal O}_M(0)\li 0\ra$ created by a heavy operator with $\Delta_M> c/12$.  This state can be viewed as obtained by performing the Euclidean CFT path integral on a disk $D$ with the operator ${\cal O}_M(0)$ placed at the origin $z=0$. We can associate a semi-classical 2D constant curvature metric to this state \cite{Nati}, via the Liouville dictionary explained in Sect. 4.1.  It takes the form (\ref{natimetric}), which is identical to the metric on a spatial slice of the BTZ black hole metric (\ref{btzadstwo}).
Surprisingly, even though the state is defined in the one-sided CFT, this semi-classical Liouville  metric (\ref{natimetric}) describes a hyperbolic cylider with {\it two} asymptotic regions, one at $\log z\bar z = \pi \ell/R$, which is where we  can choose to locate the boundary of the disk $D$, and one at $\log(z\zbar) = 0$.  The question is: what is special about the intermediate critical radius $\log(z\zbar)_{\rm crit} = 
{\pi\ell}/{2R}$ corresponding to the location of the BTZ horizon? 

Here's a simple proposal. Consider the OPE between a light operator $\hat{A}$ and the heavy operator ${\cal O}_M(0)$ (c.f. eqn (\ref{aalpha}))
\bea
\label{OPE}
\hat{A}(z,\zbar) \li M \ra \is  \sum_{\alpha}\, \frac{f^A_\alpha }{(z\bar{z})^{\ell\alpha/2}\!\!\!\!\!\!\!\!}\;\;\;\;\; \li M+\alpha\ra,
\eea
where we set $\Delta_A=0$ and  used that $\Delta_M = \frac 1 2 \ell M$, etc. Here $f^A_\alpha$ denotes the OPE coefficient. 
\begin{figure}[t]
\begin{center}
\raisebox{.8cm}{\includegraphics[scale=.38]{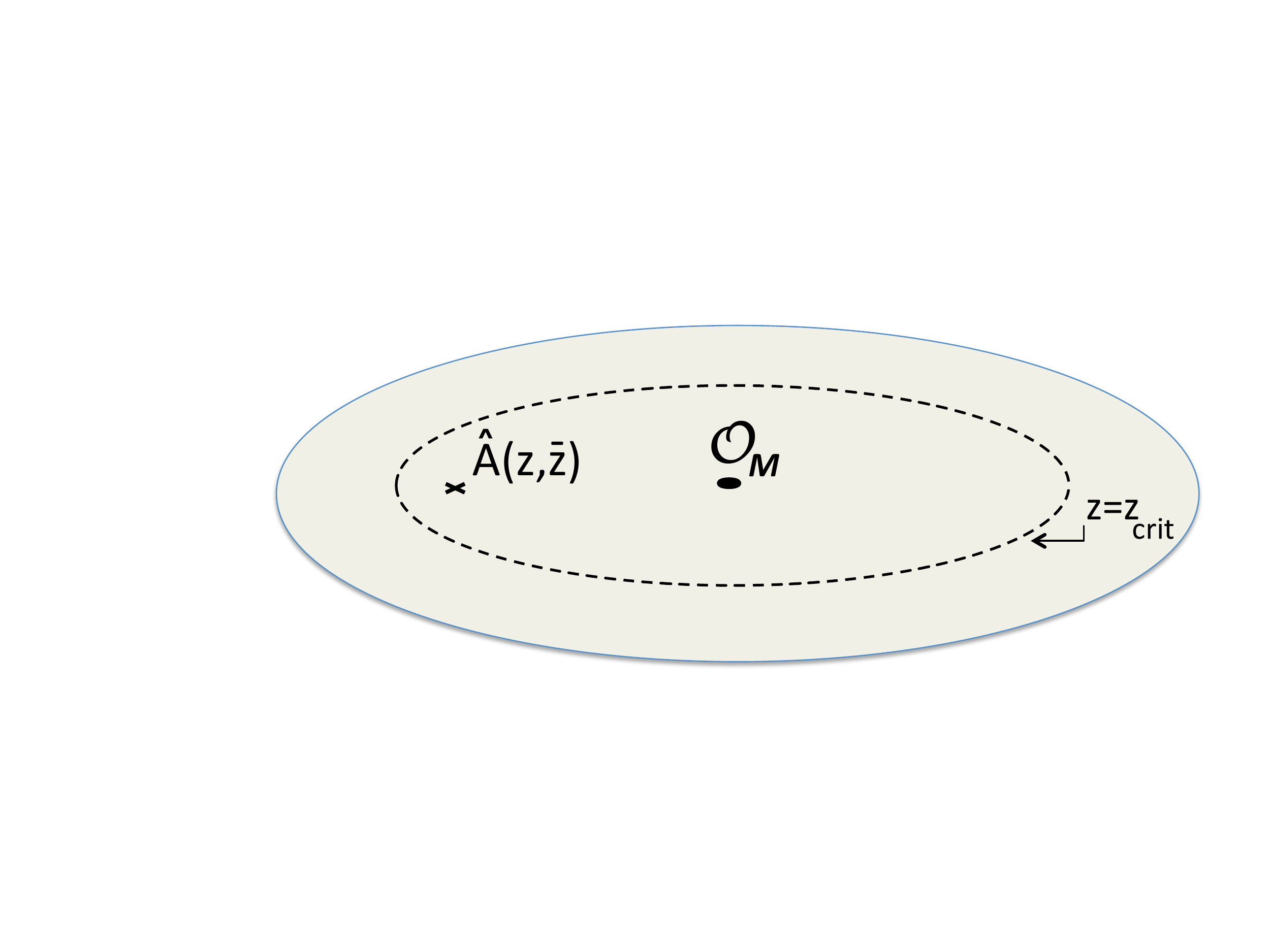}}
\vspace{-5mm}
\caption{The OPE between the light operator $\hat{A}(z,\zbar)$ and the heavy operator ${\cal O}_M(0)$ receives a dominant contribution from some final state ${\cal O}_{M+\alpha}$. The value of $\alpha$ depends on the distance between the two operators. Inside the critical radius $z= z_{\rm crit}$,  $\alpha$ is negative. Outside this radius, $\alpha$ is positive.}
\end{center}
\vspace{-5mm}
\end{figure} 
We can ask: what is the dominant term in  (\ref{OPE})? Does it occur for positive or negative $\alpha$, i.e. does the operator increase or decrease the energy of the BTZ state $\li M\ra$? To find the dominant term, we must look at the transition probability. Using Fermi's golden rule
\bea
\label{fermi}
T_{M\to M+\alpha} \is 2\pi \rho(M\! +\nspc \alpha) \, |f^A_\alpha|^2\, |z\zbar|^{-\ell \alpha},
\eea
where $\rho(M+\alpha)$ denotes the spectral density of the final state, given by the Cardy formula. 

The saddle point  value of $\alpha$, which maximizes $T_{M\to M+\alpha}$, depends on the radial location $z\zbar$. When $z\zbar$ is too small, the last factor of (\ref{fermi}) dominates and favors $\alpha$ to be arbitrarily negative. If $z\zbar$ is not too small, but still smaller than some critical radius, there  is some finite negative saddle point value for $\alpha$. So the operator $\hat{A}$ lowers the energy by a finite amount. As we argued in section 1.4, this means that $\hat{A}$ should be thought of as living behind the horizon: its phase information becomes unobservable after phase averaging over the initial states $\li M\ra$. On the other hand, when $z\zbar$ becomes larger than this critical radius, the last factor of (\ref{fermi}) favors a positive saddle point value for $\alpha$. Our proposal is that the horizon radius  $\log(z\zbar)_{\rm crit} = 
{\pi\ell}/{2R}$
 corresponds to the critical radius where the sign of $\alpha$ flips from positive (outside)  to negative (inside).

\bigskip
\bigskip

\noindent
{\large \bf Acknowledgements}

We thank Tudor Dimofte, Daniel Harlow, Thomas Hartman, Emil Martinec, Nathan Seiberg, Douglas Stanford, J{\ddoto}rg Teschner, and Erik Verlinde  for helpful discussions.
This work is supported by NSF grant PHY-0756966. L.M. is supported by a NSF GRFP under grant DGE-1148900.

\appendix

\noindent
\section{Some useful formulas}\label{appendix}

The Murakami-Yano formula \cite{Murakami-Yano} for the hyperbolic volume of a tetrahedron with dihedral angles $l_i$, $i=1,..,4,\alpha,\beta$
can be compactly written in terms of the auxiliary variables \cite{Rosly}
\bea
{\cal V}_1 \! \is\! l_{\alpha} + l_1 + l_2 \qquad  \qquad {\cal E}_1 \, = \, l_1 + l_2 + l_3 + l_4 \nonumber \\
{\cal V}_2 \! \is\! l_{\alpha} + l_3 + l_4\qquad  \qquad {\cal E}_2 \, = \, l_{\alpha}+ l_{\beta} + l_1 + l_3\nonumber \\[-3.2mm]\\[-3.2mm]
{\cal V}_3 \! \is\! l_{\beta} + l_1 + l_4 \qquad \qquad {\cal E}_3 \, = \, l_{\alpha}+ l_{\beta} + l_2 + l_4 \nonumber \\
{\cal V}_4 \! \is \! l_{\beta} + l_2 + l_3\qquad \qquad  {\cal E}_4 \, = \, 0\nonumber
\eea
\vspace{-12mm}

\noindent
as
\vspace{-3mm}
\bea
\label{tvolume}
 {\rm Vol}\Bigl(\spc T\Bigl[\mbox{\fontsize{10pt}{.5pt}$\begin{array}{ccc} \! 1\! \! &\! {2} \! &\!\nspc \raisebox{1.25pt}{\fontsize{10pt}{.5pt}$\alpha$} \! \\[-1.5mm] \! 3\! \! &\! 4 \! &\!  \nspc \raisebox{0pt}{\fontsize{10pt}{.5pt}$\!\beta$}
\!\! \end{array}$}\Bigr]\spc \Bigr)
 \is  \sum_{i=1}^4 \Bigl({\rm Li}_2\bigr( z_+ e^{{\cal V}_i/2}\bigl) \spc -\spc {\rm Li}_2\bigr( z_- e^{{\cal V}_i/2}\bigl)
\spc -\spc {\rm Li}_2\bigr( z_+ e^{{\cal E}_i/2}\bigl) \spc +\spc  {\rm Li}_2\bigr( z_- e^{{\cal E}_i/2}\bigl)\Bigr),\qquad
\eea
where 
\vspace{-5mm}
\bea
{\rm Li}_2(z) = \int_z^0 \! dz\,  \frac{\log(1-z)} z
\eea 
denotes the dilogarithm function, and $z_+$ and $z_-$ are the two solutions to the quadratic equation (note that the $z^0$ and $z^4$ terms cancel out between  both sides)
\bea
\label{quadratic}
\prod_{i=1}^4 \bigl(1 - z e^{{\cal V}_i/2}\bigr) \, = \,
\prod_{i=1}^4 \bigl(1 - z e^{{\cal E}_i/2}\bigr) .
\eea

The above formula arises as the semi-classical limit of the quantum 6j symbols of $U_q(\mathfrak{sl}(2,\mathbb{R}))$, first found by Ponsot and Teschner \cite{Ponsot}. It is expressed in terms of the
double sine function $s_b(x)$ and quantum dilogarithm $e_b(x)$. These are defined as
\bea
s_b(z) \is \prod_{m,n\geq 0} \frac{mb+nb^{-1} + \frac Q 2 - i z}{mb+nb^{-1} + \frac Q 2 + i z}, \qquad \qquad
e_b(z) = e^{\frac {iQ}2 {z^2}} s_b(z)
\eea
Both functions are symmetric under $b\to b^{-1}$.
It is useful to introduce the double Sine function  via $S_b(x) = s_b\bigl(ix-\textstyle \frac i 2 Q\bigr).$
It satisfies $S_b(x) = S_{b^{-1}}(x)$ and $S_b(x) S_b(Q - x) = 1$.
If we write $\alphaj_{ijk} = \alphaj_i + \alphaj_j + \alphaj_k$ and $\alphaj_{ijkl} = \alphaj_i + \alphaj_j + \alphaj_k + \alphaj_l$, then the quantum 6j symbols are given by the formula
\bea
\label{qsixjay}
\sixj{\alphaj_1}{\alphaj_2}{\alphaj_\alpha}{\alphaj_3}{\alphaj_4}{\alphaj_\beta}_b
\is \Delta(\alphaj_{\alpha}, \alphaj_2, \alphaj_1) 
\Delta(\alphaj_{4}, \alphaj_3, \alphaj_{\alpha}) 
\Delta(\alphaj_{\beta}, \alphaj_3, \alphaj_2) 
\Delta(\alphaj_{4}, \alphaj_{\beta}, \alphaj_1) \\[0mm]
& & \times \int_{\mc{C}} \! du \, S_b(u - \alphaj_{12\alpha}) 
S_b(u - \alphaj_{\alpha 3 4}) 
S_b(u - \alphaj_{2 3 \beta}) 
S_b(u - \alphaj_{1 \beta 4})  \nonumber \\[1mm]
& & \qquad \qquad \times S_b(\alphaj_{1234} - u) 
S_b(\alphaj_{\alpha \beta 1 3} - u) 
S_b(\alphaj_{\alpha \beta 2 4} - u) 
S_b(2Q - u)\nonumber\\[4mm]
& &  \Delta(\alphaj_3, \alphaj_2, \alphaj_1) = \Bigl( \frac{S_b(\alphaj_{123} - Q)}{\prod_{i=1}^3S_b(\alphaj_{123} - 2\alphaj_i)}\Bigr)^{1/2}.\eea 
The integral is defined for $\alphaj_k \in Q/2 + i \mbb{R}$ and a contour $\mc{C}$ which lies on the real axis in the interval $[\frac 3 2 Q, 2Q]$ then proceeds to  $2Q + i\infty$. 

Defining the Plancherel measure via (with $j = \frac{Q}{2} + ip$)
\bea\label{plancherel2} & & d\mu(j) = \rho(j)\, d\smpc j \qquad \qquad
 \rho(j)\! \equiv \! 4\sinh(2\pi b\smpc p)\sinh(2\pi b^{-1} p) .  \eea
the quantum 6j-symbols of $U_q(\mathbb{sl}(2,\mathbb{R})$ satisfy the required orthogonality conditions
\bea
\int
\! d\mu(p)\,
\sixj{j_1}{j_2}{j_p}{j_3}{j_4}{j_\alpha}_q^*
\sixj{j_1}{j_2}{j_p}{j_3}{j_4}{j_\beta}_q
\is \frac 1 {\rho(j_\alpha)
} \, \delta(j_\alpha \nspc - j_\beta),
\eea 
and polynomial equations, such as the pentagon equation 
\bea
 \int \!d\mu(p)  \,
\sixj{j_1}{j_2}{j_\alpha}{j_3}{j_\beta}{j_p}_{\! q}
\sixj{j_1}{j_p}{j_\beta}{j_4}{j_\epsilon}{j_\delta}_{\! q}
\sixj{j_2}{j_3}{j_p}{j_4}{j_\delta}{j_\gamma}_{\! q}
\is \sixj{j_\alpha}{j_3}{j_\beta}{j_4}{j_\epsilon}{j_\delta}_{\! q}
\sixj{j_1}{j_2}{j_\alpha}{j_\gamma}{j_\epsilon}{j_\delta}_{\! q},\eea
etc.

In the semi-classical limit $b\to 0$, the  double Sine function behaves as
\bea
S_b\Bigl(\frac \nu{2\pi b}\Bigr) & \simeq & e^{- \frac{i}{2\pi b^2}(\frac 1 4 \nu^2-\frac \pi 2 \nu + \frac 1 6 \pi^2)} \exp\Bigl(\frac{i}{2\pi  b^2}{\rm Li}_2(e^{i\nu})\Bigr).
\eea
Upon inserting this result into the expression (\ref{qsixjay}) for the quantum 6j-symbol, one can do the integral via the stationary phase approximation \cite{TeschnerV}. The stationary phase condition takes the form (\ref{quadratic}). One finds that for small $b$
\bea
\sixj{\alphaj_1}{\alphaj_2}{\alphaj_\alpha}{\alphaj_3}{\alphaj_4}{\alphaj_\beta}_q \! & \simeq & \! \exp\Bigl\{\frac{i}{2\pi b^2} {\rm Vol\bigl( T \Bigl[\raisebox{1pt}{\fontsize{9pt}{.5pt}$\begin{array}{ccc} \! 1\! \! &\! {2} \! &\!\nspc \raisebox{.25pt}{\footnotesize$\alpha$}  \! \\[-1.5mm] \! 3\! \! &\! 4 \! &\!  \nspc 
\raisebox{-.25pt}{\footnotesize$\beta$} 
\!\! \end{array}$}\Bigr]\spc 
\bigr)}\Bigr\},
\eea
where $T$ is the hyperbolic tetrahedron with dihedral angles $l_i = 2\pi b j_i$.

It is a straightforward calculation to specialize the formula for the quantum 6j-symbol to the kinematic regime of interest, and verify that it gives the scattering phase $S_{\alpha\beta}$ of the gravitational shock wave interaction described in the Introduction. Setting the masses of the two particles A and B to zero, we start from the exact Liouville expression 
\bea
{\cal R}_{\alpha\beta} \is e^{\pi i (\Delta_\omega - \Delta_\alpha - \Delta_\beta)}\left\{ \begin{array}{ccc} \!\! j_M \! &\! j_\omega \! &\!   \raisebox{-1pt}{\small 0} \!\! \\[.5mm]\! \!j_\alpha\! &\! j_\beta\! &\!  \raisebox{-1pt}{\small 0}\!\! \end{array}\right\}_{\! b}.
\eea
Applying the geometric dictionary and taking the semi-classical limit  gives
\bea
{\cal R}_{\alpha\beta}^2 \, =  &\!\!\! & \!\!\! \!\!\!\!\exp\bigl(\mbox{\large $\frac{i}{\hbar}$}\spc S_{\alpha\beta}\bigr), \qquad \quad
S_{\alpha\beta} = {\rm Vol}_{\cal T}(l_M,l_\omega, l_\alpha,l_\beta)\quad \\[3.5mm]
l_\omega = l_M+ \hbar\omega/\kappa,\!  \! & &  \qquad\ \; l_\alpha = l_M + \hbar \alpha/\kappa, \qquad \quad \ \; l_\beta = l_M + \hbar \beta/\kappa\quad \\[3mm]
\hbar = \pi b^2  = 4\pi/\ell, \! & &  \qquad  l_M/2\pi  = R/\ell = \kappa \smpc \ell  \ \qquad \qquad  R^2 = 8M \ell^2.\quad
\eea
We see that in the $\hbar \to 0$ limit, ${\cal T}(l_M,l_\omega, l_\alpha,l_\beta)$ approaches an ideal tetrahedron for four identical and two trivial tetrahedral angles. The Murakami-Yano formula simplifies for an ideal tetrahedron \cite{Murakami-Yano}, because one of the solutions $z_\pm$ to (\ref{quadratic}) is equal to 1. Using this simplification, it becomes an easy calculation to extract the leading order $\hbar\to 0$ limit
\bea
\label{semiclasstwo}
 \frac{1}{\hbar} S_{\alpha\beta} \! & \! \simeq \! & \! 
\frac{1}{\kappa}\Bigl\{ \spc  \alpha \log \alpha \spc + \spc \beta\log \beta
- ( \omega\nspc -\nspc \alpha) \log(\omega\nspc -\nspc \alpha) - \spc (\Et\nspc -\nspc\Ef) \log(\Et\nspc -\nspc \Ef) \nonumber\\[2mm] & &  \qquad  -\;  (\Ez + \beta - \Et\smpc ) \log\Bigl(
   2 \sinh\bigl(\mbox{\large $\frac{\pi R}{\ell}$}\bigr) \smpc (\Ez+ \Ef  - \Et \smpc) \Bigr)\Bigr\}.
\eea
This formula matches with the eqn (\ref{sabshock}) derived from the shock wave geometry.

\addtolength{\baselineskip}{-.3mm}

\end{document}